\documentclass[11pt,letter]{article}
\usepackage[top=1in, bottom=1in, left=1in, right=1in]{geometry}
\usepackage{fancyvrb}
\usepackage[utf8]{inputenc}
\usepackage{times}
\usepackage{cite}
\usepackage{calculator}
\usepackage{dsfont}
\usepackage{amsmath}    					
\usepackage{amsthm}                 			
\usepackage{amssymb}
\usepackage{graphicx}
\usepackage[inline]{enumitem}
\usepackage{booktabs}
\usepackage{longtable}
\usepackage[usenames,dvipsnames]{xcolor}
\usepackage{subfigure}  	 				
\usepackage[colorlinks=true, citecolor=blue, linkcolor=blue]{hyperref}       
\usepackage{makecell}
\usepackage{comment}
\usepackage{tablefootnote}

\usepackage{listings}
\usepackage{fancyhdr}
\fancypagestyle{plain}{
  \fancyhead{}
  
}

\definecolor{gray75}{gray}{0.75}

\newcommand{\mc}[1]{\makecell{#1}}

\theoremstyle{definition}
\newtheorem{defn}{Definition}[subsection]
\newtheorem{ex}{Example}[subsection]

\begin{document}
\vspace{-0.75in}
\begin{titlepage}

\noindent
\begin{Large}
\begin{trivlist}
\centering 
\item \textbf{The 2017 ISO New England System Operational Analysis and Renewable Energy  Integration Study (SOARES) }\\
\end{trivlist}
\end{Large}
\vspace{1in}

\noindent
\begin{Huge}
\begin{trivlist}
\centering 
\item \textbf{Final Journal Paper Manuscript}\\
\end{trivlist}
\end{Huge}
\vspace{1in}

\noindent
\begin{center}
\includegraphics[width=5.5in]{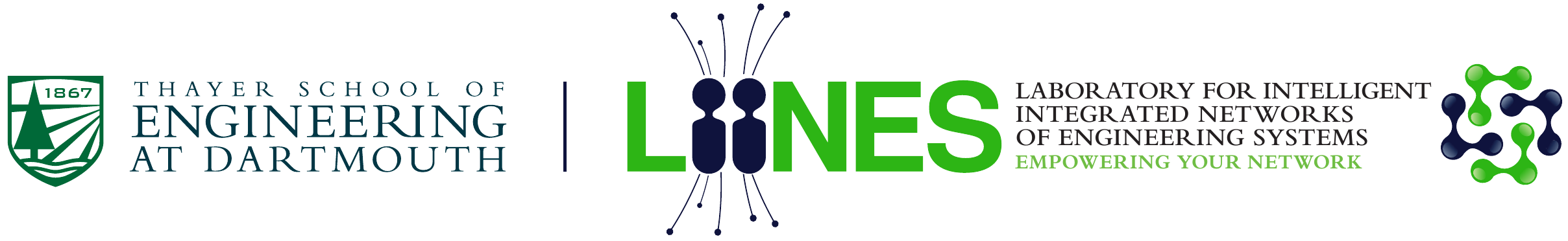}
\end{center}
\vspace{1in}

\noindent
\begin{Large}
\begin{trivlist}
\centering 
\item \textbf{by: Aramazd Muzhikyan, Steffi Muhanji, Galen Moynihan, Dakota Thompson, Zachary Berzolla, Amro M. Farid}
\end{trivlist}
\end{Large}
\vspace{1in}

\noindent
\begin{Large}
\begin{trivlist}
\centering 
\item \textbf{Modified:  \today}\\
\end{trivlist}
\end{Large}

\end{titlepage}
\newpage

\renewcommand\baselinestretch{1.1}

\renewcommand\thepage{\roman{page}}

\tableofcontents



\newpage
\setcounter{page}{1}
\renewcommand{\thepage}{\arabic{page}}

\section*{Executive Summary}
The bulk electric power system in New England is fundamentally changing. The representation of nuclear, coal and oil generation facilities is set to dramatically fall, and natural gas, wind and solar facilities will come to fill their place. The introduction of variable energy resources (VER) like solar and wind, however, necessitates fundamental changes in the power grid's dynamic operation. Such units introduce greater uncertainty and must be accurately forecasted. They also introduce greater intermittency and therefore require greater quantities of operating reserve. These new power system dynamics and their impacts on ISO New England's (ISO-NE) operations need to be systematically and rigorously assessed. To that end, ISO New England has launched the 2017 System Operational Analysis and Renewable Energy Integration Study (SOARES) as a means of assessing several scenarios of varying resource mixes to determine the impact on the load-following, ramping and regulation reserves. These scenarios were designed in consensus with ISO-NE stakeholders and reflect a set scenarios for which stakeholders requested deeper analysis but do not necessarily reflect ISO-NE's prediction of the future New England electric power system. Given their extensive publications on the topic, ISO New England has selected the Laboratory for Intelligent Networks of Engineering Systems (LIINES) at the Thayer School of Engineering at Dartmouth to conduct the study.

The heart of the project's methodology is a novel, but now extensively published, holistic assessment approach called the Electric Power Enterprise Control System (EPECS) simulator. Most fundamentally, the EPECS methodology is \emph{integrated} and \emph{techno-economic}. It characterizes a power system in terms of the physical power grid and its multiple layers of control including commitment decisions, economic dispatch, and regulation services. Consequently, it has the ability to provide clear trade-offs for any changes to the physical power systems and its associated layers of control.  

The report begins with a rationale for EPECS simulator. It argues that with respect to operations the integration of variable energy resources should not be considered as ``business-as-usual," and instead a more holistic approach is required. It lays out the requirements for such a rigorous assessment. That discussion contextualizes a review of the methodological adequacy of existing renewable energy integration studies. It highlights several key conclusions found as a consensus across the literature. Combined unit-commitment and economic dispatch (UCED) models are used to assess changes in operating costs. Statistical methods are used to assess the need for greater quantities of operating reserves. The exact degree to which these changes occur ultimately depend on individual power system properties such as generation mix and fuel cost.  They also depend on the choice of several significant but not necessarily validated methodological assumptions used in the study. Next, the report describes the EPECS simulator in detail. It provides precise definitions of how variable energy resources and operating reserves are modeled. It also includes detailed models of day-ahead resource scheduling, same-day resource scheduling, real-time balancing operations and regulation service. The report also includes the zonal-network (i.e. pipe \& bubble) model of the physical power grid. 

The key findings of this study can be summarized in the following points: 
\begin{enumerate}
\item The commitment of dispatchable resources and their associated quantities of committed load following and ramping reserves has a complex, difficult to predict, non-linear dependence on the amount of VERs and the load profile statistics. High and low levels of VERs do not necessarily correspond to high or low quantities of operating reserves respectively.  For example, during the midday hours, solar generation causes low net load conditions that will test a power system's ability to track downward using downward load following reserves.  Hours later, as solar generation wanes, net load conditions rise to their daily peak testing the power system's ability to track upward with upward load following reserves.  In the meantime, the transition hours between trough and peak conditions exhibits a sharp system ramp.  
\item For the scenarios with significant presence of VERs (2025-3, 2030-3 and 2030-6), the system may require additional amounts of upward load following reserves to effectively mitigate imbalances and maintain its reliable operations. Furthermore, these scenarios entirely exhaust their downward load following reserves; albeit for a fairly short part of the year. Despite such occurrences being rare, depletion of a resource that was assumed to be adequately available in the system for following the net load fluctuations shows the need for the procurement of both upward and downward load following reserves in the day-ahead unit commitment.
\item For the scenarios with significant presence of VERs (2025-3, 2030-3 and 2030-6), the system entirely exhausts its upward and downward ramping capabilities. Such moments coincide with power system imbalances.  These results indicate that the assumption that the generator ramping constraints in the day-ahead scheduling provide sufficient ramping capabilities to the system is inadequate. Therefore, both load following and ramping reserves should be procured in the day-ahead unit commitment.
\item Along with the load following and ramping reserves provided by dispatchable resources, the curtailment of semi-dispatchable resources becomes an integral part of balancing performance; in part to complement operating reserves and in part to mitigate the topological limitations of the system. Every scenario uses curtailment in some way at least 98.6\% of the time.  The maximum level of curtailment for all scenarios ranges from 1,605MW (in Scenario 2025-4) to 14,534MW (in Scenario 2030-2).  In all, these curtailments correspond to a loss of between 2.72\% (in Scenario 2030-4) and 41.19\% (in Scenario 2030-2) of the total semi-dispatchable energy available.    It is also important to emphasize that some of the associated topological limitations only start affecting the system performance after the integration of VERs in remote areas that replace the traditional generation units located close to the main consumption centers. Thus, VERs might have a self-limiting feature which also defines the ability of the system to accommodate them.
\item The integration of significant amounts of VERs increases the potential of congestion on several key interfaces (Orrington-South and Surowiec-South), and, therefore, require heavy curtailments of these resources. Thus, the ability of the system to accommodate more renewables is limited by its topology. A longer-term solution to accommodating large amounts of VERs while avoiding such congestions would be the construction of new transmission lines from remote areas of VER installation to the main consumption centers. 
\item For the scenarios with significant presence of VERs (Scenarios 2025-3, 2030-2, 2030-3, and 2030-6), the system experiences heavy saturations of regulation reserves and, therefore, requires additional regulation reserves to effectively respond to the residual imbalances.  Scenarios 2025-1, 2025-2, 2025-6, 2030-1 also experience moderate saturations of regulation reserves indicating the need for their increase in 8 out of the 12 scenarios studied.  
\item The scenarios with significant presence of VERs (Scenarios 2030-2, 2030-3 and 2030-6) have significantly degraded balancing performance relative to the other scenarios studied and a complementary set of new measures would be required to achieve similar performance.  It would be premature to conclude that these scenarios would result in degraded overall system reliability in real life because it is not clear at which absolute imbalance levels disruptive events might occur.  The simulated imbalance excursions in all scenarios are comparable to the historical normal operation data.  
\end{enumerate}

\clearpage
\section{Introduction}
\subsection{ISO New England's Rapidly Evolving Resource Mix}
\begin{figure}[htbp]
\centering
\includegraphics[width=6.5in]{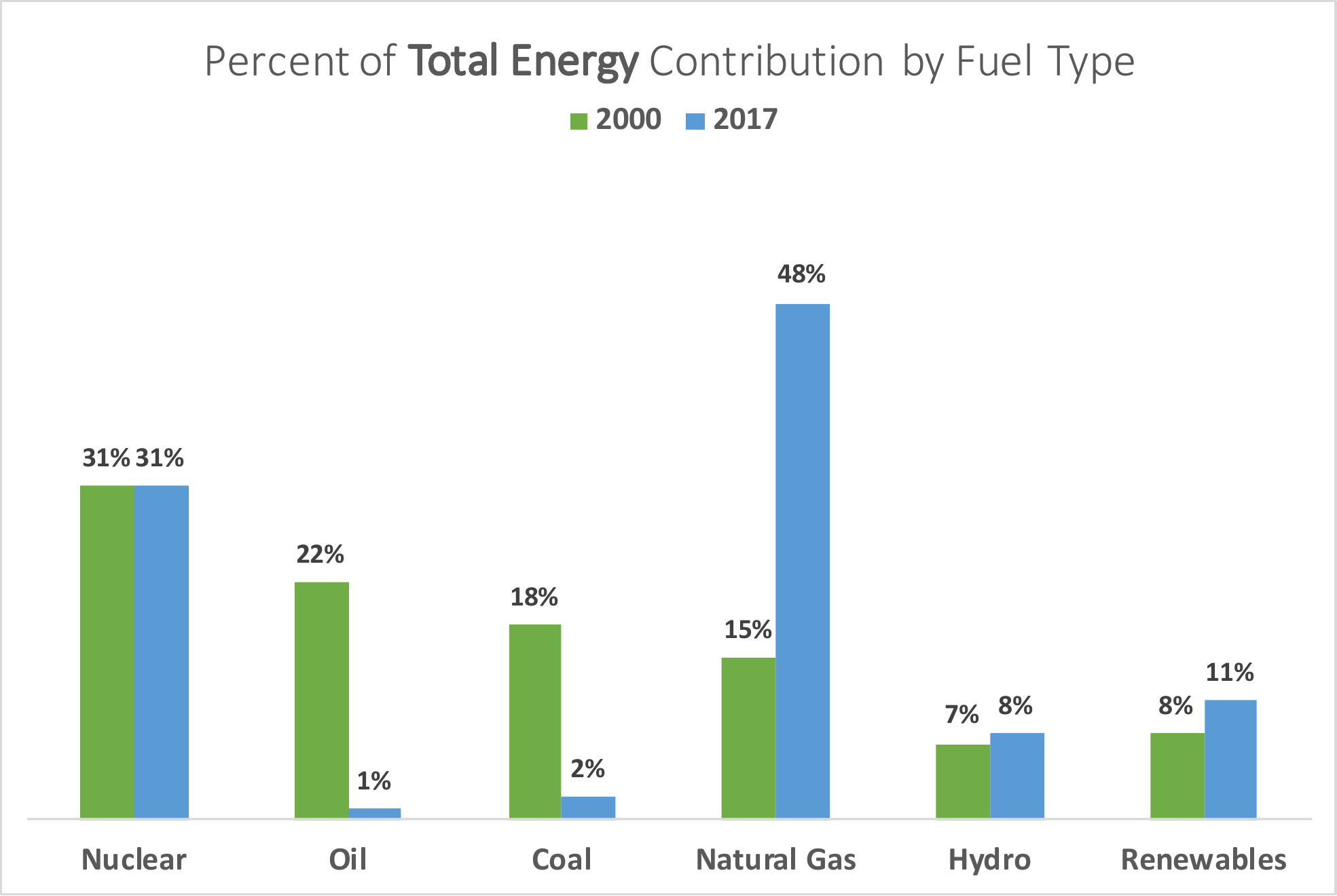}
\caption{ISO New England generation mix in 2000 and 2017 \cite{Welie:2018:00}.}
\label{Fig:PieCharts}
\end{figure}

The resource mix of ISO New England (ISO-NE) is rapidly changing. Figure~\ref{Fig:PieCharts} shows the evolution of its generation mix from 2000 to 2017 \cite{Welie:2018:00}.  As of 2015, over 9\% of the total generation came from renewable energy sources where 3.2\% was from wind and 0.9\% from solar PV \cite{ISO-NE:2017:00}. This percentage is expected to grow as the levelized cost of solar PV and wind installations continues to fall \cite{UVIG:2017:00}.  In the meantime, the representation of nuclear, coal, and oil plants in the generation portfolio is set to dramatically fall for two complementary reasons.  First, the emergence of low cost natural gas generation in recent years \cite{IEA:2017:00} has partially supplanted these facilities in the economic merit order.   Second, these facilities have an average age of over  30 years \cite{ISO-NE:2016:00} and are likely to be retired in the coming years.  For example, nuclear retirements are expected to bring down the percentage of nuclear generation to 10\% \cite{ISO-NE:2017:02} by 2025 as compared to the 31\% in 2017\cite{Welie:2018:00}. These retirements are likely to be replaced by more wind and natural gas resources in the overall resource mix. The percentage of natural gas powered generation is expected to account for over 56\% of the overall generation in 2025 \cite{ISO-NE:2017:02}. Furthermore, renewable portfolio requirements of various member states have also driven the ISO-NE resource mix to include more VERs \cite{Rourke:2015:00}. These requirements vary by state.  Some states, like Vermont, require up to 75\% of renewable energy generation including large-scale hydro\cite{Welie:2018:00}. This supply-side change in resource mix is occurring simultaneously with demand-side investment in energy efficiency measures. It is estimated that over \$7.1 billion \cite{Rourke:2015:00} will be invested in energy efficiency between 2019 and 2024 in addition to over \$4.9 billion already spent between 2009 and 2013 \cite{Welie:2018:00}. 

\begin{figure}[!t]
\centering
\includegraphics[width=6.5in]{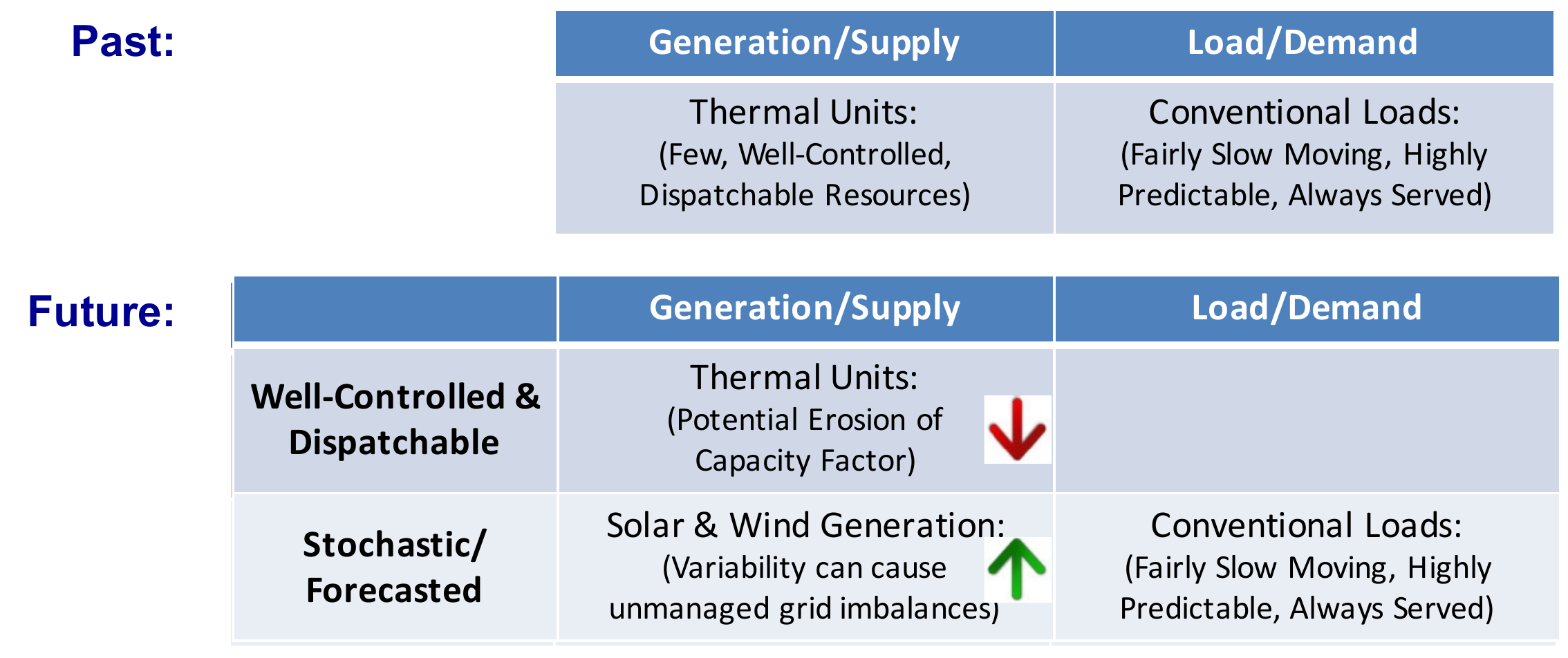}
\caption{The evolution of the power system \cite{Farid:2013:SPG-C13}}
\label{fig:portfolio_past_future}
\end{figure}

This changing resource mix, and particularly the introduction of VERs, is set to cause fundamental changes in the power grid's dynamic operations \cite{Farid:2014:SPG-J26}.  As shown in Figure \ref{fig:portfolio_past_future}, traditional power systems have often been built on the basis of an electrical energy value chain which consists of relatively few, centralized, and actively controlled thermal power generation facilities \cite{Meier:2006:00,Schavemaker:2008:00}.  These serve a relatively large number of distributed, stochastic electrical loads\cite{Meier:2006:00,Schavemaker:2008:00}.  Furthermore, the dominant operating paradigm and goal for these operators and utilities was to always serve the consumer demanded load with maximum reliability at whatever the production cost\cite{Gellings:1985:00}.  Over the years, system operators and utilities have improved their methods to achieve this task\cite{Wood:2014:00,Gomez-Exposito:2008:00}.  Generation dispatch, reserve management and automatic control has matured. Load forecasting techniques have advanced significantly to bring forecast errors to as low as a couple of percent.  System security procedures and their associated standards have evolved equally.   

The introduction of VERs evolves this status quo.  As they are added into the grid, the picture of the generation and demand portfolio gains a third quadrant as shown in the bottom half of Figure \ref{fig:portfolio_past_future}. From the perspective of dispatchability, VERs are non-dispatchable in the traditional sense: the output depends on external conditions and are not controllable by the grid operator\footnote{In recent years, significant efforts in both academic and industrial research and development have advanced the potential for variable energy resources to provide ancillary services \cite{Diaz-Gonzalez:2014:00,Mohseni:2012:00,Anonymous:2012:05}. However, these technologies have yet to become mainstream in the existing fleet of solar and wind generation facilities. This work, therefore, assumes that VERs are truly variable.} \cite{Kassakian:2011:SPG-B01}; except in a downward direction for curtailment. As VERs displace thermal generation units in the overall generation mix, the overall dispatchability of the generation fleet decreases. In regards to forecastability, VERs increase the uncertainty level in the system \cite{Kassakian:2011:SPG-B01}. Relative to traditional load, VER forecast accuracy is low, even in the short term \cite{Giebel:2011:00}. The decreased dispatchability coupled with decreased forecastability summarized by Figure \ref{fig:portfolio_past_future} calls for holistic assessment of the electric power system as it evolves.   

The integration of VERs will bring about fundamental changes that will necessitate a structured and holistic view for assessing the power system as it evolves.  While existing regulatory codes and standards will continue to apply\cite{Anonymous:2012:05,Mohseni:2012:00,Diaz-Gonzalez:2014:00}, it is less than clear how the holistic behavior of the grid will change or how reliability will be assured. Furthermore, it is important to assess the degree to which control, automation, and information technology are truly necessary to achieve the desired level of reliability. Thirdly, it is unclear what value for cost these technical integration decisions can bring.   From a societal perspective, and beyond simply variable energy integration, smart grid initiatives have been priced at several tens of billions of dollars in multiple regions\cite{Gellings:2011:00,Easton:2012:01}.  Therefore, there is a need to thoughtfully quantify and evaluate the steps taken in such a large scale technological migration of the existing power grid.
\begin{figure}[!t]
\centering
\includegraphics[width=6.5in]{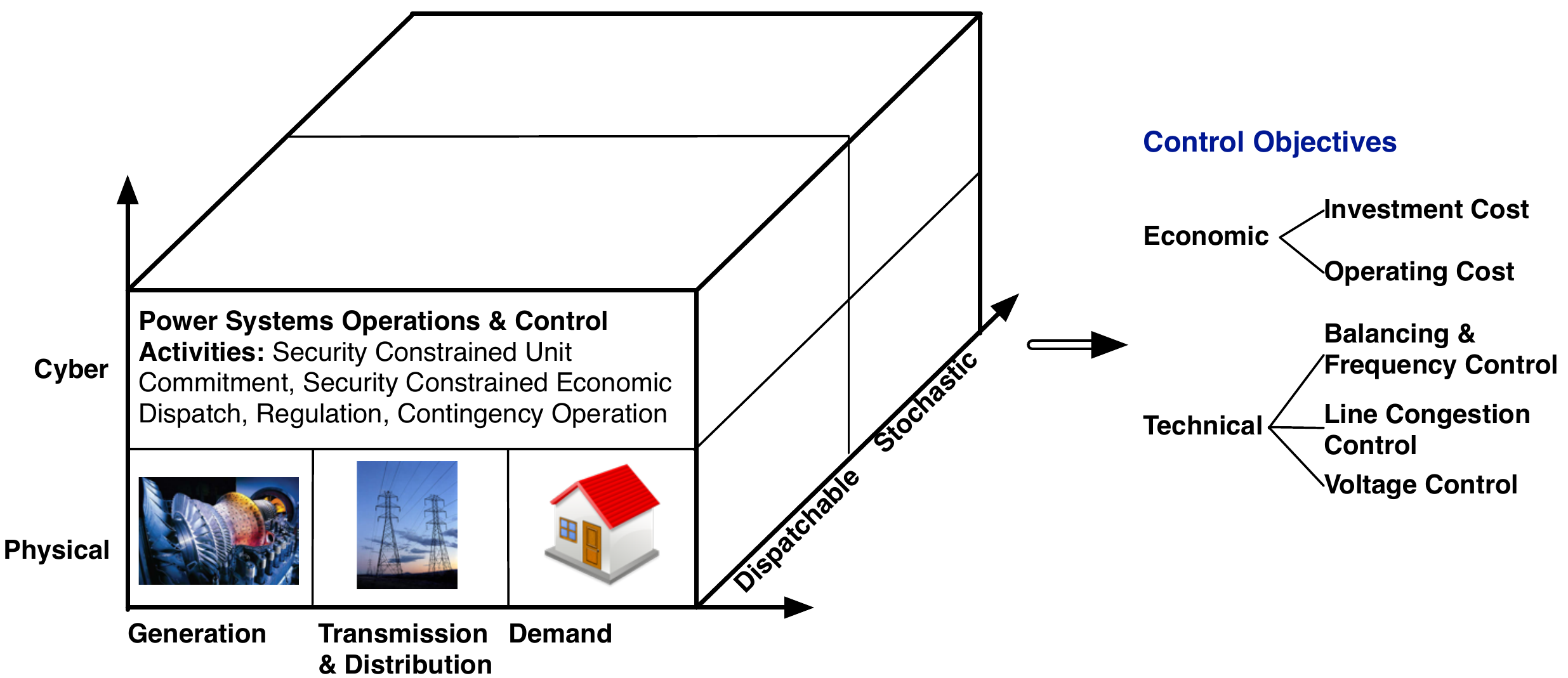}
\caption{Enterprise control as guiding assessment structure for power grids}
\label{Fig:Overview}
\end{figure}

\subsection{The Need for Holistic Techno-Economic Assessment Methods}
This work, thus, argues that a future electricity grid with a high penetration of VERs requires holistic assessment methods. This argument is structured as shown in Figure \ref{Fig:Overview}. On one axis, the electrical power grid is viewed as a cyber-physical system. That is, assessing the physical integration of VERs \emph{must} be taken in the context of the control, automation, and information technologies that would be added to mitigate and coordinate their effects. On another, it is an energy value chain spanning generation and demand. On the third axis, it contains dispatchable as well as stochastic energy resources. These axes holistically define the scope of the power grid system which must meet competing techno-economic objectives. Power grid technical objectives are often viewed as balancing operations, line congestion management and voltage management \cite{Gomez-Exposito:2008:00}. Economically speaking, the investment decision for a given technology, be it VERs or their associated control, must be assessed against the changes in reliability and operational cost. These economic and control technologies will later be viewed from the lens of dynamic properties including dispatchability, flexibility and forecastability. Naturally, such holistic assessment methods will represent an evolution of existing methods. This work thus seeks to draw from the trends and recommendations in the existing literature and frame them within the structure of Figure \ref{Fig:Overview}.

This ongoing evolution of the power grid can already be viewed through the lens of ``enterprise control". Originally, the concept of enterprise control\cite{Martin:2012:01,ANSI-ISA:2005:00} was developed in the manufacturing sector out of the need for greater agility \cite{Sanchez:2001:00,Gunasekaran:1998:00} and flexibility \cite{Beach:2000:02,De-Toni:1998:00,Pels:1997:00} in response to increased competition, mass-customization and short product life cycles.  Automation became viewed as a technology to not just manage the fast dynamics of manufacturing processes but also to integrate\cite{Lapalus:1995:00} that control with business objectives.  Over time, a number of integrated enterprise system architectures\cite{Williams:2001:00,Kosanke:1999:01} were developed coalescing in the current ISA-S95 standard\cite{ANSI-ISA:2000:00,ANSI-ISA:2005:00}. Analogously, recent work on power grids has been proposed to update operation control center architectures\cite{Wu:2005:00} and integrate the associated communication architectures \cite{Yan:2013:00}.  The recent NIST interoperability initiatives further demonstrate the trend towards integrated and holistic approaches to power grid operation\cite{Anonymous:2010:01}.  These initiatives form the foundation for further and more advanced holistic control of the grid\cite{Amin:2001:01,Amin:2006:00,Amin:2008:00,Amin:2011:00,McArthur:2012:00}.

Given the emergence of these trends in New England, ISO-NE has initiated the 2017 System Operational Analysis and Renewable Energy Integration Study (SOARES). This project serves as the last of three Phase II projects of the 2016 Economic Study \cite{Henderson:2016:00,Coste:2016:00}. Given their extensive publications on the topic, ISO-NE has selected the Laboratory for Intelligent Integrated Networks for Engineering Systems (LIINES) at the Thayer School of Engineering at Dartmouth to conduct the study.  This report describes the project's methodology as a whole emphasizing a novel, but now extensively published\cite{Muzhikyan:2015:SPG-J16,Muzhikyan:2015:SPG-J15,Muzhikyan:2015:SPG-J22,Muzhikyan:2014:SPG-C32,Muzhikyan:2015:SPG-C47, Muzhikyan:2015:SPG-C46,Muzhikyan:2016:SPG-J28, Muzhikyan:2014:SPG-C43},  holistic assessment approach called the Electric Power Enterprise Control System (EPECS) simulator.  It also situates this new approach relative to the existing renewable energy integration literature. To maintain continuity, the project specifically seeks to study ISO-NE operations in the years 2025 and 2030 for the six scenarios identified during Phase I of the 2016 Economic Study request. The study will specifically address quantifying  operating reserve requirements, ramp rates over hourly and sub-hourly periods, and identify periods of insufficient operating reserves.  

\subsection{Research Scope and Questions}

This study was commissioned by ISO New England as a means of addressing the reliability concerns presented by the evolving generation base within the region.  The scope of this study addresses six 2025 \emph{hypothetical} scenarios and six 2030 \emph{hypothetical} scenarios that were agreed upon consensually among ISO New England stakeholders.  These scenarios provide further analysis for ISO-NE stakeholders without necessarily reflecting ISO-NE's prediction of the future New England electric power systems.  They are described in Section \ref{sec:data}.  The study includes the following research questions that are answered in Section \ref{sec:results} entitled Results.  What is the impact of the 12 predefined scenarios on:  
\begin{itemize}
\item \ldots the resulting quantities of load following reserves?
\item \ldots the resulting ramping reserves?
\item \ldots the curtailment of semi-dispatchable resources?
\item \ldots the interface and tie-line performance?
\item \ldots the regulation reserves?
\item \ldots the balancing performance?
\end{itemize}
\noindent This study fits within the three critical roles ISO New England performs to ensure reliable electricity at competitive prices\cite{Welie:2018:00}:  
\begin{itemize}
\item \emph{\textbf{Grid Operation:}} Coordinate and direct the flow of electricity over the region's high voltage transmission system.
\item \emph{\textbf{Market Administration:}} Design, run, and oversee the markets where wholesale electricity is bought and sold.
\item \emph{\textbf{Power System Planning:}} Study, analyze, and plan to make sure New England's electricity needs will be met over the next 10 years.
\end{itemize}
As such, the focus of the study is to inform stakeholders in regards to these agreed upon scenarios.   

In light of the ISO New England mission, this study is \emph{not} meant to promote renewable energy resources or any other single type of energy resource.  This report does \emph{not} seek to answer resource-specific questions such as:  
\begin{itemize}
\item What is the maximum penetration rate of renewable energy resources that can be reliably integrated in the New England region?  
\item How much natural gas generation is required to achieve a desired level of system-wide flexibility (i.e. ramp rate)?  
\item How does the inflexibility of nuclear generation limit reliable balancing operations?  
\end{itemize}
Each of these questions, due to their resource-specificity, imply a certain preference for one type of energy resource over another.   Instead, this report focuses on the system-level results pertaining to the 12 scenarios mentioned above.    From such a presentation, the reader may conclude whether certain resource \emph{mixes} are more or less likely to lead to reliable operation.  

\subsection{Report Outline}
The rest of this report is structured as follows. Section~\ref{sec:background} provides a review of the methodological adequacy of existing renewable energy integration studies and the methodological characteristics of the EPECS simulator. Section~\ref{sec:methodology} presents the implementation technical details of the EPECS simulator. Section~\ref{sec:data} describes the ISO New England data used for this study, and Section~\ref{sec:results} analyzes the case study results. Finally, the report is brought to a conclusion in Section \ref{sec:conclusion}.

\clearpage
\section{Background}\label{sec:background}
This section describes the methodological characteristics of the 2016 ISO New England Economic Study, the enterprise control assessment method used in this study and other existing renewable energy integration studies found in the literature.

\subsection{Methodological Characteristics of the 2016 ISO New England Economic Study}\label{sec:2011-ISO-NE}
The 2016 ISO New England Economic Study was conducted at the request of the New England Power Pool (NEPOOL), and examines resource-expansion scenarios of the regional power system and the potential effects of these different future changes on resource adequacy, operating and capital costs, and options for meeting environmental policy goals \cite{ISO-New-England:2017:00}. The study presents a common framework for NEPOOL participants, regional electricity market stakeholders, policymakers, and consumers, information, analyses, and observations on the following:
\begin{itemize}
\item The potential impacts on the ISO New England markets of implementing public policies in the New England states
\item Projected energy market revenues, and the contribution of these revenues to the generic fixed costs of new generation, for various generation types under particular sets of assumptions
\item The potential impacts, under the status-quo forecast and compared with the public policy overlay, on system reliability and operability, resource costs and revenues, total cost of supplying load, and emissions in New England
\end{itemize}
The metrics studied include production costs, load-serving entity (LSE) energy expenses, locational marginal prices (LMPs), generic capital costs and annual carrying charges (ACCs) for each resource type, transmission- expansion costs, generation by fuel type and the emissions associated with each type, and the effects of transmission-interface constraints that may bind economic power flows.

The analyses were conducted using ABB's GridView program that calculates least-cost transmission-security-constrained unit commitment and economic dispatch under differing sets of assumptions and minimizes production costs for a given set of unit characteristics \cite{ABB:2018:00}. The program can explicitly model a full network, but the New England study model used a ``pipe and bubble" format, with ``pipes" representing transmission interfaces connecting the ``bubbles" representing the various planning areas. The ISO New England system was modeled as a constrained single area for unit commitment, and regional resources were economically dispatched in the simulations to respect the assumed transmission system security constraints under normal and contingency conditions. Depending on the case, the model dispatched up to 900 units (new and existing) in New England. For each scenario's set of resources (with their various operating characteristics), the simulation ``dispatched" power plants to meet different levels of customer demand in every hour of the year being analyzed. These simulations established a wide array of hypothetical data about how the electric power system ``performed" in terms of reliability, economics, and environmental indicators and the effects of transmission system constraints.

\subsection{Methodological Characteristics of Existing Renewable Energy Integration Studies}
A review of existing renewable energy integration studies is conducted from the perspective of the guiding structure found in Figure \ref{Fig:Overview}. Collectively, the renewable energy integration studies have many similarities \cite{Ela:2009:00,Holttinen:2012:01,Holttinen:2013:00, Brouwer:2014:00}. They generally apply combined unit-commitment and economic dispatch (UCED) models to assess the additional operating costs of renewable energy integration. Fewer studies add a model of regulation as a separate ancillary service. These three enterprise control layers are conducted primarily to assess the additional operating cost of renewable energy integration and are not integrated with a model of the physical grid to calculate technical variables such as potential power grid imbalances \cite{Brooks:2002:01,Ummels:2009:01,Brouwer:2014:00}. One often cited concern is that these simulations do not correspond to the existing enterprise control practice. For example, time steps, market structure and physical constraints should correspond to the operating reality \cite{Georgilakis:2008:00,Soder:2008:00,Holttinen:2012:01,Holttinen:2013:00,Brouwer:2014:00}. In the case of market time step, it has been confirmed both numerically\cite{Bird:2012:00,Holttinen:2012:01,Holttinen:2013:00, Muzhikyan:2015:SPG-J15,Muzhikyan:2015:SPG-J16} as well as analytically\cite{Muzhikyan:2016:SPG-J28,Muzhikyan:2014:SPG-C32,Muzhikyan:2015:SPG-C46,Muzhikyan:2015:SPG-C47} to affect power grid imbalances and costs.  Such a conclusion inextricably ties power system operation and control to their associated policies and regulations.

In contrast, the assessment of additional operating reserve requirements is mostly done by using statistical methods \cite{Ela:2009:00,Holttinen:2012:01,Holttinen:2013:00,Brouwer:2014:00} that are generally some variation on the theme found in \cite{Holttinen:2008:00}.  The differences between these approaches has been classified by Brouwer et al\cite{Brouwer:2014:00}.  In general, the standard deviation $\sigma$ of potential imbalances is calculated using the probability distribution of net load \emph{or} forecast error.  The load following and regulation reserve requirements are then defined to cover appropriate confidence intervals of the distribution based on the experience of power system operators and existing standards. A detailed discussion on the definition and types of operating reserves is provided in Section \ref{Sec:DefReserves}.  Normally, load following is taken to equal to $2\sigma$ \cite{Holttinen:2008:00,Robitaille:2012:00} to comply with the North American Electric Reliability Corporation (NERC) balancing requirements: NERC defines the minimum score for Control Performance Requirements 2 (CPS2) equal to $90\%$ \cite{NERC:2012:00}. Other integration studies have used a $3\sigma$ confidence interval \cite{Aigner:2012:00,Ummels:2007:00}  to correspond to the industry standard of $95\%$ \cite{Halamay:2011:00}. Based on the experience of power system operators, regulation is normally taken to be between $4\sigma$ and $6\sigma$ \cite{Holttinen:2008:00,Robitaille:2012:00,Hansen:2012:00}.

With respect to timescales, not all studies consider multiple timescales of operation. However, in order to characterize a power system's imbalances accurately, it is necessary to use a multi-timescale analysis. A single timescale would only capture part of the variability of the net load and leave out either slower or faster phenomena. For example, reference \cite{Halamay:2011:00} does not consider regulation because the available data has 10 minute resolution. References \cite{Luickx:2009:00,Albadi:2011:00} implement only unit commitment models, according to the assumption that wind integration has the biggest impact on unit commitment.  Furthermore, another concern is the usage and treatment of different power system timescales in the integration studies. Load following and regulation reserves operate at different but overlapping timescales. Net load variability as a property exists in all timescales, although with changing magnitudes. Forecast error, on the other hand, appears in two timescales: 1 hour (day-ahead forecast error) and 5-15 minutes (short term forecast error). Thus, VER intra-hour variability and day-ahead forecast error are relevant to load following reserve requirements. Meanwhile, 5-15 minute variations and short-term forecast error are relevant to regulation reserve requirements. This division of impacts is not carefully addressed in the literature.

In conclusion, renewable energy integration studies, as a collective body of literature, give a much more holistic understanding of the power grid and its potential evolution in the future. While these studies continue to evolve, they may require incorporation of certain methodological changes to better reflect the current need for more holistic assessment methods. Particularly, in regards to balancing operation, they use statistical methods for which there is a lack of consensus and which are based upon questionable assumptions. It is likely that the assessment of reserves will ultimately shift to simulation-based and analytical methods. UCED simulations form an integral piece of most integration studies and are likely to remain so. However, several authors have already advocated for the need to maintain the coherence between market operating procedures and the simulations. 

\subsection{Methodological Characteristics of Enterprise Control Assessment}
The methodological limitations of the existing renewable energy integration literature described in the previous section can be addressed by a framework for holistic power grid enterprise control assessment. In such a way, the variability of renewable energy resources can be viewed as an input disturbance which the (enterprise) power system systematically manages to deliver attenuated power system imbalances. Consequently, the power from renewable energy sources is modeled in terms of its key characteristics, namely penetration level, forecast errors, and variability. Such an approach is in agreement with several recommendations in the literature for integrated approaches\cite{Ummels:2009:01,Soder:2008:00,Holttinen:2012:01,Holttinen:2013:00}. Furthermore, one work advocates the role of custom-built simulators to assess the future electricity grid\cite{Podmore:2010:00}. Gathering the discussions from the previous section, such an approach fulfills the following requirements:
\begin{itemize}
\item allows for an evolving mixture of generation and demand as energy resources; be they dispatchable, semi-dispatchable, variable, or must-run.  
\item allows for the simultaneous study of generation, transmission and load
\item allows for the time domain simulation of the convolution of relevant grid enterprise control functions
\item allows for the time domain simulation of changes to the power grid topology in the operations time scale
\item specifically addresses the holistic dynamic properties of dispatchability, flexibility and forecastability
\item represents potential changes in enterprise grid control functions as impacts on these dynamic properties
\item accounts for the consequent changes in operating cost and the required investment costs
\end{itemize}
The first four of these requirements are basically associated with the nature of the power grid itself as it evolves. In the meantime, the next two are associated with the behavior of the power grid in the operations time scale. Finally, the last requirement contextualizes the simulation with cost accounting. 

The EPECS simulator used for this study is developed in accordance with such an enterprise control assessment framework. While it is not feasible to incorporate all power system operation processes within a single model, the EPECS simulator captures the ones most relevant to ISO New England balancing operations, namely day-ahead resource scheduling, same-day resource scheduling, real-time balancing operations and regulation service. Most fundamentally, the EPECS methodology is \emph{integrated} and \emph{techno-economic}.  Consequently, it has the ability to provide clear techno-economic trade-offs for any changes to the physical power system and its associated layers of control. The detailed description of the EPECS simulator different control layers is presented in Section~\ref{sec:methodology}. 

\clearpage
\section{Methodology: Electric Power Enterprise Control System Simulator for ISO New England}\label{sec:methodology}
\subsection{Overview of Electric Power Enterprise Control System Simulation}\label{Sec:Framework}
\begin{figure}[!ht]
\centering
\includegraphics[width=6.5in]{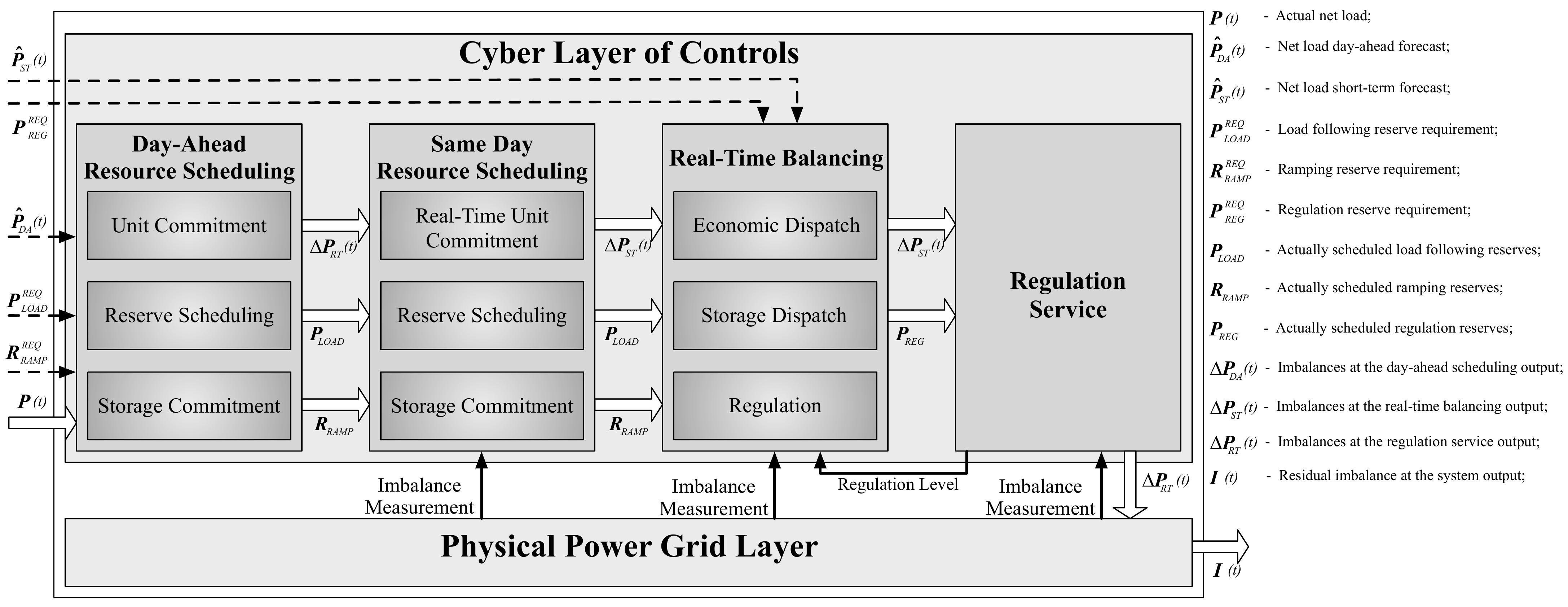}
\caption{Architecture of the Electric Power Enterprise Control System (EPECS) simulator customized for ISO New England operations}
\label{Fig:EPECS}
\end{figure}

This section introduces the Electric Power Enterprise Control System (EPECS) simulator customized for ISO New England's operations.  Its architecture is graphically depicted in Figure \ref{Fig:EPECS} and may be viewed as an extension of several enterprise control works\cite{Farid:2014:SPG-J26,Farid:2013:SPG-C13} involving variable energy integration\cite{Muzhikyan:2015:SPG-J15,Muzhikyan:2015:SPG-J16,Muzhikyan:2016:SPG-J28,Muzhikyan:2013:SPG-C17,Muzhikyan:2013:SPG-C18,Muzhikyan:2014:SPG-C32,Muzhikyan:2015:SPG-C46,Muzhikyan:2015:SPG-C47,Muzhikyan:2016:SPG-C55}, energy storage\cite{Muzhikyan:2015:SPG-J22,Muzhikyan:2014:SPG-C43}, and demand response\cite{Jiang:2016:SPG-J30,Jiang:2015:SPG-J21,Jiang:2015:SPG-C51,Jiang:2015:SPG-C44,Jiang:2015:SPG-C45}.  The simulator includes a physical power grid layer and several layers of primary, secondary, and tertiary enterprise control functions as shown in Figure \ref{Fig:EPECS}.  These include day-ahead resource scheduling, same-day resource scheduling, real-time balancing, and the regulation service.  Such an approach has several advantages.  First, the net load $P(t)$ may be viewed as a system disturbance which is systematically rejected by forecasting and relevant enterprise control functions to give a highly attenuated system imbalance time domain signal $I(t)$.  Second, it can address the recommendations in the literature\cite{Georgilakis:2008:00} to assess the impact of variable generation on operating reserve requirements.  Such an approach helps lay the methodological foundation for understanding renewable energy integration independent of the particularities of a physical power system in a given region\cite{Holttinen:2013:00}.  Finally, the EPECS simulator is quite flexible.  Its layers are modular and may be modified as necessary to assess the impact of a given control function or technology on the time domain simulation.  

This section now explains each of the layers in EPECS simulator in detail; focusing on the specific characteristics of ISO-NE's operations.   First, Sections \ref{Sec:DefVER} and \ref{Sec:DefReserves} introduce several fundamental definitions in order to facilitate the usage of the EPECS simulator across different power systems and introduce greater objectivity in this study's methodology.  Section \ref{Sec:SCUC} describes the day-ahead resource scheduling at ISO-NE in the form of a Security Constrained Unit Commitment (SCUC).  Second, Section \ref{Sec:RTUC} then describes same-day resource scheduling in the form of a Real-Time Unit Commitment (RTUC).  Section \ref{Sec:SCED} then describes real-time balancing operations in the form of a Security Constrained Economic Dispatch (SCED).  Section \ref{Sec:RegRes} describes a pseudo-steady state model of the regulation service. Finally, Section~\ref{Sec:physgrid} describes the physical power grid model. 

\subsection{Fundamental Definitions on Variable Energy Resources}\label{Sec:DefVER}
The EPECS simulator has several types of energy resources; including variable, dispatchable, semi-dispatchable, and must-run resources.  

\begin{defn}
Variable Resources: Resources that have a stochastic and intermittent power output.  Normally, these include wind, solar, run-of-river hydro, and tie-lines are assumed to be variable resources. In this study, all variable resources served as semi-dispatchable resources.
\end{defn}
\begin{defn}
Semi-Dispatchable Resources: Energy resources that can be dispatched downwards (i.e curtailed) from their uncurtailed power injection value. When curtailment is allowed for variable resources, they become dispatchable.  In this study, wind, solar, run-of-river hydro, and tie-lines are assumed to be semi-dispatchable resources.
\end{defn}
\begin{defn}
Must-Run Resources: Energy resources that must run all the time at their maximum output. In this study, nuclear generation units are assumed to be must-run resources.
\end{defn}
\begin{defn}
Dispatchable Resources:  Energy resources that can be dispatched up and down from their current value of power injection. In this study, all other resources are assumed to be dispatchable.
\end{defn}

Within the EPECS simulator, variable energy resources are modeled as a time-dependent exogeneous spatially-distributed quantity that contributes directly to the net load.  They are described in terms of a number of non-dimensional quantities.  
\begin{defn}
\emph{Penetration Level ($\pi$)}:  The (aggregated) installed VER capacity $P_V^{max}$ normalized by the system peak load $P^{peak}_{L}$ \cite{Wang:2012:00}:
\begin{equation}\label{eq:defpen}
\pi = \dfrac{P_V^{max}}{P_L^{peak}}
\end{equation}
\end{defn}
\begin{defn}
\emph{VER Capacity Factor ($\gamma$):} The average VER power output $\overline{P_V(t)}$ (e.g., over 1 year period) per installed capacity \cite{Muzhikyan:2014:SPG-C32}:
\begin{equation}\label{eq:defcap}
\gamma = \frac{\overline{P_V(t)}}{P_V^{max}}
\end{equation}
\end{defn}

Next, it is important to introduce the concept of variability as it is applied to the VERs, the load, and/or the net load. The variability of each of these plays a significant role in balancing operations. Intuitively speaking, variability is associated with the change rates of a given output. In this paper, it is defined as:
\begin{defn}
\emph{Variability (A)}: Given the choice of the output $P(t)$ (e.g. the VER generation, the load, the net load), the variability is the root-mean-square of that output's rate normalized by the root-mean-square of that output \cite{Muzhikyan:2014:SPG-C32}:
\begin{equation}\label{eq:defvar}
A = \frac{rms\left(\mathrm{dP(t)}/\mathrm{dt}\right)}{rms\left(P(t)\right)}
\end{equation}
\end{defn}

Since the power spectra of the VER and load have distinctive shapes \cite{Apt:2007:01,Curtright:2008:00}, the way to change the variability of the profile without distorting its spectral shape is temporal scaling \cite{Muzhikyan:2014:SPG-C32}. Assume that a default profile $P_0(t)$ has a variability $A_0$ and $P(t)$ is related to it in the following way:
\begin{equation}\label{eq:defvar1}
P(t) = P_0(\alpha t)
\end{equation}
According to (\ref{eq:defvar}), the variability of $P(t)$ is:
\begin{align}
A &= \frac{rms\left(\mathrm{dP_0(\alpha t)}/\mathrm{dt}\right)}{rms\left(P_0(\alpha t)\right)} = \alpha\cdot\frac{rms\left(\mathrm{dP_0(\alpha t)}/\mathrm{d(\alpha t)}\right)}{rms\left(P_0(\alpha t)\right)} = \alpha A_0
\end{align}
Thus, $\alpha$ can be viewed as a scaling factor between the given profile and the default profile variabilities:
\begin{equation}\label{eq:defvar2}
\alpha = \frac{A}{A_0}
\end{equation}

The definitions for the forecast and forecast error are introduced next. Fundamentally speaking, while the net load is a continuously varying function in time, the forecast has a specific value resolved with each day ahead market time block (e.g. 1 hour). Therefore, the two are inherently different types of quantities. To address this issue, the concept of a ``Best Forecast" is introduced as:
\begin{defn}
\emph{The Best Forecast} \cite{Muzhikyan:2014:SPG-C32}: Given the output $P(t)$ (e.g. the VER generation, the load, the net load), the best forecast $\bar{P}_k$ is equivalent to the average value of that output during the $k^{th}$ market time block of duration $T$:
\begin{equation}
\bar{P}_k = \frac{1}{T}\int\limits_{kT}^{(k+1)T}P(t)\mathrm{dt}
\end{equation}
\end{defn}
\noindent Similarly, the forecast error defines the deviation between the actual and best forecasts, which in turn may have various measures such as mean absolute error (MAE) and mean square error (MSE)\cite{Monteiro:2009:00}. Here, the VER forecast error is normalized by the installed capacity.
\begin{defn}
\emph{VER Forecast Error ($\varepsilon$)} \cite{Muzhikyan:2014:SPG-C32}:
The standard deviation of the difference between the best ($\bar{P}_k$) and actual VER forecasts ($\hat{P}_k$) is normalized by the installed capacity:
\begin{equation}\label{eq:deferr}
\varepsilon = \frac{\sqrt{\frac{1}{n}\sum\limits_{k = 0}^{n}\left(\bar{P}_k - \hat{P}_k\right)^2}}{P_V^{max}}
\end{equation}
\end{defn}

The above definitions are used to simulate different integration scenarios.  More specifically, in developing sensitivity cases, the VER model systematically changes five main parameters:  penetration level, capacity factor, variability, day-ahead and short-term forecast errors.  First, the definitions of VER penetration level and capacity factor in (\ref{eq:defpen}) and (\ref{eq:defcap}) respectively can be used to define the actual VER output.
\begin{align}
\label{eq:pen}
P_V(t) &= \frac{P_{V}(t)}{\overline{P_V(t)}}\frac{\overline{P_V(t)}}{P_V^{max}}\cdot\frac{P_V^{max}}{P_L^{peak}}\cdot P_L^{peak} = p_V(t)\cdot\gamma\cdot\pi\cdot P_L^{peak}
\end{align}
where $p_V(t)$ is VER power normalized to a unit capacity factor. Equation (\ref{eq:pen}) shows that if a single $p_V(t)$ is taken as a default profile, the actual VER output can be systematically adjusted with the values of $\pi$ and $\gamma$.  Next, the definition of VER forecast error in Equation (\ref{eq:deferr}) can be used to define the actual VER forecast error. Two types of forecasts (and their errors) are used in the power system simulations, day-ahead and short-term. The day-ahead forecast is used in the SCUC model for day-ahead resource scheduling. It normally has a 1 hour resolution and up to 48 hour forecast horizon. The short-term forecast is used in the RTUC model for the same-day resource scheduling and the SCED model for real-time balancing operations. It has a ten minute time resolution and up to six hour time horizon \cite{Giebel:2011:00,Moreno-Munoz:2008:00}. The VER forecast can be expressed as:
\begin{equation}
\hat{P}_V(t) = P_V(t) - E(t)
\end{equation}
where $\hat{P}_V(t)$ is the forecasted VER profile, and $E(t)$ is the error term. Using the definition of the forecast error in (\ref{eq:deferr}), the error term can be written as:
\begin{align}\label{eq:err}
E(t) &= \frac{E(t)}{std\left(E(t)\right)}\cdot\frac{std\left(E(t)\right)}{P_V^{max}}\cdot\frac{P_V^{max}}{P_L^{peak}}\cdot P_L^{peak} =\nonumber\\
&= e(t)\cdot\varepsilon\cdot\pi\cdot P_L^{peak}
\end{align}
where $e(t)$ is the error term normalized to the unit standard deviation. Equation~(\ref{eq:err}) shows that if a single $e(t)$ is taken for each type of market as a default profile, the actual error profile can be systematically adjusted with the values of $\pi$ and $\varepsilon$. It is important to emphasize that the error term $e(t)$ is different for the day-ahead and short-term applications. They may have different probability distributions and power spectra. Additionally, the forecast error ranges are generally different with the short-term forecast having higher accuracy as compared to the day-ahead forecast.  Finally, the actual variability can be similarly adjusted with the value of $\alpha$.  Using Equations (\ref{eq:pen}) and (\ref{eq:err}) and the properties of variability in Equations (\ref{eq:defvar1}) and (\ref{eq:defvar2}), the VER model can be expressed as follows:
\begin{align}
&P_V(t) = p_V(\alpha t)\cdot\gamma\cdot\pi\cdot P_L^{peak} \\
&\hat{P}_V(t) = \left(\gamma\cdot p_V(\alpha t) - \varepsilon\cdot e(\alpha t)\right)\cdot\pi\cdot P_L^{peak}\\
&\alpha = A/A_0
\end{align}
This set of equations defines the VER model used in this study. As an input, it requires the actual VER profile $p_V(t)$ normalized to unit capacity factor, and the error term profile $e(t)$, normalized to unit standard deviation. The model explicitly includes the five major parameters of VER.

\subsection{Fundamental Definitions of Operating Reserves}\label{Sec:DefReserves}
In addition to the definitions associated with variable energy resources, a number of definitions related to operating reserves are provided. The challenge here is that the taxonomy and definition of operating reserves from one power system geography to the next varies\cite{Holttinen:2012:00}.  Furthermore, this taxonomy and definition is often different from the methodological foundations found in the literature\cite{Holttinen:2012:00}.   There is even significant differences in the definitions found within the literature itself\cite{Holttinen:2012:00,Ela:2011:00,Rebours:2007:00,CIGRE:2010:00}.  Therefore, this report first introduces the definitions of operating reserves in the EPECS simulator in Section \ref{Sec:DefReservesLit}, then introduces the definitions used in ISO New England in Section \ref{Sec:DefReservesISONE}, and then concludes by reconciling these concepts in Section \ref{Sec:ReconcileReserves}.  
\subsubsection{Operating Reserves in the EPECS Simulator Methodology}\label{Sec:DefReservesLit}
\begin{figure}[!t]
\centering
\includegraphics[width=5.5in]{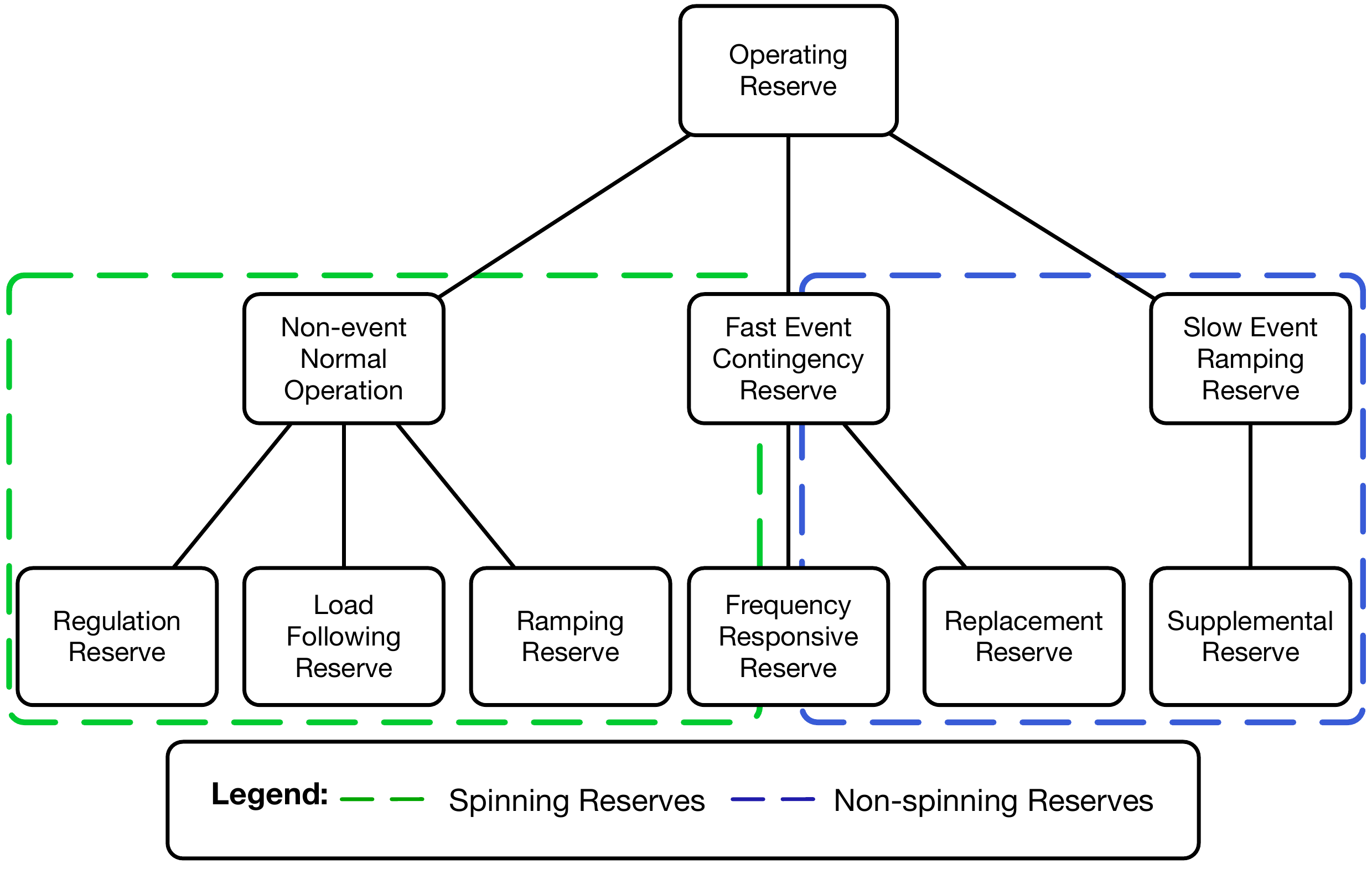}
\vspace{-0.1in}
\caption{Operating reserves classification (adapted from \cite{Holttinen:2012:00})}
\label{Fig:OperResTaxonomy}
\end{figure}

\begin{figure}[!t]
\centering
\includegraphics[width=5.5in]{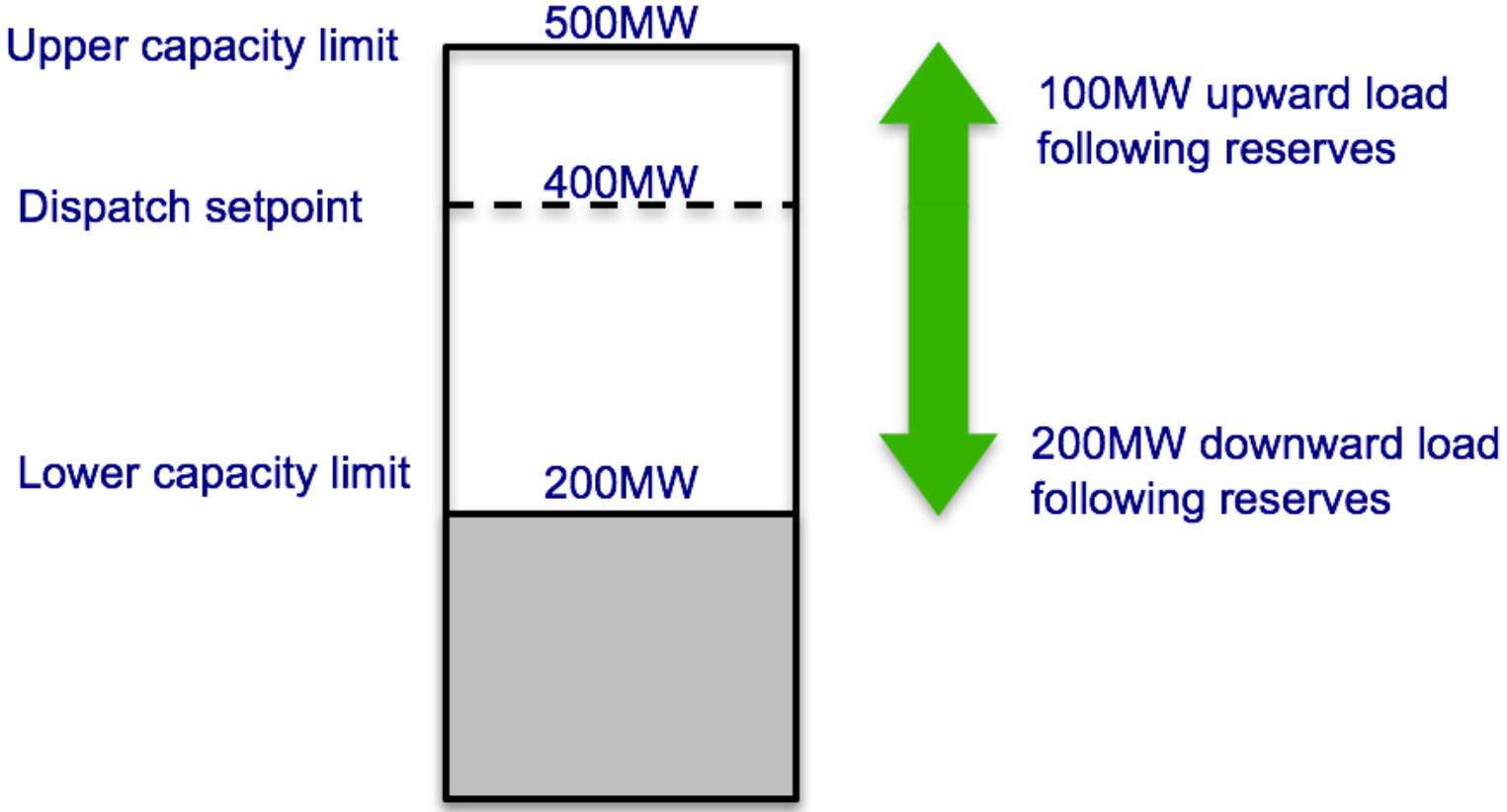}
\caption{Load following reserves example}
\label{Fig:LFR}
\end{figure}

The EPECS simulator methodology adopts the operating reserves concepts found in \cite{Holttinen:2012:00,Ela:2010:00} with minor differences.  Figure \ref{Fig:OperResTaxonomy} shows the taxonomy of the various types of operating reserves. 
The primary distinction is between the operating reserves used to respond to contingency events and those used during normal operation to respond to forecast errors and variability in the net load.  Since the outage of any individual wind or solar generation facility has a much smaller impact on the system than the largest thermal plant, solar and wind integration will not increase contingency reserves requirements \cite{Holttinen:2012:00}. The exception to this general rule is when a transmission line transports a large amount of power from variable energy resources in a remote area (e.g. off-shore wind).  In such a case, the loss of the transmission line could be comparable in size to the loss of a large thermal power plant.  In spite of this exception, the focus of most renewable energy integration has primarily been on normal operating reserves.  They are further classified as load following, ramping, and regulation reserves depending on the mechanisms by which they are acquired and activated.  
\begin{defn}
\emph{Load Following Reserves} \cite{Ela:2010:00,Holttinen:2012:00}: Power capacity available during normal operations for assistance in active power balance to correct the future anticipated imbalances upward or downward.  The actual quantity of upward load following reserves is given by:
\begin{equation}
\sum_{k=1}^{N_G}\big(w_{kt}P_k^{max}-P_{kt}\big)
\end{equation}
where $N_G$ is the number of generators, $w_{kt}$ is the (binary) online state of the $k^{th}$ generator at time $t$, $P_k^{max}$ is the maximum capacity of the $k^{th}$ generator, and $P_{kt}$ is the value at which it is currently generating.  Similarly, the actual quantity of downward load following reserves is given by:  
\begin{equation}
\sum_{k=1}^{N_G}\big(P_{kt}-w_{kt}P_k^{min}\big)
\end{equation}
where $P_k^{min}$ is the minimum capacity of the $k^{th}$ generator.  Within ISO-NE, load following reserves are often called economic surplus reserves.  
\end{defn}
\begin{ex}
Consider Figure \ref{Fig:LFR} as an example.   It consists of a single generator generating at 400MW.  It has a maximum capacity of 500MW and a minimum capacity of 200MW.  It provides 100MW of upward load following reserves and 200MW of downward load following reserves.  
\end{ex}
\noindent Returning back to Figure \ref{Fig:EPECS}, load following reserves are acquired during the day-ahead and same-day resource scheduling steps in the EPECS simulator.  Furthermore, they are utilized during the real-time balancing operation.  Note that this definition of load following reserves is purely a property of the physical system.  This is entirely independent of whether some system operators monetize this property in the form of a \emph{reserve product} or not.  

\begin{figure}[!b]
\centering
\includegraphics[width=5.5in]{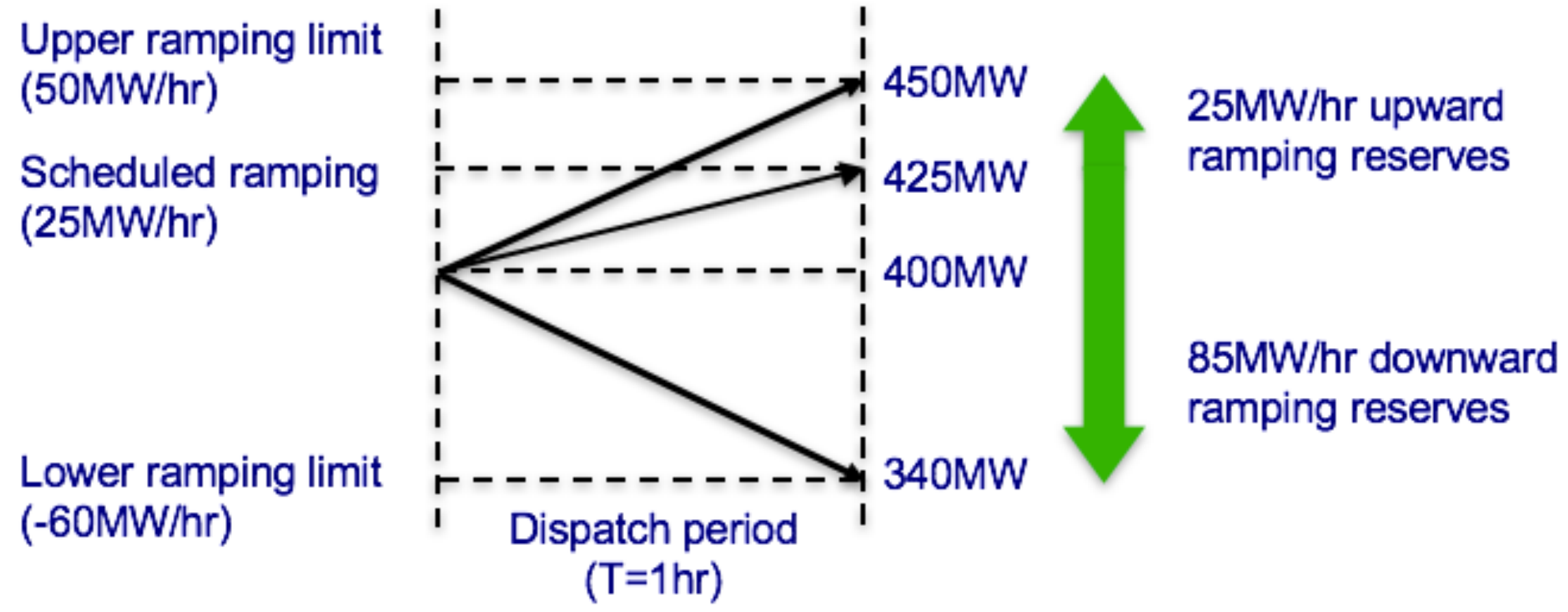}
\caption{Ramping reserves example} 
\label{Fig:RampR}
\end{figure}

\begin{defn}
\emph{Ramping Reserves} \cite{Ela:2010:00,Holttinen:2012:00}: Ramp rate capacity available during normal operations for assistance in active power balance to correct the future anticipated imbalances upward or downward.  The actual quantity of upward ramping reserves is given by:
\begin{equation}
\sum_{k=1}^{N_G}\left(w_{kt}R_k^{max} - \dfrac{P_{kt} - P_{k,t-1}}{\Delta T}\right)
\end{equation}
where $R_k^{max}$ is the maximum upward ramp rate of the $k^{th}$ generator, and $\Delta T$ is duration of a time step between the generator levels $P_{kt}$ and $P_{k,t-1}$.  Normally, $\Delta T$ is equal to one hour.  Similarly, the actual quantity of downward ramping reserves is given by:  
\begin{equation}
\sum_{k=1}^{N_G}\left(w_{kt}R_k^{max} - \dfrac{P_{kt} - P_{k,t-1}}{\Delta T}\right)
\end{equation}
where $R_k^{min}$ is the maximum downward ramp rate of the $k^{th}$ generator.  
\end{defn}
\begin{ex}
Consider Figure \ref{Fig:RampR} as an example.   It consists of a single generator that is scheduled to ramp from 400MW to 425MW within a given period $\Delta T$ equal to one hour.   It has the ability to ramp up at 50MW/hr and ramp down at 60MW/hr.  It provides 25MW/hr of upward ramping reserves and 85MW/hr of downward ramping reserves. 
\end{ex}
\noindent Returning back to Figure \ref{Fig:EPECS}, ramping reserves, much like load following reserves, are acquired during the day-ahead and same-day resource scheduling steps in the EPECS simulator.  Furthermore, they are utilized during the real-time balancing operation.  Note that this definition of ramping reserves is purely a property of the physical system.  This is entirely independent of whether some system operators monetize this property in the form of a \emph{reserve product} or not.  

\begin{figure}[!t]
\centering
\includegraphics[width=5.5in]{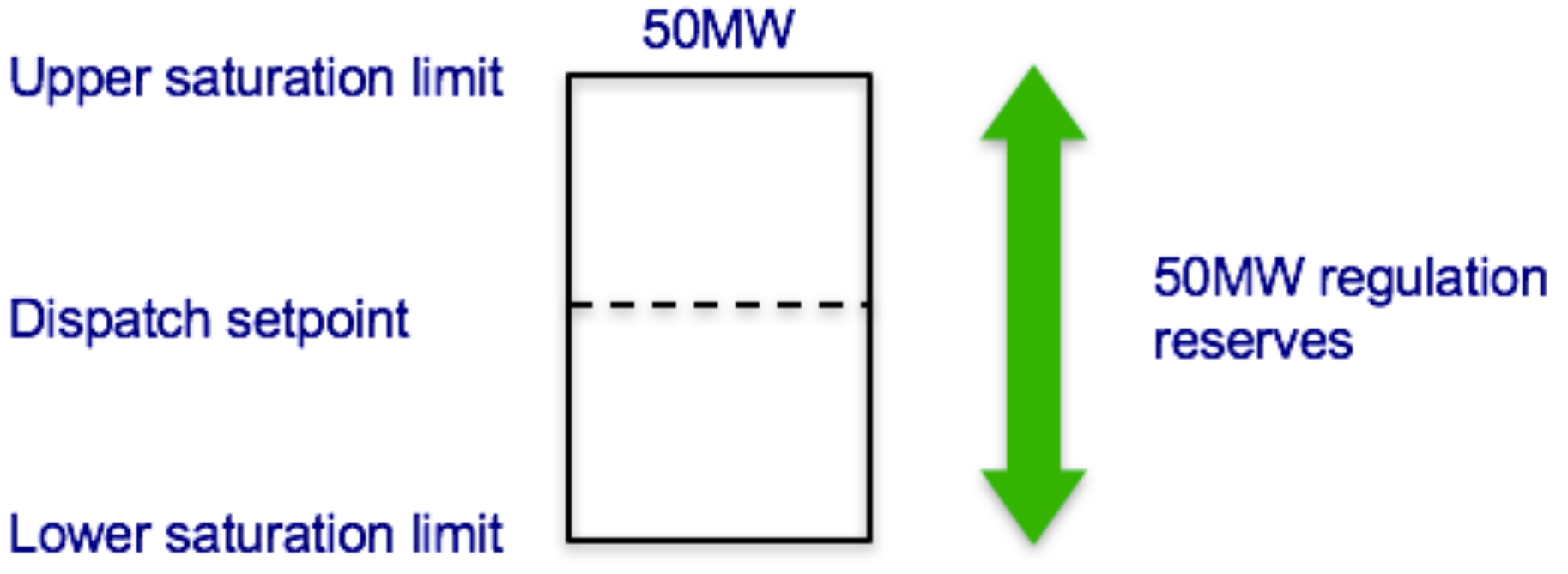}
\caption{Regulating reserves example}
\label{Fig:RegR}
\end{figure}

\begin{defn}
\emph{Regulation Reserves} \cite{Ela:2010:00,Holttinen:2012:00}: Power capacity available during normal conditions for assistance in active power balance to correct the current imbalance that requires a fast, real-time, automatic response.  The regulation reserve requirement up or down is given by $P_{REG}^{REQ}$.  The regulation level at a given time $t$ is given by $G_t$.  Its absolute value must remain less than the requirement.  
\end{defn}
\noindent Returning back to Figure \ref{Fig:EPECS}, the regulation reserve requirement is taken as an input and is utilized in the automatic generation control (AGC) algorithm of the regulation service (See Section \ref{Sec:RegRes} for further details).  It is a physical property of the saturation limits on the AGC.  In most power systems, this quantity is monetized.  
\begin{ex}
Consider Figure \ref{Fig:RegR} for example.   It consists of a single generator that is dispatched to an arbitrary level.  Its automatic generation control has saturation limits of 50MW upward and downward.  Consequently, it provides 50MW of regulation reserves. 
\end{ex}

Together, these three types of operating reserves are used to respond to forecast errors and variability in the net load during normal operation.  In all cases, the actual quantities of these reserves are physical properties of the power system.  They exist regardless of whether the system operator places requirements on these physical quantities or whether they incentivize generators to provide these reserve quantities in the form of reserve products.  
\subsubsection{Operating Reserve Requirements in ISO New England}\label{Sec:DefReservesISONE}
\begin{figure}[!t]
\centering
\includegraphics[width=5.5in]{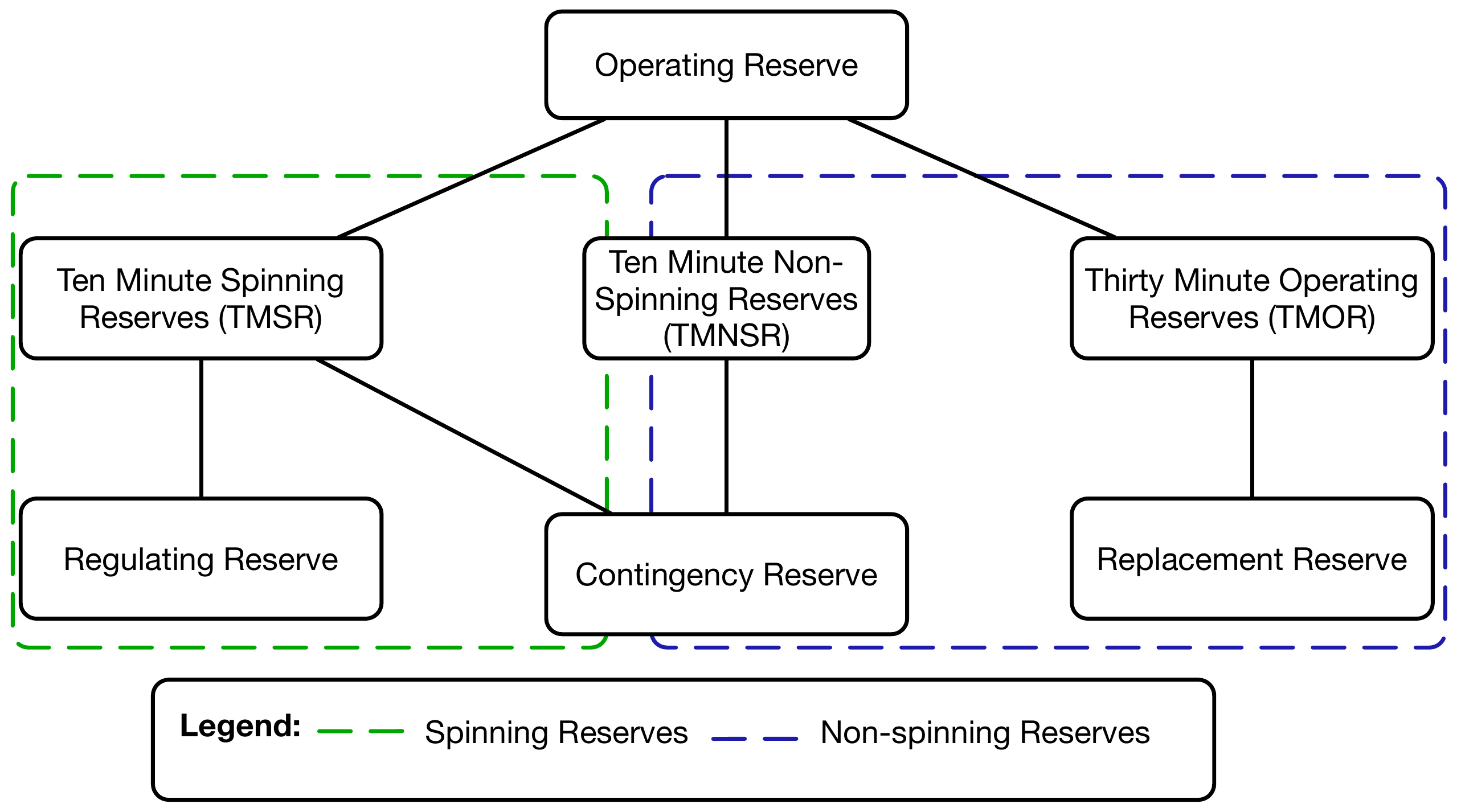}
\caption{Operating reserve classification in ISO New England}
\label{Fig:ISO-NERes}
\end{figure}

In contrast to the above, ISO-NE maintains three types of operating reserve requirements\cite{ISO-NE:2017:04}.  
\begin{defn}
\emph{Ten-Minute Spinning Reserve (TMSR)\cite{ISO-NE:2017:04}:} The TMSR is the largest reserve product that is provided by \textit{on-line resources} able to increase their output within ten minutes.  It is currently set to the largest contingency on the system.  
\end{defn}
\begin{defn}
\emph{Ten-Minute Nonspinning Reserve (TMNSR)\cite{ISO-NE:2017:04}:} The TMNSR is the second largest reserve quantity that is provided by \textit{off-line units} that can successfully synchronize to the grid and ramp up within ten minutes.  It is currently set to one half of the second largest contingency on the system.  
\end{defn}
\begin{defn}
\emph{Thirty-Minute Operating Reserve (TMOR)\cite{ISO-NE:2017:04}:} TMOR is the lowest reserve quantity that is provided by \textit{on-line resources} that can ramp up within 30 minutes and \textit{off-line units} that synchronize to the grid and ramp up within 30 minutes.  Furthermore, there exist Local TMOR requirements for three reserve zones:  Connecticut (CT), Southwest Connecticut (SWCT), and NEMA/Boston (NEMABSTN).  Until recently, it was set equal to the sum of the two ten-minute operating reserve requirements.  As of October 2013, an additional replacement reserve requirement of 160MW in the summer and 180MW in the winter was added to the TMOR\cite{ISO-NE:2017:04}.
\end{defn}
The above definitions imply a taxonomy of operating reserves shown in Figure \ref{Fig:ISO-NERes}.  Note that all three of the reserve products are defined in an upward direction as result of their focus on contingency events and because historically downward reserves have not been difficult to obtain in day-to-day operations. Furthermore, the ten-minute spinning reserve includes regulation reserves but also serves as a fast-event contingency reserve.

\subsubsection{Reconciliation of Operating Reserve Definitions for the SOARES Project}\label{Sec:ReconcileReserves}
In order to apply the EPECS simulator methodology to the ISO New England region, the two taxonomies of operating reserves summarized in Figures  \ref{Fig:OperResTaxonomy} and \ref{Fig:ISO-NERes} must be reconciled.  First, it is important to recognize that the EPECS operating reserves definitions reflect physical quantities while the ISO-NE operating reserves definitions reflect requirements.  Furthermore, it is beyond the scope of this study to define new types of operating reserve requirements.   Therefore, this project makes the following reconciliation:

\paragraph{Regulation Reserves:}
For regulation reserves, there appears to be no conceptual discrepancy. The maximum and minimum quantities of regulating reserves are equated to the regulating reserve requirement.  

\paragraph{Ten-Minute Spinning Reserves \& Load Following Reserves:}
For the ten-minute spinning reserves, we observe that this requirement is imposed on the quantity of load following reserves.  While the system will continue to require a TMSR of at least the largest contingency on the system, a high penetration of variable energy resources might require this quantity to be significantly increased.

\begin{ex}
Consider a hypothetical scenario in New England on a year where the peak load is 25GW.  A 40\% penetration of variable energy resources would equate to 10GW. If 50\% of these VERS were to drop out suddenly (beyond the forecast)\footnote{Note that a 50\% forecast error is highly unlikely for a system with 20\% penetration rate. The choice of values is purely illustrative in nature.}, there would be a 5GW shortfall.  This is significantly larger than the largest single-facility contingency in the system. Therefore, there would need to be a load following reserve requirement to address such a situation. In the absence of a new reserve requirement, the TMSR can be increased so as to respond to both single-facility contingencies as well as the variability and forecast error of variable energy resources.  
\end{ex}
\noindent Therefore, this study sets the TMSR requirement equal to the greater of two quantities: 1.) the size of the largest contingency 2.) the load following reserve requirement. The determination of the latter is part of the central objective of this work. In this context, the TMSR needs to be understood in both an upward as well as a downward direction.   

\paragraph{Non-Spinning Reserves:}
The two non-spinning reserve requirements will remain unchanged. VER integration is fundamentally a normal operation phenomena. Non-spinning reserves only protect the system in the event of a loss of generation but do not protect the system in the event of an excess of generation. Furthermore, the variability of renewable energy generation means that a system with a negative imbalance can quickly switch to a system with a positive imbalance. Therefore, it is inadvisable to try to protect the power system from VER variability and forecast error with non-spinning reserves.  

\paragraph{Ramping Reserves:}
Finally, in the case of ramping reserves, currently there is no requirement in ISO New England that provides an effective equivalent. This study will determine the ramping reserve requirements for the scenarios described in Section \ref{sec:scenarios}. Such results may motivate the need for the implementation of a ramping reserve requirement.

\subsection{Day-Ahead Resource Scheduling at ISO New England}\label{Sec:SCUC}
Power system balancing operations start with day-ahead resource scheduling implemented as a security-constrained unit commitment (SCUC). The goal of the SCUC problem is to choose the right set of generation units that are able to meet the real-time demand at minimum cost. In the original formulation, the SCUC problem is formulated as a mixed integer nonlinear optimization program with integrated power flow equations and system security requirements \cite{Frank:2012:00}. However, the optimization constraints are often linearized, as in \cite{Muzhikyan:2015:SPG-J15,Muzhikyan:2015:SPG-J16}, to avoid potential convergence issues. The SCUC formulation in \cite{Muzhikyan:2015:SPG-J15,Muzhikyan:2015:SPG-J16} has been further modified to reflect ISO-NE operations. In particular: 
\begin{enumerate}
\item Constraints reflecting minimum up time, minimum down time and maximum number of daily start-ups of the generators are added, which also take the initial online hours into account.
\item The outages are incorporated into the model.
\item The optimization program models pumped-storage units to reflect operating parameters, including the maximum daily energy constraints, the maximum draw down, and the reservoir limitations.
\item Constraints ensuring procurement of system-wide ten-minute and 30-minute reserve requirements are added to the SCUC model.
\item A zonal network model is implemented.
\item External transactions with proper interface limits are modeled.
\end{enumerate}  
The generation cost curves are modeled as quadratic functions of heat rates. The total operation cost is a combination of the generation cost, generator startup and shutdown costs, and the ``supergeneration"\footnote{Mathematically speaking, ``super-generators" implement a penalty factor in the objective function so that the hard power balance constraint can be turned into a soft one.  This provides a robust solution that protects against infeasible optimization solutions.  Physically speaking, negative values of super-generation indicates the need for curtailment of semi-dispatchable resources.  Positive values of super-generation indicates a short-fall of dispatchable generation which rarely occurs in operations timescale studies.} cost:
\begin{align}
&\min \sum_{t = 1}^{T}\left(\sum_{k = 1}^{N_G}C_{Fkt}\left(w_{Gkt}H_{Fk}+H_{Lk}P_{kt}+H_{Qk}P_{kt}^{2}+u_{Gkt}H_{Uk}+v_{Gkt}H_{Dk}\right)+\sum_{m = 1}^{N_D}C_{m}P_{mt}\right.\nonumber\\
&\left.+\sum_{x = 1}^{N_X}C_{Fxt}H_{Lx}\left(P_{xt}+N_{xt}\right) + \sum_{\mathcal{L}=1}^{N_\mathcal{L}}C_\mathcal{L}d_\mathcal{L}w_{\mathcal{L}t}\tilde{P}_{\mathcal{L}t} + \sum_{\mathcal{W}=1}^{N_\mathcal{W}}C_\mathcal{W}\big(1 - d_\mathcal{W}w_{\mathcal{W}t}\big)\tilde{P}_{\mathcal{W}t} + \right.\nonumber\\
&\left.+ \sum_{\mathcal{V}=1}^{N_\mathcal{V}}C_\mathcal{V}\big(1 - d_\mathcal{V}w_{\mathcal{V}t}\big)\tilde{P}_{\mathcal{V}t} + \sum_{\mathcal{H}=1}^{N_\mathcal{H}}C_\mathcal{H}\big(1 - d_\mathcal{H}w_{\mathcal{H}t}\big)\tilde{P}_{\mathcal{H}t} + \sum_{\mathcal{T}=1}^{N_\mathcal{T}}C_\mathcal{T}\big(1 - w_{O\mathcal{T}t}\big)\big(1 - d_\mathcal{T}w_{\mathcal{T}t}\big)\tilde{P}_{\mathcal{T}t}\right) \label{eq:SCUCobj}
\end{align}
The optimization program is subject to the following constraints:
\begin{align}
& \sum_{k=1}^{N_G}A_{nk}P_{kt} + \sum_{s=1}^{N_S}A_{ns}(P_{st} - S_{st}) + \sum_{x=1}^{N_X}A_{nx}(P_{xt} - N_{xt})                     && \nonumber\\
& \hspace{0.5in} + \sum_{m=1}^{N_D}A_{nm}P_{mt} - (1 + \gamma)\sum_{\mathcal{L}=1}^{N_\mathcal{L}}A_{n\mathcal{L}}\big(1 - d_\mathcal{L}w_{\mathcal{L}t}\big)\tilde{P}_{\mathcal{L}t}&& \nonumber\\
& \hspace{0.5in} + (1 + \gamma)\sum_{\mathcal{W}=1}^{N_\mathcal{W}}A_{n\mathcal{W}}\big(1 - d_\mathcal{W}w_{\mathcal{W}t}\big)\tilde{P}_{\mathcal{W}t}&& \nonumber\\
& \hspace{0.5in} + (1 + \gamma)\sum_{\mathcal{V}=1}^{N_\mathcal{V}}A_{n\mathcal{V}}\big(1 - d_\mathcal{V}w_{\mathcal{V}t}\big)\tilde{P}_{\mathcal{V}t}&& \nonumber\\
& \hspace{0.5in} + (1 + \gamma)\sum_{\mathcal{H}=1}^{N_\mathcal{H}}A_{n\mathcal{H}}\big(1 - d_\mathcal{H}w_{\mathcal{H}t}\big)\tilde{P}_{\mathcal{H}t}&& \nonumber\\
& \hspace{0.5in} + \sum_{\mathcal{T}=1}^{N_\mathcal{T}}A_{n\mathcal{T}}\big(1 - w_{O\mathcal{T}t}\big)\big(1 - d_\mathcal{T}w_{\mathcal{T}t}\big)\tilde{P}_{\mathcal{T}t} = \sum_{l = 1}^{N_L}B_{nl}F_{lt}	&& n = 1:N_B; t = 1:T  \label{eq:SCUCbal} \\
& \sum_{l=1}^{N_L}K_{il}F_{lt} \leq I_i^{max}	&& i = 1:N_I; t = 1:T                  \label{eq:SCUCint} \\
& w_{Gkt}P_{k}^{min}\leq P_{kt}\leq w_{Gkt}\big(1 - w_{Okt}\big)P_{k}^{max}                 && k = 1:N_G; t = 1:T   \label{eq:SCUCplim}     \\
& P_{m}^{min}\leq P_{mt}\leq P_{m}^{max}                                            && m = 1:N_D; t = 1:T   \label{eq:SCUCpmlim}  
\end{align}
\begin{align}
& w_{Pst}P_{s}^{min}\leq P_{st}\leq w_{Pst}P_{s}^{max}                              && s = 1:N_S; t = 1:T   \label{eq:SCUCpslim}    \\
& w_{Sst}S_{s}^{min}\leq S_{st}\leq w_{Sst}S_{s}^{max}                              && s = 1:N_S; t = 1:T   \label{eq:SCUCsslim}  \\  
& R_k^{min} - \frac{P_k^{max}}{T_h}v_{Gkt}\leq \dfrac{P_{kt}-P_{k,t-1}}{T_h} \leq R_k^{max} + \frac{P_k^{max}}{T_h}u_{Gkt}  && k = 1:N_G; t = 1:T   \label{eq:SCUCrlim}     \\ 
& E_{st} = E_{s,t-1} + (\eta_sS_{st} - P_{st})\cdot T_h                             && s = 1:N_S; t = 1:T   \label{eq:SCUCstor}     \\
& E_{s}^{min}\leq E_{st}\leq E_{s}^{max}                                            && s = 1:N_S; t = 1:T   \label{eq:SCUCeslim}\\
& P_{k0} = \xi_k                                                                    && k = 1:N_G            \label{eq:SCUCrlim2}    \\
& E_{s0} = \varepsilon_s                                                            && s = 1:N_S            \label{eq:SCUCstor2}    \\
& w_{Gk,t-1} + u_{Gkt} - v_{Gkt} = w_{Gkt}                                          && k = 1:N_G; t = 1:T   \label{eq:SCUCbin}      \\
& u_{Gkt} + v_{Gkt} \leq 1                                                          && k = 1:N_G; t = 1:T   \label{eq:SCUCbin2}     \\
& w_{Pst} + w_{Sst} \leq 1                                                          && s = 1:N_S; t = 1:T   \label{eq:SCUCbin3}     \\
& w_{Pst} + w_{Ss,(t-1)} \leq 1                                                     && s = 1:N_S; t = 1:T   \label{eq:SCUCbin4}     \\
& w_{Ps,(t-1)} + w_{Sst} \leq 1                                                     && s = 1:N_S; t = 1:T   \label{eq:SCUCbin5}     \\
& w_{Ps0} = \omega_{Ps0}                                                            && s = 1:N_S            \label{eq:SCUCbin6}     \\
& w_{Ss0} = \omega_{Ss0}                                                            && s = 1:N_S            \label{eq:SCUCbin7}     \\
& w_{Gkt} \geq u_{Gk,(t - \tau)}	                                                && k = 1:N_G; t = 1:T, \tau = 1:T_{u}-1 \label{eq:SCUCminup}    \\	
& 1 - w_{Gkt} \geq v_{Gk,(t - \tau)}	                                                && k = 1:N_G; t = 1:T, \tau = 1:T_{d}-1 \label{eq:SCUCmindown}  \\	
& \sum_{t=1}^Tu_{Gkt} \leq u_{Gk}^{max}                                                 && k = 1:N_G                            \label{eq:SCUCmaxup}    \\	
& C_{1t} \geq A_{nk}w_{Gkt}P_k^{max}                  && t = 1:T          \label{eq:SCUCCG1}     \\
& C_{1t} \geq A_{n\mathcal{T}}\big(1 - w_{O\mathcal{T}t}\big)\big(1 - d_\mathcal{T}w_{\mathcal{T}t}\big)\tilde{P}_{\mathcal{T}t} && t = 1:T \label{eq:SCUCCT1}     \\
& P_{Gkt}^{TMSR} \leq w_{Gkt}P_k^{max} - P_{kt}                                     && k = 1:N_G; t = 1:T   \label{eq:SCUCTMSR1} \\
& P_{Gkt}^{TMSR} \leq R_k^{max}\cdot T_{10}                                         && k = 1:N_G; t = 1:T   \label{eq:SCUCTMSR2} \\
& P_{nt}^{TMSR} = \sum_{k = 1}^{N_G}A_{nk}P_{Gkt}^{TMSR}                            && n = 1:N_B; t = 1:T   \label{eq:SCUCTMSR3} \\
& P_{nt}^{TMSR} \geq \alpha_n^{TMSR}\cdot\alpha_{sys}^{TMR}\cdot C_{1t}             && n = 1:N_B; t = 1:T   \label{eq:SCUCTMSR4} \\
& \sum_{n=1}^{N_B}P_{nt}^{TMSR} \geq \alpha_{sys}^{TMSR}\cdot\alpha_{sys}^{TMR}\cdot C_{1t}     && t = 1:T  \label{eq:SCUCTMSR5} \\
& P_{Gkt}^{TMOR} \leq \big(1 - w_{Gkt}\big)P_k^{max}                                && k = 1:N_G; t = 1:T   \label{eq:SCUCTMOR1} \\
& P_{Gkt}^{TMOR} \leq R_k^{max}\cdot T_{30}                                         && k = 1:N_G; t = 1:T   \label{eq:SCUCTMOR2} \\
& P_{nt}^{TMOR} = \sum_{k = 1}^{N_G}A_{nk}P_{Gkt}^{TMOR}                            && n = 1:N_B; t = 1:T   \label{eq:SCUCTMOR3} \\
& P_{nt}^{TMSR} + P_{nt}^{TMOR} \geq \alpha_n^{TMOR}\cdot\alpha_{sys}^{TMR}\cdot C_{1t}     && n = 1:N_B; t = 1:T   \label{eq:SCUCTMOR4}\\
& \sum_{n=1}^{N_B}\Big(P_{nt}^{TMSR} + P_{nt}^{TMOR}\Big) \geq \alpha_{sys}^{TMOR}\cdot\alpha_{sys}^{TMR}\cdot C_{1t}   && t = 1:T  \label{eq:SCUCTMOR5}
\end{align}
where the following notations are used:
\begin{align*}
& k, m, x, s, n, l, i, t                && \text{generator, active DR, supergenerator, storage, bubble, branch, interface}\\
&                                       && \text{and time indices respectively;}                             \\
& N_G, N_D, N_X, N_S, N_B, N_L, N_I     && \text{number of generators, active DR's, supergenerators, storages, bubbles,} \\
&                                       && \text{branches and interfaces respectively;}\\
& \mathcal{L}, \mathcal{W}, \mathcal{V}, \mathcal{H}, \mathcal{T}   && \text{load, wind, solar, hydro, tie line indices respectively;}\\
& N_\mathcal{L}, N_\mathcal{W}, N_\mathcal{V}, N_\mathcal{H}, N_\mathcal{T}   && \text{number of loads, winds, solars, hydros, tie lines respectively;}\\
& A_{n,(k,s,x,\mathcal{L},\mathcal{W},\mathcal{V},\mathcal{H},\mathcal{T})}		&& \text{incidence matrix of (generators, storages, supergenerators,}\\
&                                                                               && \text{loads, winds, solars, hydros, tie lines) to bubbles;}                                     \\
& B_{nl}					            && \text{incidence matrix of branches to bubbles;}                                    \\
& K_{il}					            && \text{incidence matrix of branches to interfaces;}                                    \\
& T_h, T                                		&& \text{SCUC time step and horizon;}                                                           \\
& H_{Fk}, H_{Lk}, H_{Qk}, H_{Uk}, H_{Dk} && \text{fixed, linear, quadratic, startup and shutdown heat rates for generator $k$;}     \\
& C_{Fkt}                               && \text{fuel cost of generator $k$ at time $t$;}                                           \\
& C_{m}                                 && \text{linear cost of active DR unit $m$;}                                           \\
& C_\mathcal{L}, C_\mathcal{W}, C_\mathcal{V}, C_\mathcal{H}, C_\mathcal{T}     && \text{load, wind, solar, hydro and tie line curtailment threshold prices respectively;}\\                 & P_k^{max},P_k^{min}                   && \text{minimum/maximum power outputs of generator $k$;}          \\
& R_k^{max},R_k^{min}                   && \text{maximum/minimum ramping rate of generator $k$;}                                        \\
& P_s^{max},P_s^{min}                   && \text{minimum/maximum power outputs of storage $s$;}                                       \\
& S_s^{max},S_s^{min}                   && \text{minimum/maximum pumping rate of storage $s$;}                                       \\
& E_s^{max},E_s^{min}                   && \text{minimum/maximum energy capacity of storage $s$;}                                       \\
& T_{u}, T_{d}, u_{Gk}^{max}            && \text{minimum up time, minimum down time and maximum startups}                               \\
&                                       && \text{in a day for generator $k$;}                                                           \\
& d_\mathcal{L}, d_\mathcal{W}, d_\mathcal{V}, d_\mathcal{H}, d_\mathcal{T}     && \text{curtailable fractions of load, wind, solar, hydro and tie line respectively;}                               \\
& w_{Gkt}, u_{Gkt}, v_{Gkt}             && \text{ON/OFF, startup and shutdown statuses of generator $k$ at time $t$;}              \\
& w_{\mathcal{L}t}, w_{\mathcal{W}t}, w_{\mathcal{V}t}, w_{\mathcal{H}t}, w_{\mathcal{T}t}  && \text{fractions of curtailable load, wind, solar, hydro and tie line curtailed at time $t$;}              \\
& w_{Okt}, w_{O\mathcal{T}t}            && \text{fractions of generator $k$ and tie-line $\mathcal{T}$ under outage at time $t$;} \\
& w_{Pst}, w_{Sst}                      && \text{generation and pumping mode indicators of storage $s$ at time $t$;}              \\
& P_{kt}, \xi_k                         && \text{power output of generator $k$ at time $t \geq 1$ and $t = 0$;}                         \\
& P_{st}, S_{st}                        && \text{generation and pumping rates of storage $s$ at time $t$;}                         \\
& E_{st}, \varepsilon_s                 && \text{reservoir level of storage $s$ at time $t \geq 1$ and $t = 0$;}                         \\
& P_{xt}, N_{xt}                        && \text{positive and negative components of supergenerator $x$ output at time $t$;}    \\
& F_{lt}                                && \text{power flow through branch $l$ at time $t$;}                                     \\
& \tilde{P}_{\mathcal{L}t}, \tilde{P}_{\mathcal{W}t}, \tilde{P}_{\mathcal{V}t}, \tilde{P}_{\mathcal{H}t}, \tilde{P}_{\mathcal{T}t},                         && \text{load, wind, solar, hydro and tie line forecasts at time $t$;}          \\
& \gamma                                && \text{transmission losses as a percentage of the total demand;}\\
\end{align*}
\begin{align*}
& C_{1t}                                && \text{largest contingency at time $t$;}\\
& P_{Gkt}^{TMSR}, P_{Gkt}^{TMOR}        && \text{amount of TMSR and TMOR obtained from generator $k$ at time $t$;}\\
& P_{nt}^{TMSR}, P_{nt}^{TMOR}          && \text{amount of TMSR and TMOR available at bubble $n$ at time $t$;}\\
& \alpha_n^{TMSR}, \alpha_n^{TMR}, \alpha_n^{TMOR} && \text{TMSR, TMR and TMOR requirements at bubble $n$ as percentages of the}\\
&                                       && \text{largest contingency;}\\
& \alpha_{sys}^{TMSR}, \alpha_{sys}^{TMR}, \alpha_{sys}^{TMOR} && \text{system-wide TMSR, TMR and TMOR requirements as percentages of the}\\
&                                       && \text{largest contingency.}
\end{align*}
Constraint (\ref{eq:SCUCbal}) is the DC power flow equation with incorporated loss term. Constraint (\ref{eq:SCUCint}) sets the interface limits. Constraints (\ref{eq:SCUCplim})--(\ref{eq:SCUCsslim}) set generator, active demand response and storage power output maximum and minimum limits. Constraint (\ref{eq:SCUCrlim}) places limits on the generator up and down ramping rates. Constraints~(\ref{eq:SCUCstor})--(\ref{eq:SCUCeslim}) set storage energy limits. Constraints (\ref{eq:SCUCbin})--(\ref{eq:SCUCbin7}) logically bind the status binary variables of generators and storage units. Constraints (\ref{eq:SCUCminup}) and (\ref{eq:SCUCmindown}) set the generator minimum up and minimum down times respectively. Constraint (\ref{eq:SCUCmaxup}) limits the maximum number of generator startups in a day. Constraints~(\ref{eq:SCUCCG1})--(\ref{eq:SCUCCT1}) calculate the largest generator and tie line contingencies respectively. Constraints~(\ref{eq:SCUCTMSR1})--(\ref{eq:SCUCTMSR5}) procure ten-minute spinning reserves (TMSR) from online units. Similarly, constraints (\ref{eq:SCUCTMOR1})--(\ref{eq:SCUCTMOR5})  procure thirty-minute operating reserves (TMOR) from offline fast-start units. 

\subsection{Same-Day Resource Scheduling at ISO-NE}\label{Sec:RTUC}
The same-day resource scheduling uses an optimization program similar to that of the SCUC. The optimization program, called real-time unit commitment (RTUC), is modified in the following ways to reflect ISO-NE operations:
\begin{enumerate}
\item The optimization considers 16 15-minute time intervals, spanning a 4-hour period.
\item This optimization program is run once every hour rather than once a day (in the case of the day-ahead resource scheduling). \item The process only commits and de-commits fast-start units.
\item The commitment is based upon short-term load and VER forecasts (a couple of hours look-ahead).
\item This optimization model enforces system  reserve requirements.
\end{enumerate}
The formulation of the RTUC is similar to the SCUC. The objective function is written as:
\begin{align}
& \min \sum_{t = 1}^{T}\left(\sum_{k = 1}^{N_G}C_{Fkt}\left(w_{Gkt}H_{Fk}+H_{Lk}P_{kt}+H_{Qk}P_{kt}^{2}+u_{Gkt}H_{Uk}+v_{Gkt}H_{Dk}\right)+\sum_{m = 1}^{N_D}C_{m}P_{mt}\right.\nonumber\\
&\left.+\sum_{x = 1}^{N_X}C_{Fxt}H_{Lx}\left(P_{xt}+N_{xt}\right) + \sum_{\mathcal{L}=1}^{N_\mathcal{L}}C_\mathcal{L}d_\mathcal{L}w_{\mathcal{L}t}\tilde{P}_{\mathcal{L}t} + \sum_{\mathcal{W}=1}^{N_\mathcal{W}}C_\mathcal{W}\big(1 - d_\mathcal{W}w_{\mathcal{W}t}\big)\tilde{P}_{\mathcal{W}t}\right.\nonumber\\
& \left.+ \sum_{\mathcal{V}=1}^{N_\mathcal{V}}C_\mathcal{V}\big(1 - d_\mathcal{V}w_{\mathcal{V}t}\big)\tilde{P}_{\mathcal{V}t} + \sum_{\mathcal{H}=1}^{N_\mathcal{H}}C_\mathcal{H}\big(1 - d_\mathcal{H}w_{\mathcal{H}t}\big)\tilde{P}_{\mathcal{H}t} + \sum_{\mathcal{T}=1}^{N_\mathcal{T}}C_\mathcal{T}\big(1 - w_{O\mathcal{T}t}\big)\big(1 - d_\mathcal{T}w_{\mathcal{T}t}\big)\tilde{P}_{\mathcal{T}t}\right)\label{eq:RTUCobj}
\end{align}
The optimization program is subject to the following constraints:
\begin{align}
& \sum_{k=1}^{N_G}A_{nk}P_{kt} + \sum_{s=1}^{N_S}A_{ns}(\mathcal{P}_{st} - \mathcal{S}_{st}) + \sum_{x=1}^{N_X}A_{nx}(P_{xt} - N_{xt})                     && \nonumber\\
& \hspace{0.5in}+ \sum_{m=1}^{N_D}A_{nm}P_{mt} - (1 + \gamma)\sum_{\mathcal{L}=1}^{N_\mathcal{L}}A_{n\mathcal{L}}\big(1 - d_\mathcal{L}w_{\mathcal{L}t}\big)\tilde{P}_{\mathcal{L}t} && \nonumber\\
& \hspace{0.5in}+ (1 + \gamma)\sum_{\mathcal{W}=1}^{N_\mathcal{W}}A_{n\mathcal{W}}\big(1 - d_\mathcal{W}w_{\mathcal{W}t}\big)\tilde{P}_{\mathcal{W}t} && \nonumber\\
& \hspace{0.5in}+ (1 + \gamma)\sum_{\mathcal{V}=1}^{N_\mathcal{V}}A_{n\mathcal{V}}\big(1 - d_\mathcal{V}w_{\mathcal{V}t}\big)\tilde{P}_{\mathcal{V}t} && \nonumber\\
& \hspace{0.5in} + (1 + \gamma)\sum_{\mathcal{H}=1}^{N_\mathcal{H}}A_{n\mathcal{H}}\big(1 - d_\mathcal{H}w_{\mathcal{H}t}\big)\tilde{P}_{\mathcal{H}t} && \nonumber\\
& \hspace{0.5in} + \sum_{\mathcal{T}=1}^{N_\mathcal{T}}A_{n\mathcal{T}}\big(1 - w_{O\mathcal{T}t}\big)\big(1 - d_\mathcal{T}w_{\mathcal{T}t}\big)\tilde{P}_{\mathcal{T}t} = \sum_{l = 1}^{N_L}B_{nl}F_{lt}	&& n = 1:N_B; t = 1:T  \label{eq:RTUCbal} \\
& \sum_{l=1}^{N_L}K_{il}F_{lt} \leq I_i^{max}	&& i = 1:N_I; t = 1:T                  \label{eq:RTUCint}\\
& w_{Gkt}P_{k}^{min}\leq P_{kt}\leq w_{Gkt}\big(1 - w_{Okt}\big)P_{k}^{max}                              && k = 1:N_G; t = 1:T   \label{eq:RTUCplim}     \\
& P_{m}^{min}\leq P_{mt}\leq P_{m}^{max}                              && m = 1:N_D; t = 1:T   \label{eq:RTUCpmlim}     \\
& R_k^{min} - \frac{P_k^{max}}{T_r}v_{Gkt} \leq \dfrac{P_{kt}-P_{k,t-1}}{T_r} \leq R_k^{max} + \frac{P_k^{max}}{T_r}u_{Gkt}                      && k = 1:N_G; t = 1:T   \label{eq:RTUCrlim}     \\ 
& P_{k0} = \xi_k                                                                    && k = 1:N_G            \label{eq:RTUCrlim2}    \\
& w_{Gk,t-1} + u_{Gkt} - v_{Gkt} = w_{Gkt}                                          && k = 1:N_G; t = 1:T   \label{eq:RTUCbin}      \\
& w_{G\kappa t} = \omega_{G\kappa t}                                          && \kappa = 1:\mathcal{N}_G; t = 1:T   \label{eq:RTUCbin2}      \\
& u_{Gkt} + v_{Gkt} \leq 1                                                          && k = 1:N_G; t = 1:T   \label{eq:RTUCbin3}     \\
& w_{Gkt} \geq u_{Gk,(t - \tau)}	                                                && k = 1:N_G; t = 1:T, \tau = 1:T_{u}-1 \label{eq:RTUCminup}    \\	
& 1 - w_{Gkt} \geq v_{Gk,(t - \tau)}	                                                && k = 1:N_G; t = 1:T, \tau = 1:T_{d}-1 \label{eq:RTUCmindown}  \\	
& n_{Gk} + \sum_{t=1}^Tu_{Gkt} + m_{Gk} \leq u_{Gk}^{max}                                                 && k = 1:N_G                            \label{eq:RTUCmaxup}    \\	
& C_{1t} \geq A_{nk}w_{Gkt}P_k^{max}                  && t = 1:T          \label{eq:RTUCCG1}     \\
& C_{1t} \geq A_{n\mathcal{T}}\big(1 - w_{O\mathcal{T}t}\big)\big(1 - d_\mathcal{T}w_{\mathcal{T}t}\big)\tilde{P}_{\mathcal{T}t} && t = 1:T \label{eq:RTUCCT1}     \\
& P_{Gkt}^{TMSR} \leq w_{Gkt}P_k^{max} - P_{kt}                                     && k = 1:N_G; t = 1:T   \label{eq:RTUCTMSR1} \\
& P_{Gkt}^{TMSR} \leq R_k^{max}\cdot T_{10}                                         && k = 1:N_G; t = 1:T   \label{eq:RTUCTMSR2} \\
& P_{nt}^{TMSR} = \sum_{k = 1}^{N_G}A_{nk}P_{Gkt}^{TMSR}                            && n = 1:N_B; t = 1:T   \label{eq:RTUCTMSR3} 
\end{align}
\begin{align}
& \hspace{-0.5in} P_{nt}^{TMSR} \geq \alpha_n^{TMSR}\cdot\alpha_{sys}^{TMR}\cdot C_{1t}             && n = 1:N_B; t = 1:T   \label{eq:RTUCTMSR4} \\
& \hspace{-0.5in} \sum_{n=1}^{N_B}P_{nt}^{TMSR} \geq \alpha_{sys}^{TMSR}\cdot\alpha_{sys}^{TMR}\cdot C_{1t}     && t = 1:T  \label{eq:RTUCTMSR5} \\
& \hspace{-0.5in} P_{Gkt}^{TMOR} \leq \big(1 - w_{Gkt}\big)P_k^{max}                                && k = 1:N_G; t = 1:T   \label{eq:RTUCTMOR1} \\
& \hspace{-0.5in} P_{Gkt}^{TMOR} \leq R_k^{max}\cdot T_{30}                                         && k = 1:N_G; t = 1:T   \label{eq:RTUCTMOR2} \\
& \hspace{-0.5in} P_{nt}^{TMOR} = \sum_{k = 1}^{N_G}A_{nk}P_{Gkt}^{TMOR}                            && n = 1:N_B; t = 1:T   \label{eq:RTUCTMOR3} \\
& \hspace{-0.5in} P_{nt}^{TMSR} + P_{nt}^{TMOR} \geq \alpha_n^{TMOR}\cdot\alpha_{sys}^{TMR}\cdot C_{1t}     && n = 1:N_B; t = 1:T   \label{eq:RTUCTMOR4}\\
& \hspace{-0.5in} \sum_{n=1}^{N_B}\Big(P_{nt}^{TMSR} + P_{nt}^{TMOR}\Big) \geq \alpha_{sys}^{TMOR}\cdot\alpha_{sys}^{TMR}\cdot C_{1t}   && t = 1:T  \label{eq:RTUCTMOR5}
\end{align}
where the following notations are used in addition to the ones introduced in the previous section:
\begin{align*}
& \kappa        && \text{indices not-fast-start generators;}    \\
& \mathcal{N}_G && \text{number not-fast-start generators;}     \\
& T_r, T        && \text{RTUC time step and horizon;}           \\
& n_{Gk}, m_{Gk}   && \text{number of startups during the day before and after the current RTUC time} \\
&                   && \text{block respectively;}                           \\
& \omega_{G\kappa t}    && \text{the commitment schedules of not-fast-start units obtained from SCUC}                            \\
& \mathcal{P}_{st}, \mathcal{S}_{st}                                          && \text{storage generation and pumping schedules obtained from the SCUC.} 
\end{align*}

\subsection{Real-Time Balancing Operations at ISO-NE}\label{Sec:SCED}
The real-time balancing operations move available generator outputs to new setpoints (dispatch) in the most cost-efficient way. In its original formulation, generation dispatch is implemented as a non-linear optimization model, called AC optimal power flow (ACOPF) \cite{Carpentier:1962:00}. Due to problems with convergence and computational complexity \cite{Frank:2012:00}, most of the U.S. independent system operators (ISO) moved from ACOPF to linear optimization models. The most commonly used model is called security-constrained economic dispatch (SCED) \cite{Stott:2009:00}. This SCED formulation has been further modified to reflect ISO-NE operations. In particular:
\begin{enumerate}
\item The modified SCED adopts a 10-min look-ahead window, and considers the initial state of a unit (UCM code) and its start-up and shut-down instruction from the RTUC.
\item Area interchanges are honored.
\end{enumerate}
The objective function is written as:
\begin{align}
& \hspace{-1in}\min\left(\sum_{k = 1}^{N_G}C_{Fk}\left(H_{Lk}P_{k}+H_{Qk}P_{k}^{2}\right)+\sum_{m = 1}^{N_D}C_{m}P_{m}+\sum_{x = 1}^{N_X}C_{Fx}H_{Lx}\left(P_{x}+N_{x}\right)  \right.\nonumber\\
& \hspace{-0.5in}\left.+ \sum_{\mathcal{L}=1}^{N_\mathcal{L}}C_\mathcal{L}d_\mathcal{L}w_{\mathcal{L}}\tilde{P}_{\mathcal{L}} + \sum_{\mathcal{W}=1}^{N_\mathcal{W}}C_\mathcal{W}\big(1 - d_\mathcal{W}w_{\mathcal{W}}\big)\tilde{P}_{\mathcal{W}} + \sum_{\mathcal{V}=1}^{N_\mathcal{V}}C_\mathcal{V}\big(1 - d_\mathcal{V}w_{\mathcal{V}}\big)\tilde{P}_{\mathcal{V}}\right.\nonumber\\
& \hspace{-0.5in}\left. + \sum_{\mathcal{H}=1}^{N_\mathcal{H}}C_\mathcal{H}\big(1 - d_\mathcal{H}w_{\mathcal{H}}\big)\tilde{P}_{\mathcal{H}} + \sum_{\mathcal{T}=1}^{N_\mathcal{T}}C_\mathcal{T}\big(1 - w_{O\mathcal{T}}\big)\big(1 - d_\mathcal{T}w_{\mathcal{T}}\big)\tilde{P}_{\mathcal{T}}\right)\label{eq:SCEDobj}
\end{align}
The optimization program is subject to the following constraints:
\begin{align}
& \sum_{k=1}^{N_G}A_{nk}P_{k} + \sum_{s=1}^{N_S}A_{ns}(\mathcal{P}_{s} - \mathcal{S}_{s}) + \sum_{x=1}^{N_X}A_{nx}(P_{x} - N_{x})                      && \nonumber\\
& \hspace{0.5in} + \sum_{k=1}^{N_G}A_{nm}P_{m} - (1 + \gamma)\sum_{\mathcal{L}=1}^{N_\mathcal{L}}A_{n\mathcal{L}}\big(1 - d_\mathcal{L}w_{\mathcal{L}}\big)\tilde{P}_{\mathcal{L}} && \nonumber\\
& \hspace{0.5in} + (1 + \gamma)\sum_{\mathcal{W}=1}^{N_\mathcal{W}}A_{n\mathcal{W}}\big(1 - d_\mathcal{W}w_{\mathcal{W}}\big)\tilde{P}_{\mathcal{W}} && \nonumber\\
& \hspace{0.5in} + (1 + \gamma)\sum_{\mathcal{V}=1}^{N_\mathcal{V}}A_{n\mathcal{V}}\big(1 - d_\mathcal{V}w_{\mathcal{V}}\big)\tilde{P}_{\mathcal{V}} && \nonumber\\
& \hspace{0.5in} + (1 + \gamma)\sum_{\mathcal{H}=1}^{N_\mathcal{H}}A_{n\mathcal{H}}\big(1 - d_\mathcal{H}w_{\mathcal{H}}\big)\tilde{P}_{\mathcal{H}} && \nonumber\\
& \hspace{0.5in} + \sum_{\mathcal{T}=1}^{N_\mathcal{T}}A_{n\mathcal{T}}\big(1 - w_{O\mathcal{T}}\big)\big(1 - d_\mathcal{T}w_{\mathcal{T}}\big)\tilde{P}_{\mathcal{T}} = \sum_{l = 1}^{N_L}B_{nl}F_{l}	&& n = 1:N_B  \label{eq:SCEDbal}  \\
& \sum_{l=1}^{N_L}K_{il}F_{l} \leq I_i^{max}	&& i = 1:N_I                \label{eq:SCEDint} \\
& \omega_{Gk}P_k^{min} \leq P_{k} \leq \omega_{Gk}\big(1 - w_{Ok}\big)P_k^{max}                                       && k = 1:N_G  \label{eq:SCEDplim} \\
& P_m^{min} \leq P_{m} \leq P_m^{max}                                       && m = 1:N_D  \label{eq:SCEDmplim} \\
& R_k^{min} - \frac{P_k^{max}}{T_m}v_{Gk}\leq \dfrac{P_{k}-P_{k}^0}{T_m} \leq R_k^{max} + \frac{P_k^{max}}{T_m}u_{Gk}                      && k = 1:N_G   \label{eq:SCEDrlim}
\end{align}
where the following notations are used in addition to the ones introduced in previous sections:
\begin{align*}
& T_m       && \text{real-time balancing time step;}    \\
& \omega_{Gk}, u_{Gk}, v_{Gk}  && \text{generator state, startup and shutdown indicators for the given time step} \\
&                               && \text{obtained from the RTUC;} \\
& P_{k}^0                       && \text{current power output of generator $k$;}
\end{align*}
Constraint (\ref{eq:SCEDbal}) is the DC power flow equation with incorporated loss term. Constraint (\ref{eq:SCEDint}) sets the interface limits. Constraints (\ref{eq:SCEDplim}) and (\ref{eq:SCEDmplim}) set generator and active demand response power output limits respectively. Constraint (\ref{eq:SCEDrlim}) places limits on the generator up and down ramping rates.

\subsection{Regulation Service Model}\label{Sec:RegRes}
\begin{figure}[!t]
\centering
\includegraphics[width=5.5in]{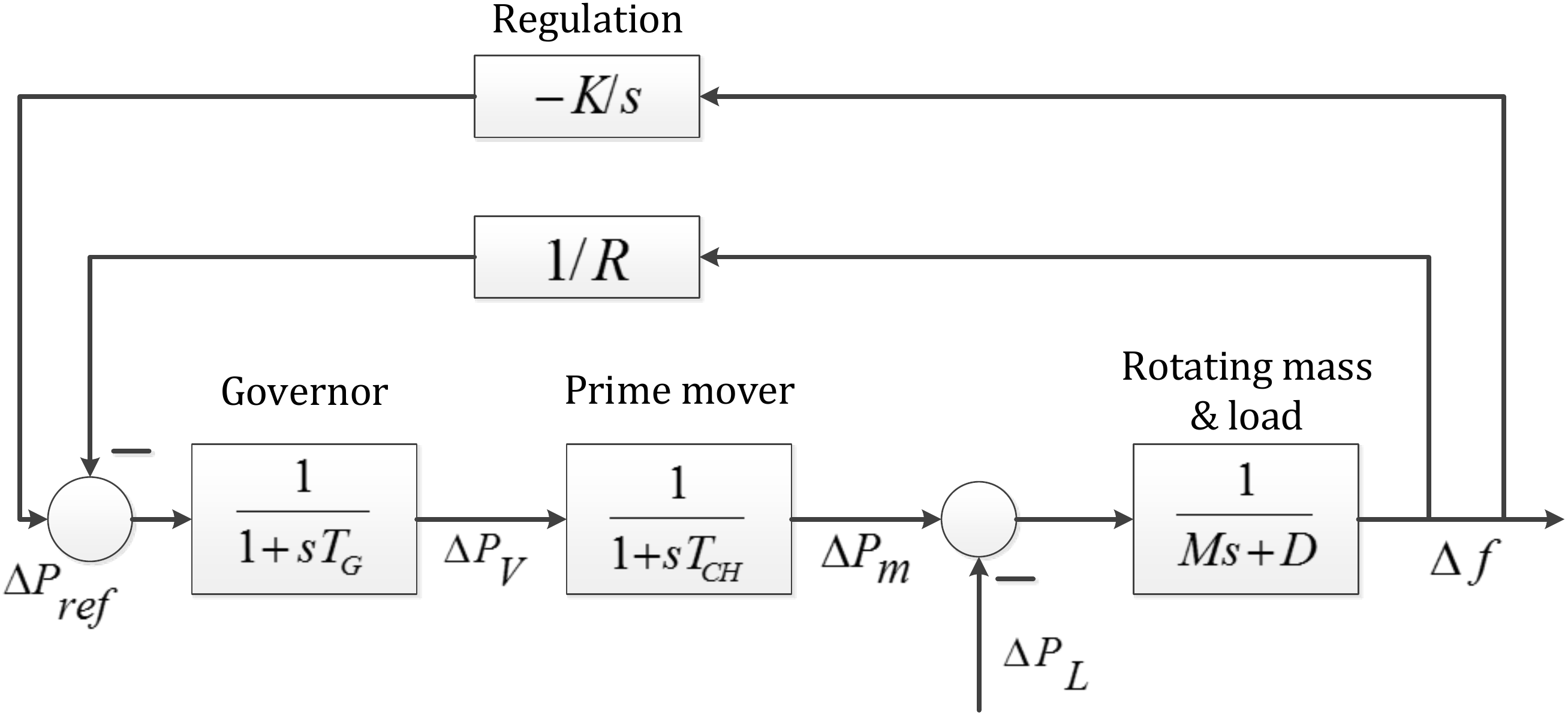}
\caption{Power system automatic generation control \cite{Wood:2014:00}}
\label{fig:dynamics}
\end{figure}

The regulation service is provided by generation units that are fully or partially controlled by the dynamic AGC model described in Fig.~\ref{fig:dynamics}. This study uses one minute increments as its finest time scale resolution. In the meantime, the cycle time of slow transient stability phenomena is approximately ten seconds. Given the 6x difference, the transfer function shown in Fig.~\ref{fig:dynamics} can be replaced with the steady-state equivalent of a gain with saturation limits. Furthermore, this work allows for the regulation service to be rate limited so as to have an ``automatic-response-rate".  In this work, the automatic response rate is set to 10\% of the regulation service saturation limits.  These, in turn, are defined by the percentage of the capacity in the corresponding generation unit controlled by AGC. In implementation, the regulation service responds to the imbalances by moving the regulation units in the opposite direction according to their predefined participation factors. The regulation units change their outputs until imbalances are mitigated or regulation service reaches saturation.

\subsection{Physical Power Grid Model}\label{Sec:physgrid}
The pseudo-steady-state approximation of the regulation service model ties directly to a power flow analysis model of the physical power grid.  Normally, the imbalances at the output of the regulation service model would be represented in frequency changes.  However, for steady-state simulations, the concept of frequency is not applicable. Instead, a designated \emph{virtual} swing bus consumes the mismatch of generation and consumption to make the steady-state power flow equations solvable. Therefore, for steady state simulations, the power system imbalance is measured at the slack generator output \cite{Gomez-Exposito:2008:00}.
\begin{figure}[!t]
\centering
\includegraphics[width=5.5in]{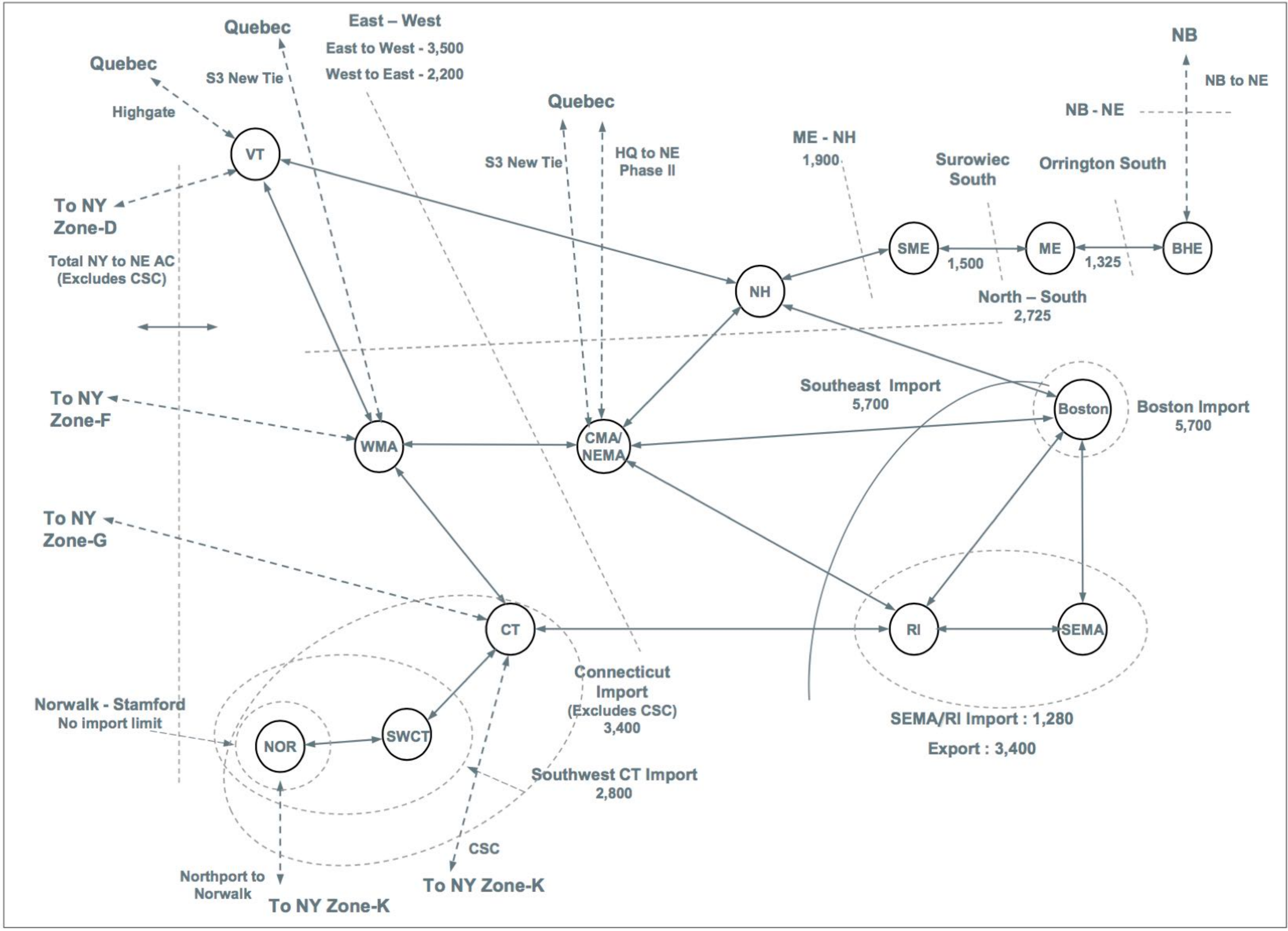}
\caption{Topology of the ISO-NE zonal network model \cite{Coste:2016:00}}
\label{fig:topology}
\end{figure}

In the SOARES study, the full AC topology of ISO-NE is replaced by the zonal network (i.e. pipe and bubble) model shown in Figure \ref{fig:topology}.  It consists of 13 bubbles, their interfaces and external tie-lines with neighboring ISOs.  This model is represented by a DC power flow analysis with each zone-bubble represented as a bus and each zone-interface is represented as a line.  In order to recognize that ISO New England is part of the Eastern Interconnect, the swing bus is added to represent power imbalances exchanged with New York ISO.  This swing bus is connected to the Vermont, Western Massachusetts, Connecticut, and Norwalk bubbles but is distinct from the tie-lines to these bubbles.  In such a way, the power flows to and from this New York swing bus also represent the deviations away from scheduled tie-line flows.  

In the normal operating mode, the regulation service and the real-time balancing operations are able to keep the system balanced. However, a sudden line or generator outage can create a large imbalance that the real-time market and regulation service are unable to mitigate. The EPECS simulator is able to address forced outage events by switching from a normal operations to an emergency operations mode.   In the event of a forced outage, the ISO-NE contingency operations are assumed to run a RTUC in the same time step.  The simulator then continues to run the regulation and SCED models until a time that is evenly divisible by 15 minutes at which point the RTUC is called as in normal operations.  

\clearpage
\section{Data: Characteristics of the ISO New England Case Study}\label{sec:data}
This section describes the six scenarios analyzed for this study and the ISO New England data used for each scenario.

\subsection{Study Scenarios}\label{sec:scenarios}
A total of 12 scenarios are studied for the years 2025 and 2030; six scenarios for each year. Each scenario is described by different characteristics of load profiles, renewable energy integration and the generation base as shown in Table~\ref{tab:scenarios}. These scenarios are described in more detail in \cite{ISO-New-England:2017:00}.

\subsubsection{Scenario 1 -- ``RPSs + Gas"}
Scenario 1 uses the generation base expected for 2019/2020. The gross demand, the solar PV and the energy efficiency values are based on the ISO New England 2016 report on capacity, energy, load, and transmission (2016 CELT report) \cite{ISO-New-England:2016:00}. The amounts of renewable energy sources in the system, such as wind, are chosen according to ISO New England 2016 Renewable Portfolio Standards (RPS) \cite{ISO-New-England:2016:01}. Half of the oldest oil and coal generation units are planned to retire by 2025, while the other half by 2030. The retired units are replaced by natural gas combined-cycle (NGCC) units at the same locations. The amount of new NGCC generation is planned to meet the net Installed Capacity Requirement (NICR). The historical profiles are used for imports from Hydro-Quebec (HQ) and New Brunswick (NB).

\subsubsection{Scenario 2 -- ``ISO Queue"}
Scenario 2 is identical to Scenario 1 in terms of generation base, planned retirements, gross demand and energy imports from HQ and NB being based on forecasts in 2016 CELT report. However, for Scenario 2, the retired oil and coal units are replaced by renewable energy sources instead of NGCC. Similar to Scenario 1, the addition of renewable energy sources meets the assumed NICR. The locations of the renewable energy sources are according to ISO Interconnection Queue \cite{:2017:00}.

\subsubsection{Scenario 3 -- ``Renewables Plus"}
Scenario 3 uses additional renewable energy sources, such as behind-the-meter (BTM) and utility-scale solar PV and wind units, to replace the retiring units and meet or exceed the existing RPSs. In addition to the new renewable energy source, Scenario 3 adds battery energy systems, energy efficiency and plug-in hybrid electric vehicles (PHEV) to the system. Moreover, two new tie lines are added to increase the hydroelectric imports. The added resources exceed the assumed NICR.

\subsubsection{Scenario 4 -- ``No Retirements beyond FCA \#10"}\label{sec:scenario4}
Scenario 4 uses the same data as Scenario 1 in terms of gross demand, energy efficiency and solar PV integration being based on the 2016 CELT report. The historical imports data is also used similar to Scenario 1.  However, the renewable energy integration is done according to ``I.3.9" approval to meet the RPSs \cite{:2017:00}. Additionally, in contrast to other scenarios, no generation units are retired beyond known FCA resources which are replaced by NGCC located at the Hub to meet meet the NICR.

\subsubsection{Scenario 5 -- ``ACPs + Gas"}
Scenario 5 uses the same data as Scenario 4 in terms of gross demand, energy efficiency and renewable energy integration being based on the 2016 CELT report. The historical imports data is also used similar to Scenario 4. However, half of the oldest oil and coal generation units are planned to retire by 2025, while the other half by 2030 which are replaced by new NGCC units to meet the NICR. 

\subsubsection{Scenario 6 -- ``RPSs + Geodiverse Renewables"}
Scenario 6 is identical to Scenario 2 in terms of gross demand, energy efficiency, generation base and retirement schedules being based on 2016 CELT report. The HQ and NB imports are also based on historical data. Also, the addition of renewable energy sources are used to meet the RPSs and the NICR. However, the renewable energy sources are split into three equal groups: the first group consists of solar PV units located mainly in Southern New England area, the second group is the onshore wind power located in Maine, and the third group is the offshore wind located connected to southeastern Massachusetts/Newport, Rhode Island, and Rhode Island bordering Massachusetts (SEMA/RI) and Connecticut. Thus, the solar PV and offshore wind units are located closer to the main load centers while the onshore wind in Maine is in a remote area.

\begin{table}[!t]
\centering
\resizebox{\textwidth}{!}{
\begin{tabular}{| c | c | c | c | c | c | c | c |}
\hline			
Scenario                            
& Retirements                       
& \mc{Gross\\Demand}                
& PV                                        
& \mc{Energy\\Efficiency}           
& Wind                              
& \mc{New NG\\Units}                
& \mc{HQ and NB\\External Ties \&\\Transfer Limits} \\\hline
1                                   
& \mc{1/2 in 2025\\1/2 in 2030}     
& \mc{Based on 2016\\CELT forecast} 
& \mc{Based on 2016\\CELT forecast}   
& \mc{Based on 2016\\CELT forecast} 
& \mc{As needed to\\meet RPSs}      
& NGCC                              
& \mc{Based on historical\\profiles}   \\\hline        
2                                   
& \mc{1/2 in 2025\\1/2 in 2030}     
& \mc{Based on 2016\\CELT forecast} 
& \mc{BTM Based on 2016\\CELT forecast; non-\\BTM same as wind}   
& \mc{Based on 2016\\CELT forecast} 
& \mc{Used to satisfy\\net ICR}      
& None                              
& \mc{Based on historical\\profiles}   \\\hline        
3                                   
& \mc{1/2 in 2025\\1/2 in 2030}     
& \mc{Based on 2016\\CELT forecast} 
& \mc{8,000MW (2025)\\12,000MW (2030)\\ \\BTM PV 4,000MW (2025)\\6,000MW (2030)\\ \\Utility PV 4,000MW (2025)\\6,000MW (2030)}
& \mc{4,844MW (2025)\\7,009MW (2030)} 
& \mc{5,733MW (2025)\\7,283MW (2030)}      
& None                              
& \mc{Based on historical\\profiles plus\\additional imports}\\\hline        
4                                   
& \mc{No retirements\\beyond FCA \#10}     
& \mc{Based on 2016\\CELT forecast} 
& \mc{Based on 2016\\CELT forecast} 
& \mc{Based on 2016\\CELT forecast} 
& \mc{Existing plus those\\with I.3.9 approval}      
& NGCC                              
& \mc{Based on historical\\profiles}\\\hline        
5                                   
& \mc{1/2 in 2025\\1/2 in 2030}     
& \mc{Based on 2016\\CELT forecast} 
& \mc{Based on 2016\\CELT forecast} 
& \mc{Based on 2016\\CELT forecast} 
& \mc{Existing plus those\\with I.3.9 approval}      
& NGCC                              
& \mc{Based on historical\\profiles}\\\hline        
6                                   
& \mc{1/2 in 2025\\1/2 in 2030}     
& \mc{Based on 2016\\CELT forecast} 
& \mc{381MW (2025)\\1,611MW (2030)} 
& \mc{Based on 2016\\CELT forecast} 
& \mc{Onshore wind:\\381MW (2025)\\1,611MW (2030)\\ \\Offshore wind:\\381MW (2025)\\1,611MW (2030)}      
& None                              
& \mc{Based on historical\\profiles}\\\hline        
\end{tabular}}
\caption{The six scenarios of the ISO New England SOARES project}
\label{tab:scenarios}
\end{table}

\clearpage
\subsection{Load Profiles}
This section describes the statistical characteristics of the system load for each of 12 scenarios. As shown in Table~\ref{tab:scenarios}, all scenarios use the gross demand based on 2016 CELT forecast. Therefore, the gross demand profiles for the scenarios from the same year are identical. However, the combined value of gross demand, energy efficiency and electric vehicles charging loads are studied here as a better representation of the actual load in the system than needs to be served. This introduces some differences between load profiles for different scenarios as discussed below.

\begin{figure}[!h]
\centering
\includegraphics[width=6.5in]{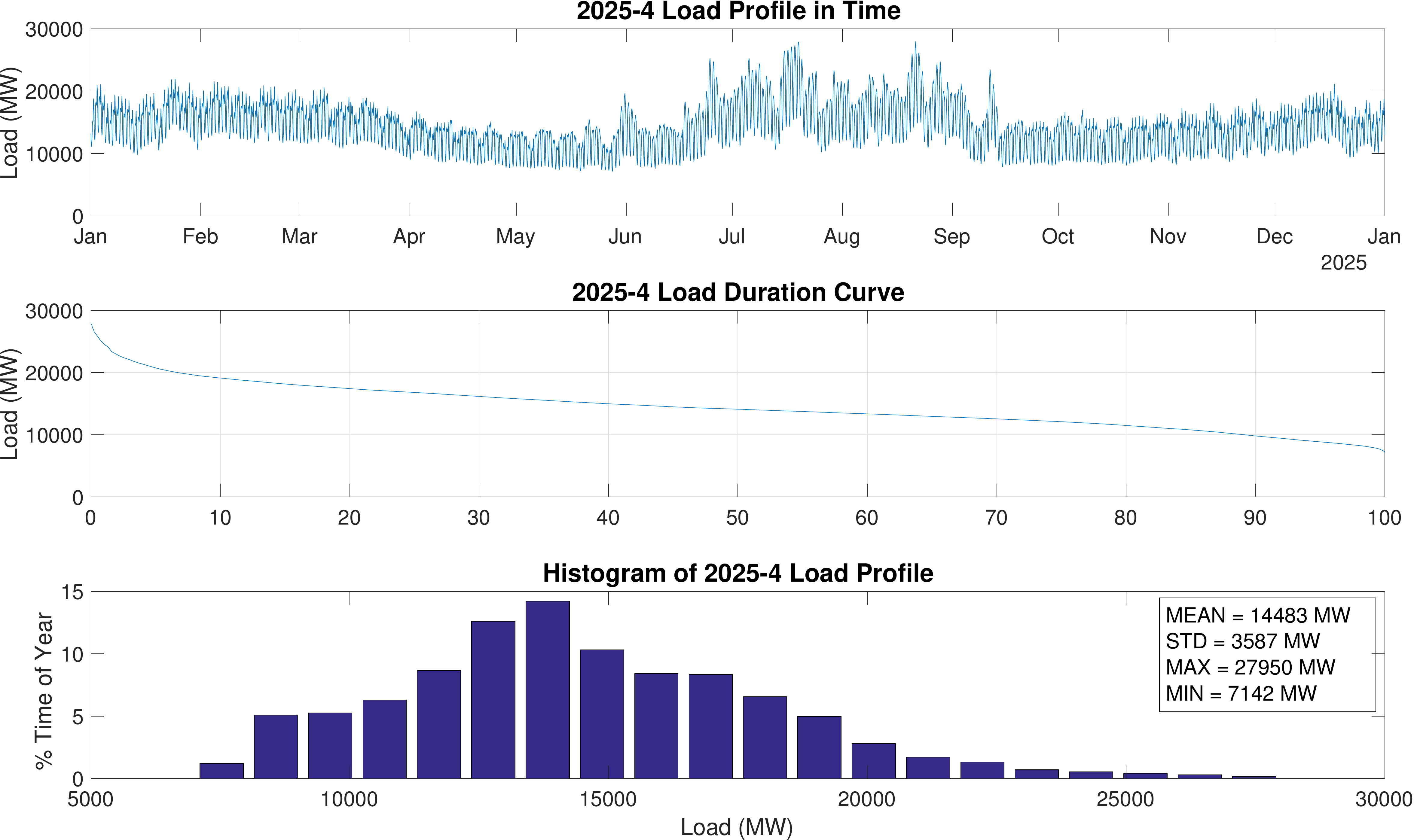}
\caption{The aggregated load profile for Scenario 2025-4}
\label{fig:2025-4LoadProfile}
\end{figure}

Load data can be represented as time profiles, duration curves and histograms, where each representation carries different information about its statistical characteristics. The aggregated load data for Scenario 2025-4 is shown in Figure~\ref{fig:2025-4LoadProfile} as an example. The choice of this scenario is due to the fact that it follows the established evolution pattern of the ISO New England generation base as described in Section~\ref{sec:scenario4}, and is often referred to as ``business-as-usual'' case. The graphs in Figure~\ref{fig:2025-4LoadProfile} show that the aggregated system load varies in a wide range during the year, reaching the summer peak value of 27,950MW and dropping to the lowest 7,142MW value during spring months. The average load during the year is 14,483MW with a standard deviation of 3,587MW.

While the ability to serve the system peak load is one of the most important components of power system reliable operations, within the scope of this study the periods of the year with lowest aggregated system load represent a bigger interest due to the following two main factors. First, some of the scenarios described in Table~\ref{tab:scenarios} assume integration of large amount of solar and wind resources into the system. This may lead to significantly low, or even negative, net load that the system generation needs to serve. Second, a significant portion of the system generation base are nuclear units and they are assumed to operate in a ``must-run" mode. This further increases the possibility of having excess generation in the system and the need of curtailment of renewables.

\begin{figure}[!t]
\centering
\includegraphics[width=6.5in]{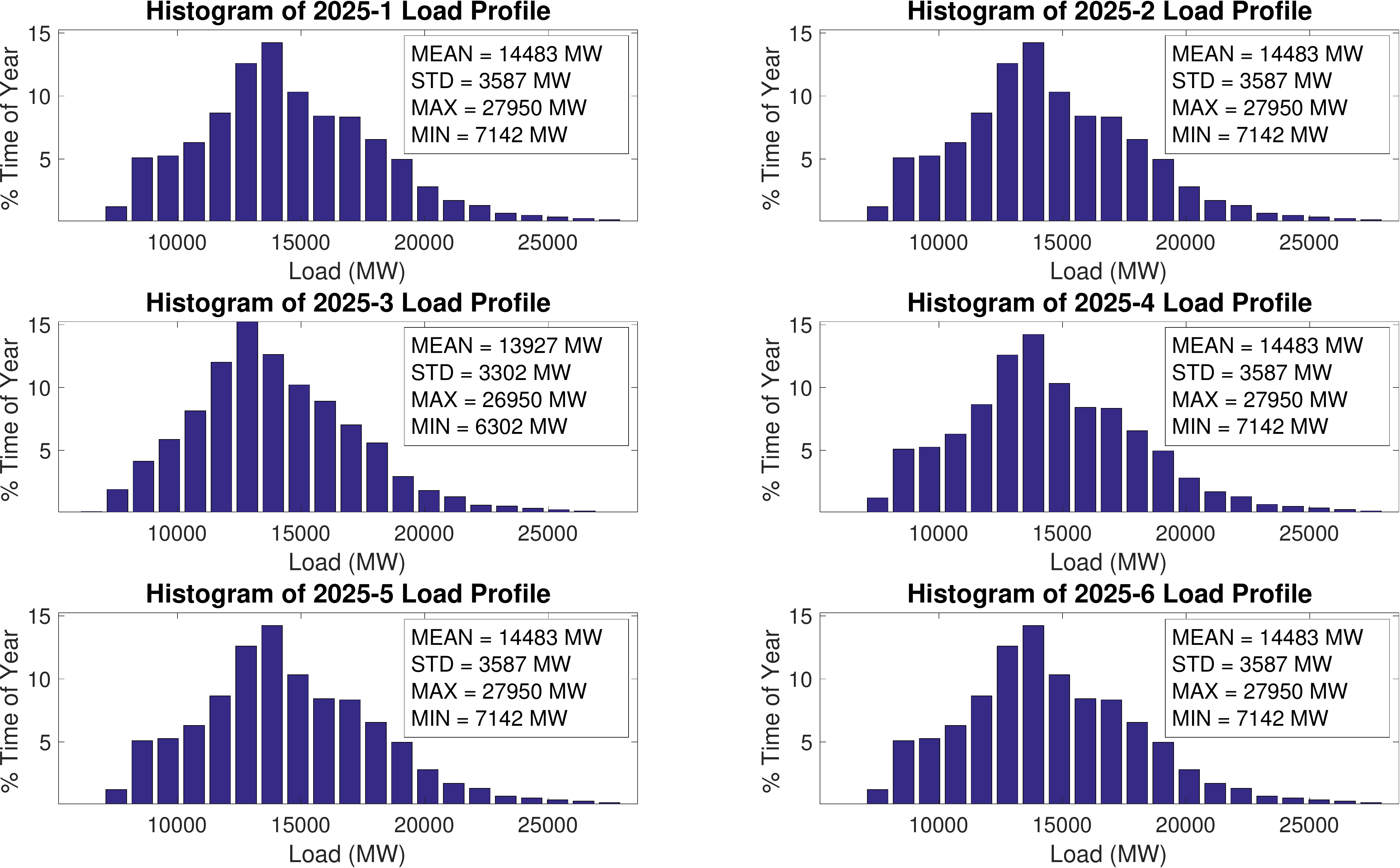}
\caption{Load profile histograms for 2025 scenarios}
\label{fig:2025LoadProfiles}
\end{figure}

The histograms of load profiles for the year 2025 are presented in Figure~\ref{fig:2025LoadProfiles}. The load distributions exhibit the same statistical characteristics, except for Scenario 3 due to the addition of energy efficiency and electric vehicle charging loads mentioned above. As a result, Scenario 3 has slightly lower peak load and minimum load levels. Table~\ref{tab:2025LoadStatistics} summarizes the statistical characteristics of the load data for the six scenarios of the year 2025. 
\begin{table}[!t]
\caption{Load profile statistics for 2025 scenarios}
\begin{footnotesize}
\begin{center}
\begin{tabular}{lrrrrrr}\toprule
& \textbf{2025-1}		& \textbf{2025-2}	& \textbf{2025-3}	& \textbf{2025-4}	& \textbf{2025-5}	& \textbf{2025-6}	\\ \toprule
\textbf{Max} (MW) 		& 27,950 	& 27,950 	& 26,950 	& 27,950 	& 27,950 	& 27,950 	\\\midrule
\textbf{Min} (MW) 		&   7,142 	&  7,142 	&   6,302 	&   7,142  &   7,142 	&   7,142 	\\\midrule
\textbf{Energy} (TWh) 	&   	127 	&     127 	&      122 	&      127 	&      127 	&      127 	\\\midrule
\textbf{Mean} (MW)		& 14,483 	& 14,483 	& 13,927 	& 14,483 	& 14,483 	& 14,483 	\\\midrule
\textbf{STD}  (MW)		&   3,587 	&   3,587 	&   3,302 	&   3,587 	&   3,587 	&   3,587 	\\\bottomrule
\end{tabular}
\end{center}
\end{footnotesize}

\label{tab:2025LoadStatistics}
\end{table}

\begin{figure}[!t]
\centering
\includegraphics[width=6.5in]{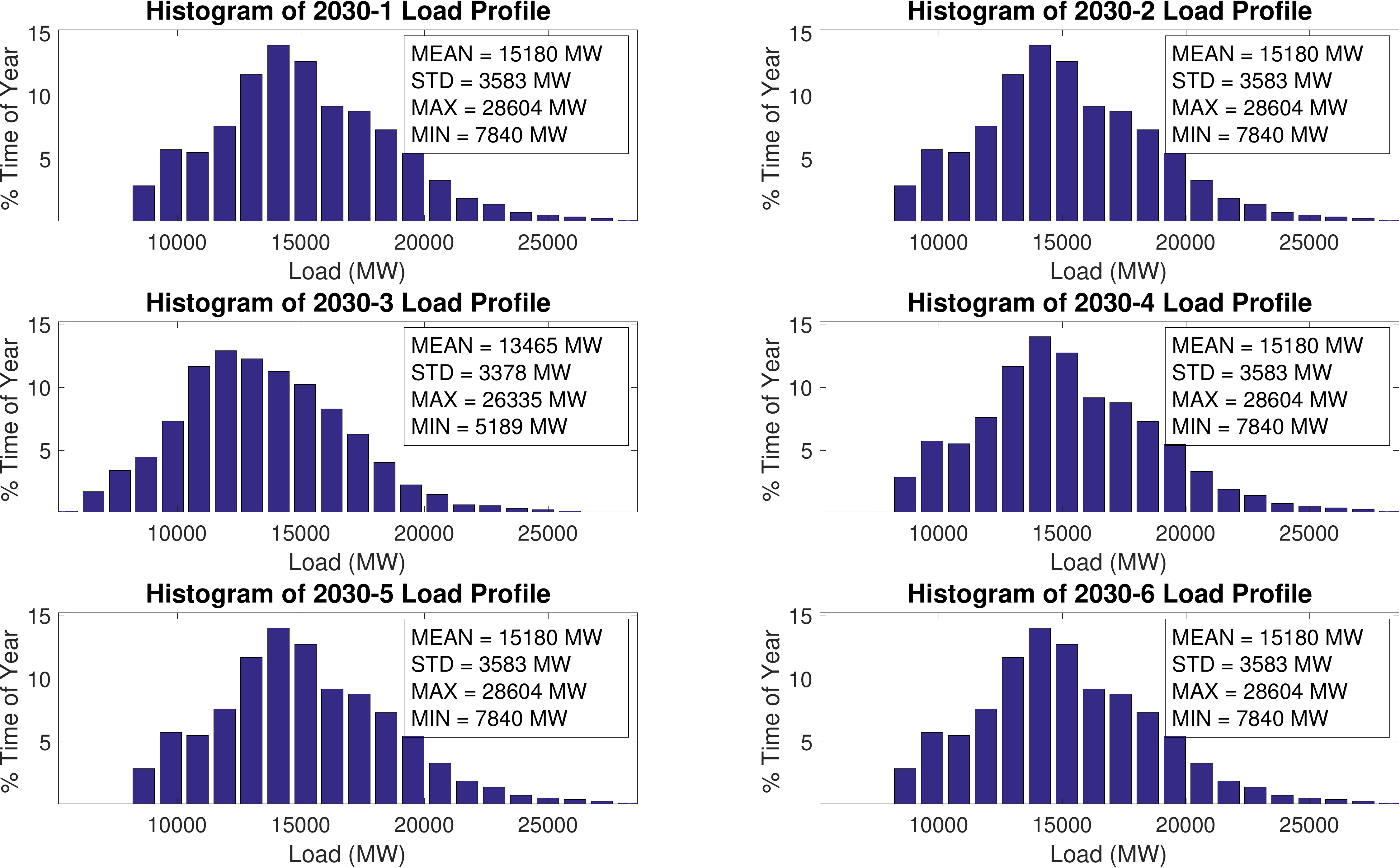}
\caption{Load profile histograms for 2030 scenarios}
\label{fig:2030LoadProfiles}
\end{figure}

Next, the load profiles for 2030 scenarios are studied. Similar to the 2025 case, the profiles for all scenarios have identical distributions except for Scenario 3 due to incorporation of energy efficiency and electric vehicle charging loads as shown in Figure~\ref{fig:2030LoadProfiles}. The statistical characteristics of the load profiles for 2030 scenarios are presented in Table~\ref{tab:2030LoadStatistics}. A pattern similar to 2025 scenarios is observed here too when Scenario 3 has slightly lower power and energy indicators. The following observation should be made that while the overall consumption, the peak load, the minimum load experience slight increase for all scenarios compared to 2025, Scenario 3 shows the opposite trend. This is explained by increased amounts of energy efficiency and electric vehicle penetration compared to 2025 as shown in Table~\ref{tab:scenarios}. 
\begin{table}[!t]
\caption{Load profile statistics for 2030 scenarios}
\begin{footnotesize}
\begin{center}
\begin{tabular}{lrrrrrrrrrrrr}\toprule
& \textbf{2030-1}	& \textbf{2030-2}	& \textbf{2030-3}	& \textbf{2030-4}	& \textbf{2030-5}	& \textbf{2030-6}	\\\toprule
\textbf{Max} (MW) 		& 28,604 	& 28,604 	& 26,335 	& 28,604 	& 28,604 	& 28,604 	\\\midrule
\textbf{Min} (MW) 		&   7,840 	&   7,840 	&   5,189 	&   7,840 	&   7,840 	&   7,840 	\\\midrule
\textbf{Energy} (TWh) 	&      133 	&      133 	&      118 	&      133 	&      133 	&      133 	\\\midrule
\textbf{Mean} (MW)		& 15,180 	& 15,180 	& 13,465 	& 15,180 	& 15,180 	& 15,180 	\\\midrule
\textbf{STD}  (MW)		&   3,583 	&   3,583 	&   3,378 	&   3,583 	&   3,583 	&   3,583 	\\\bottomrule
\end{tabular}
\end{center}
\end{footnotesize}

\label{tab:2030LoadStatistics}
\end{table}

\clearpage
\subsection{Net Load Profiles}
\begin{figure}[!h]
\centering
\includegraphics[width=6.5in]{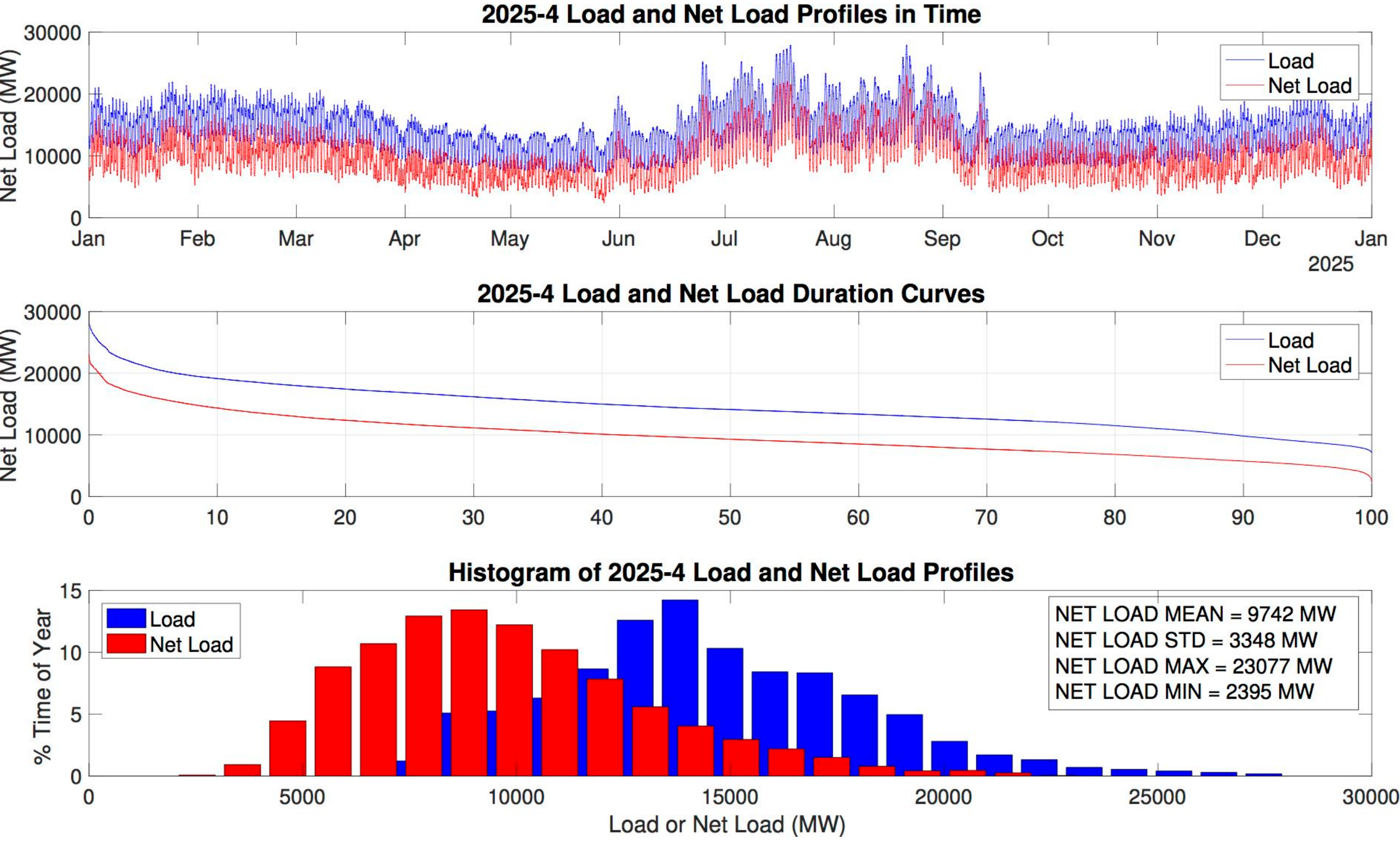}
\caption{Comparison of load and net load for Scenario 2025-4}
\label{fig:2025-4NetLoadProfile}
\end{figure}

Net load is the difference between the aggregated system load and the total generation produced by the renewable energy sources. The shape of the net load profile is more relevant when studying the ability of the system to maintain balance as it represents the actual amount ofMW that needs to be supplied by dispatchable resources, such as generators, pumped storage units and demand response. The comparison of the system load and the corresponding net load for Scenario 2025-4 is presented in Figure~\ref{fig:2025-4NetLoadProfile}. The graphs show that the incorporation of renewable energy alters the power demand pattern significantly. The overall system demand decreases, and the shape of the histogram shifts to the left. This indicates that the system may need less generation capacity to meet the demand. On the other hand, it is uncertain whether the system is prepared to effectively harvest the power generated by renewable energy units given the limitations of the associated generation fleet.  

\begin{figure}[!h]
\centering
\includegraphics[width=6.5in]{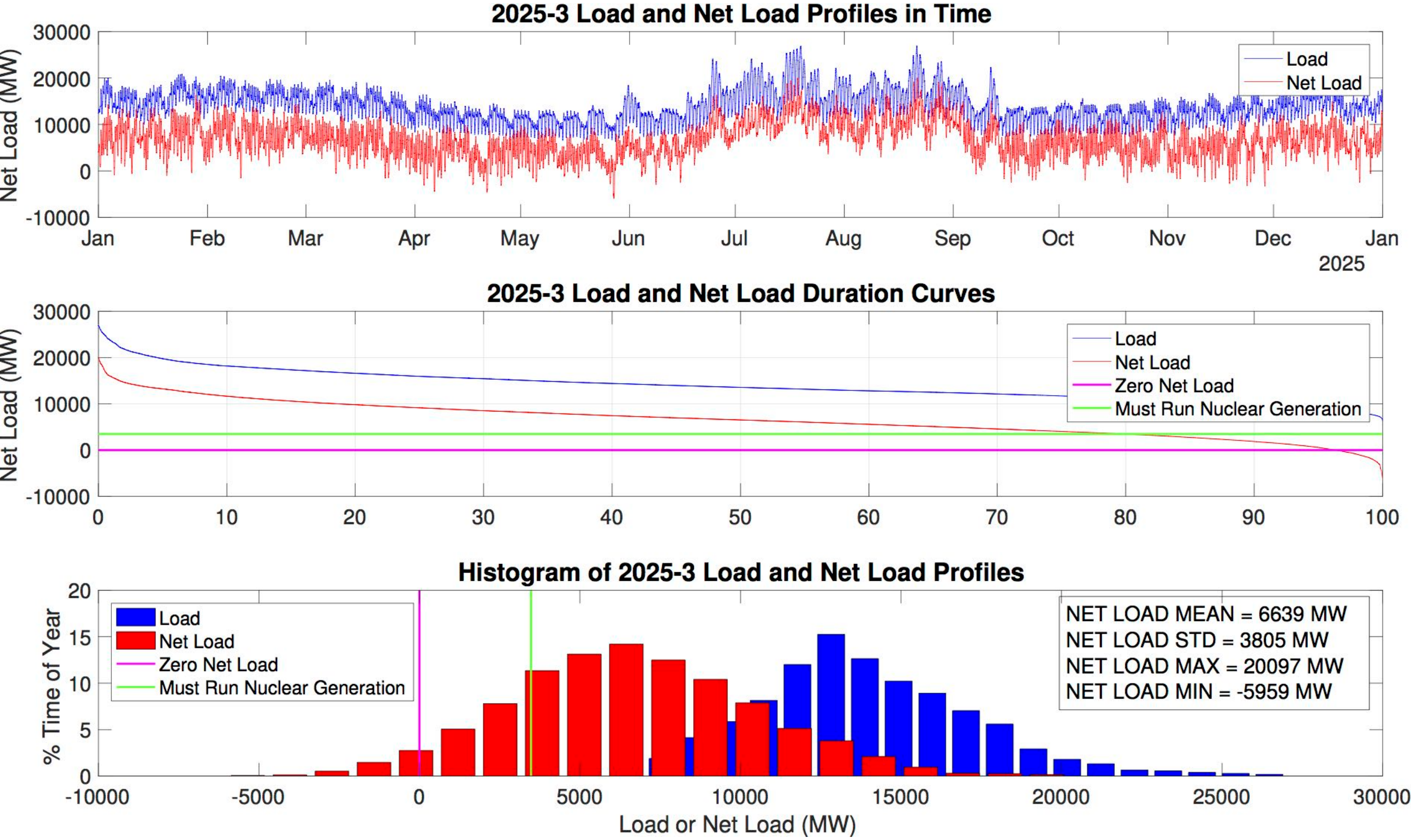}
\caption{Comparison of load and net load for Scenario 2025-3}
\label{fig:2025-3NetLoadProfile}
\end{figure}

The relevance of this question is particularly emphasized for Scenario 2025-3 where the net load drops below zero at different instances throughout the year as shown in Figure~\ref{fig:2025-3NetLoadProfile}. Negative net load is an indication of excess renewable energy generation in the system which supports the statement above that the system will necessarily be unable to harvest it all. This is a challenge for a system with a large presence of ``must-run" nuclear units. The matter is further complicated when considering that most renewable energy generation is located in remote areas of Maine, relatively far from major consumption areas, such as Massachusetts. 

\begin{figure}[!h]
\centering
\includegraphics[width=6.5in]{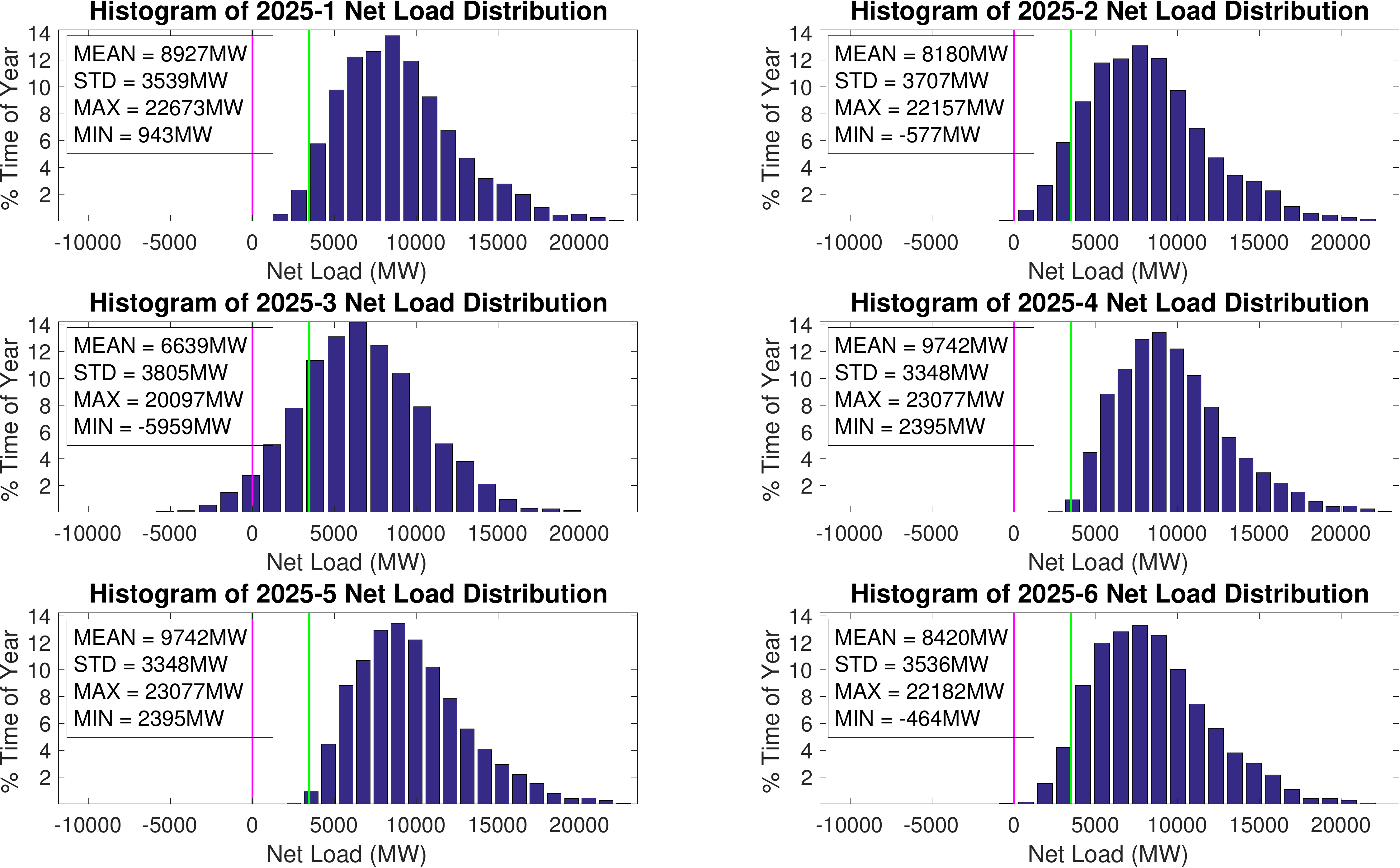}
\caption{Net load profile histograms for 2025 scenarios}
\label{fig:2025NetLoadDistributions}
\end{figure}

Net load distributions for the six scenarios of 2025 are presented in Figure~\ref{fig:2025NetLoadDistributions}. Unlike the load data studied above, net load distributions differ from each other due to differences in renewable energy quantities present in each scenario. The graphs show that three out of six scenarios reach negative net load at some point during the year with Scenario 3 being the most severe example; its net load drops to the minimum of --5,959MW due to the heavy presence of renewable energy shown in Table~\ref{tab:2025NetLoadStatistics}. Moreover, when the presence of ``must-run" nuclear units is taken into account, it becomes obvious that all six scenarios of 2025 have excess generation in the system.
\begin{table}[!h]
\caption{Net load profile statistics for 2025 scenarios}
\begin{footnotesize}
\begin{center}
\begin{tabular}{lrrrrrr}\toprule
& \textbf{2025-1}	& \textbf{2025-2}	& \textbf{2025-3}	& \textbf{2025-4}	& \textbf{2025-5}	& \textbf{2025-6}	\\\toprule
\textbf{Max} (MW)		& 22,673 	& 22,157 	& 20,097 	& 23,077 	& 23,077 	& 22,182 	\\\midrule
\textbf{Min} (MW)		&      943 	&     -577 	&  -5,959 	&   2,395 	&   2,395 	&     -464 	\\\midrule
\textbf{Energy} (TWh)	&        78 	&        72 	&        58 	&        85 	&        85  &         74 	\\\midrule
\textbf{Mean} (MW)		&   8,927 	&   8,180 	&   6,639 	&   9,742 	&   9,742 	&    8,420 	\\\midrule
\textbf{STD} (MW) 		&   3,539 	&   3,707 	&   3,805 	&   3,348 	&   3,348 	&    3,536 	\\\midrule
\textbf{\% Time Excess Gen.} 	&  3.12 	&  8.33 	& 20.13 	&  0.27 	&  0.27 	&  5.09 	\\\midrule
\textbf{\% Time Neg Net Load} 	&  0.00 	&  0.05 	&   3.68 	&  0.00 	&  0.00 	&  0.03 	\\\bottomrule
\end{tabular}
\end{center}
\end{footnotesize}

\label{tab:2025NetLoadStatistics}
\end{table}

\begin{figure}[!h]
\centering
\includegraphics[width=6.5in]{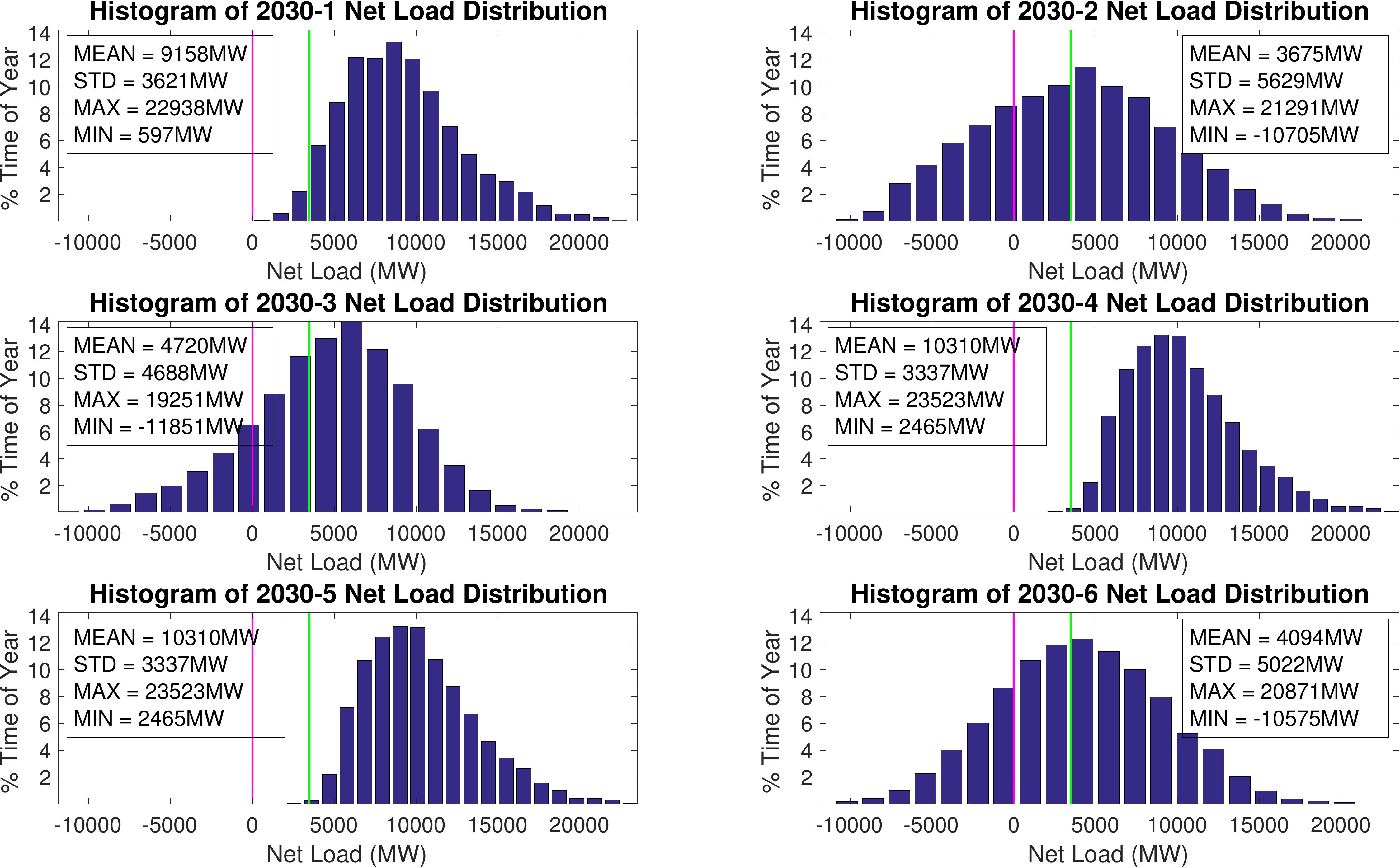}
\caption{Net load profile histograms for 2030 scenarios}
\label{fig:2030NetLoadDistributions}
\end{figure}

The challenges described above are exacerbated for the year 2030 as shown in Figure~\ref{fig:2030NetLoadDistributions}. Significant parts of net load distributions are now below zero. Moreover, Table~\ref{tab:2030NetLoadStatistics} shows that net load values for three scenarios drop below --10,000MW. For a system with less than 30,000MW peak load this is a significant challenge. In order to maintain reliable operations under such conditions, the power system needs to make a comprehensive use of its demand response resources, renewable energy curtailment and pumped storage units. Some scenarios with heavy curtailment of renewables may reveal their infeasibility under the current system configuration, implying the need for more critical changes in the system, such as the construction of new transmission lines or availability of more storage or demand response capabilities.
\begin{table}[!h]
\caption{Net load profile statistics for 2030 scenarios}
\begin{footnotesize}
\begin{center}
\begin{tabular}{lrrrrrrrrrrrr}\toprule
 & \textbf{2030-1}& \textbf{2030-2}& \textbf{2030-3}& \textbf{2030-4}& \textbf{2030-5}& \textbf{2030-6} \\\toprule
\textbf{Max} (MW)& 22,938 & 21,291 & 19,251 & 23,523 & 23,523 & 20,871 \\\midrule
\textbf{Min} (MW)&   597 & -10,705 & -11,851 &  2,465 &  2,465 & -10,575 \\\midrule
\textbf{Energy} (TWh)&    80 &    32 &    41 &    90 &    90 &    36 \\\midrule
\textbf{Mean} (MW)&  9,158 &  3,675 &  4,720 & 10,310 & 10,310 &  4,094 \\\midrule
\textbf{STD} (MW) &  3,621 &  5,629 &  4,688 &  3,337 &  3,337 &  5,022 \\\midrule
\textbf{\% Time Excess Gen.}&  2.91 & 48.11 & 37.02 &  0.09 &  0.09 & 45.74 \\\midrule
\textbf{\% Time Neg. Net Load} &  0.00 & 27.49 & 15.79 &  0.00 &  0.00 & 21.38 \\\bottomrule
\end{tabular}
\end{center}
\end{footnotesize}

\label{tab:2030NetLoadStatistics}
\end{table}

\clearpage
\subsection{Net Load Ramping Characteristics}
\begin{figure}[!h]
\centering
\includegraphics[width=6.5in]{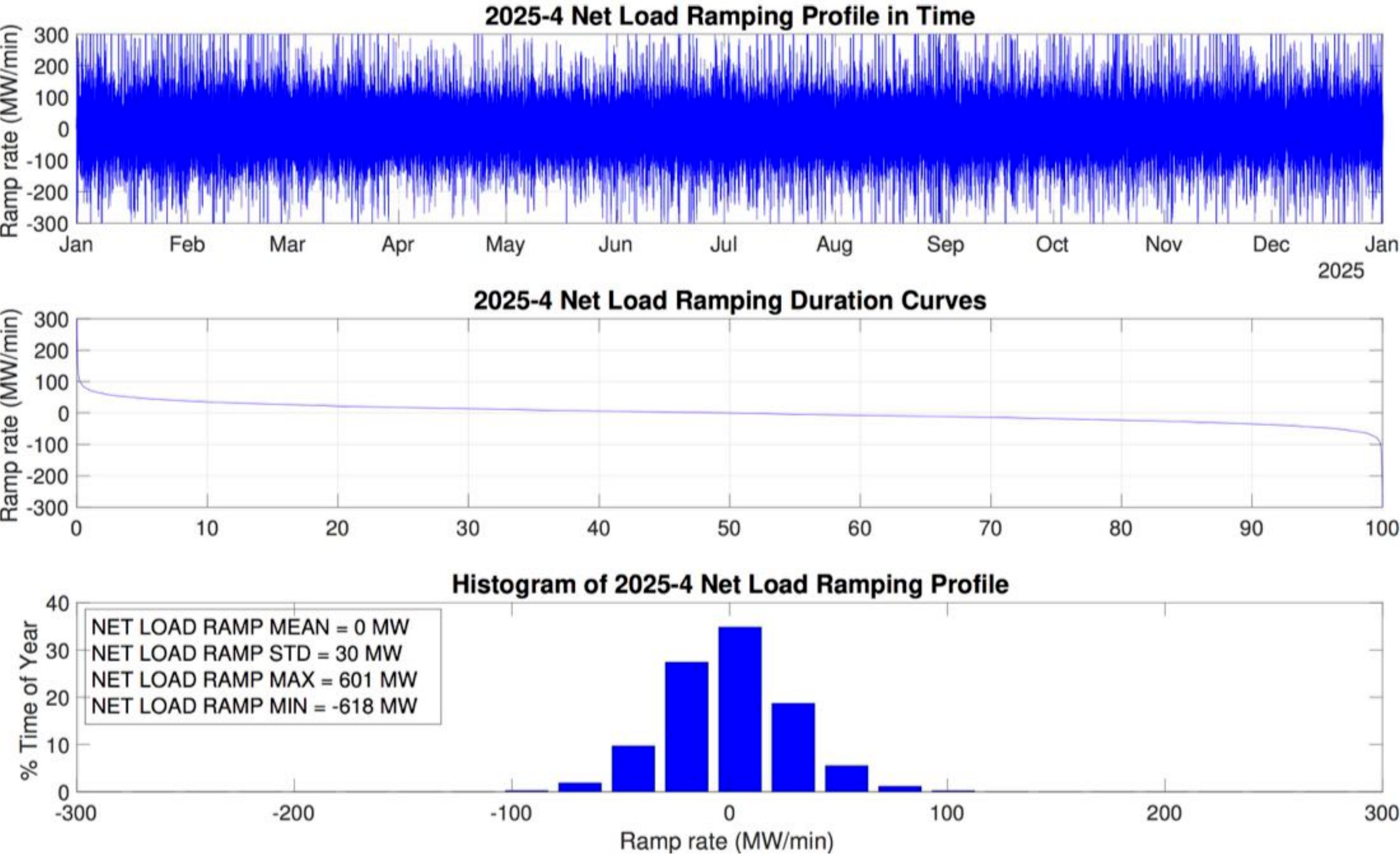}
\caption{Net load ramping profile for Scenarios 2025-4}
\label{fig:2025-4NetLoadRampProfile}
\end{figure}

In addition to the net load variations, another important characteristic that defines the dynamics of the system is the rate at which the consumption profile changes (i.e. the system ramping rate).  It is shown in Figure~\ref{fig:2025-4NetLoadRampProfile} for Scenario 2025-4. Considering ramping dynamics is particularly important when the study deals with a significant integration of highly volatile renewable energy sources. A system with a significant presence of nuclear power (especially when assumed to must-run at full capacity) may lack the flexibility to follow such a net load profile;  creating another challenge for the full utilization of available renewable generation by the system. However, emergence of more flexible natural gas units may compensate for their lack of flexibility.

Ramping rates for four different time resolutions are defined and studied here, namely 1-minute, 10-minute, 1-hour and 4-hour, to capture net load dynamics at various layers of control, such as real-time, SCED, RTUC and SCUC.  They indicate whether the system will be able to schedule the necessary generation, dispatch them and supply the demand in the real-time.
 The definitions of these ramping rates with different time resolutions are defined below.
\begin{defn}
Inter-1-minute ramping rate: The difference between consecutive points on the net load profile with one minute resolution. 
\end{defn}
\begin{defn}
Inter-10-minute ramping rate: The difference between consecutive points on the net load profile after it has been averaged into 10 minute time blocks.  
\end{defn}
\begin{defn}
Inter-1-hour ramping rate: The difference between consecutive points on the net load profile after it has been averaged into one hour time blocks.  
\end{defn}
\begin{defn}
Inter-4-hour ramping rate: The average \emph{sustained} ramp within a four hour window that covers the minimum and maximum net load values of that time period.   
\end{defn}

\begin{figure}[!h]
\centering
\includegraphics[width=6.5in]{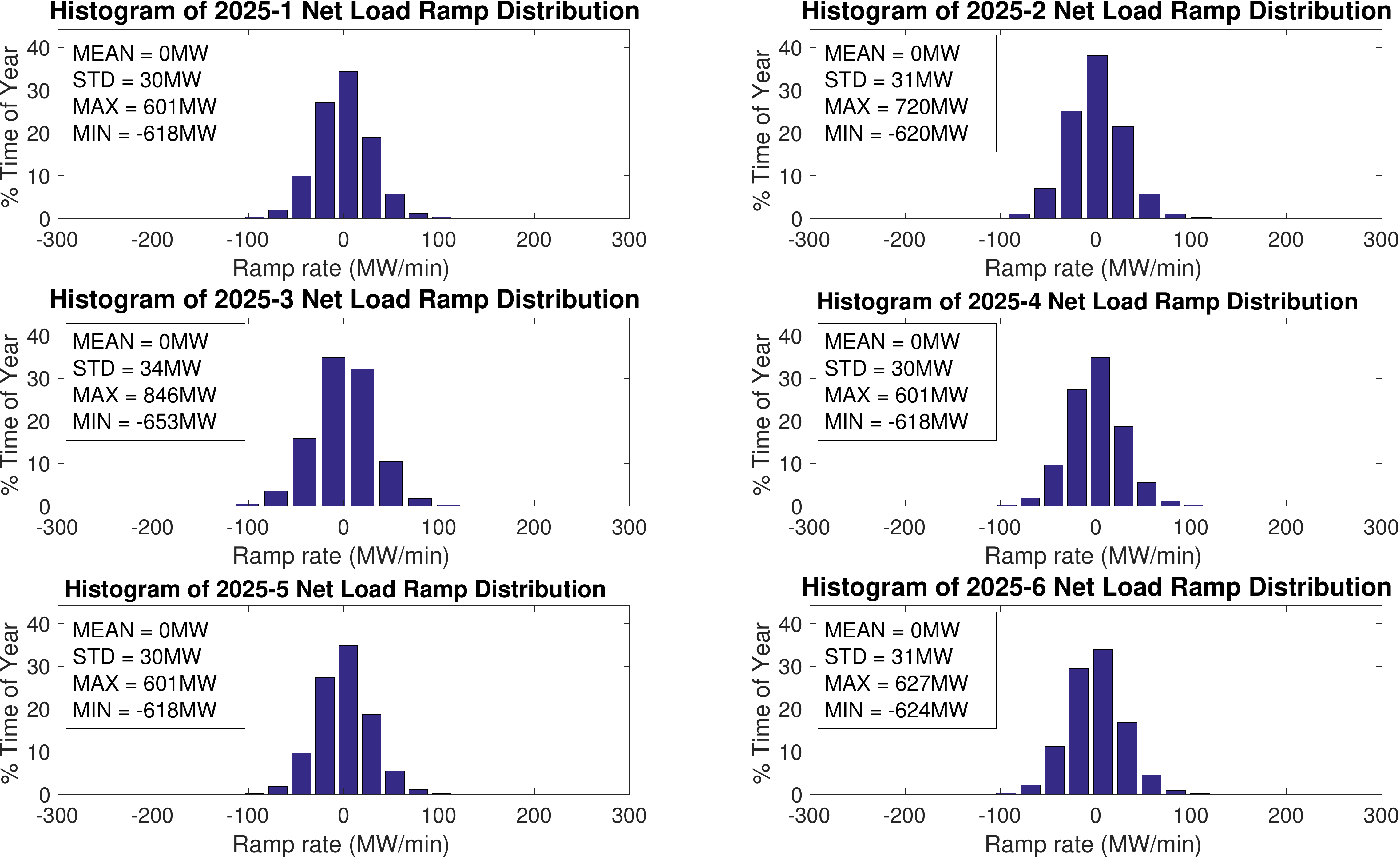}
\caption{Net load ramping histograms for 2025 scenarios}
\label{fig:2025NetLoadRampDistributions}
\end{figure}

Comparison of 1-minute ramping rate distributions for different scenarios of 2025 in Figure~\ref{fig:2025NetLoadRampDistributions} shows that they have comparably similar ramping characteristics. Table~\ref{tab:2025NetLoadRampStatistics} summarizes net load ramping characteristics for the six scenarios of 2025. The data in Table~\ref{tab:2025NetLoadRampStatistics} shows that ramping rates are the highest when calculated with 1-minute resolution and generally decreases with a coarser resolution. This is due to the fact that ramping rates calculated with coarser time resolution are equivalent to averaging ramping rates with finer resolution which narrows the range of their maximum and minimum values. This is a key observation, since it indicates that generation scheduling programs with coarser time steps always underestimate the need for ramping capabilities. This issue becomes more relevant as more renewable energy is integrated into the system, and, therefore, present an argument in favor of the procurement of ramping reserves. 
\begin{table}[!h]
\caption{Net load ramping statistics for 2025 scenarios}
\vspace{-0.1in}
\begin{footnotesize}
\begin{center}
\begin{tabular}{lrrrrrr}\toprule
 & \textbf{2025-1}& \textbf{2025-2}& \textbf{2025-3}& \textbf{2025-4}& \textbf{2025-5}& \textbf{2025-6}\\\toprule
\textbf{Max 1-Min-Up}$^1$ (MW/min)&   601 &   720 &   846 &   601 &   601 &   627 \\\midrule
\textbf{Max 1-Min-Down$^1$} (MW/min)&   618 &   620 &   653 &   618 &   618 &   624 \\\bottomrule
\textbf{Max 10-Min-Up$^2$} (MW/min)&   184 &   251 &   312 &   126 &   126 &   220 \\\midrule
\textbf{Max 10-Min-Down$^2$} (MW/min)&    81 &    84 &    78 &    73 &    73 &    78 \\\bottomrule
\textbf{Max 1h-Up$^2$} (MW/min)&    49 &    52 &    73 &    49 &    49 &    57 \\\midrule
\textbf{Max 1h-Down$^2$} (MW/min)&    46 &    45 &    60 &    40 &    40 &    44 \\\bottomrule
\textbf{Max 4h-Up$^3$} (MW/min)&    30 &    33 &    49 &    29 &    29 &    37 \\\midrule
\textbf{Max 4h-Down$^3$} (MW/min)&    38 &    40 &    42 &    36 &    36 &    38 \\\bottomrule
\end{tabular}
\end{center}
\end{footnotesize}

\label{tab:2025NetLoadRampStatistics}
\vspace{-0.1in}
\begin{footnotesize}
$^1$-- Inter 1-minute ramps are calculated as the difference between consecutive points on the net load profile with 1-minute resolution.  $^2$-- Inter 10 minute and Inter 1h ramps are calculated as the difference between consecutive points on the net load profile after it has been averaged into 10 minute or 1h blocks respectively.  $^3$  -- Intra 4 hour ramps are calculated as the average \emph{sustained} ramp within a four hour window that covers the minimum and maximum net load values of that time period.  
\end{footnotesize}
\end{table}

Table~\ref{tab:2025NetLoadRampStatistics} also shows that scenarios 2025-2 and 2025-3 exhibit the greatest net load ramp up at 1-minute, 10 minute and 4 hour resolution.  Also, the maximum 1 minute down ramp and the maximum 10 minute down ramp are similar for all 2025 scenarios. Scenario 2025-3 exhibits the greatest net load ramp up and down at the 1-hour resolution. Scenarios 2025-3 and 2025-6 exhibit the largest intra 4-hour ramp in an upward direction.  

\begin{figure}[!t]
\centering
\includegraphics[width=6.5in]{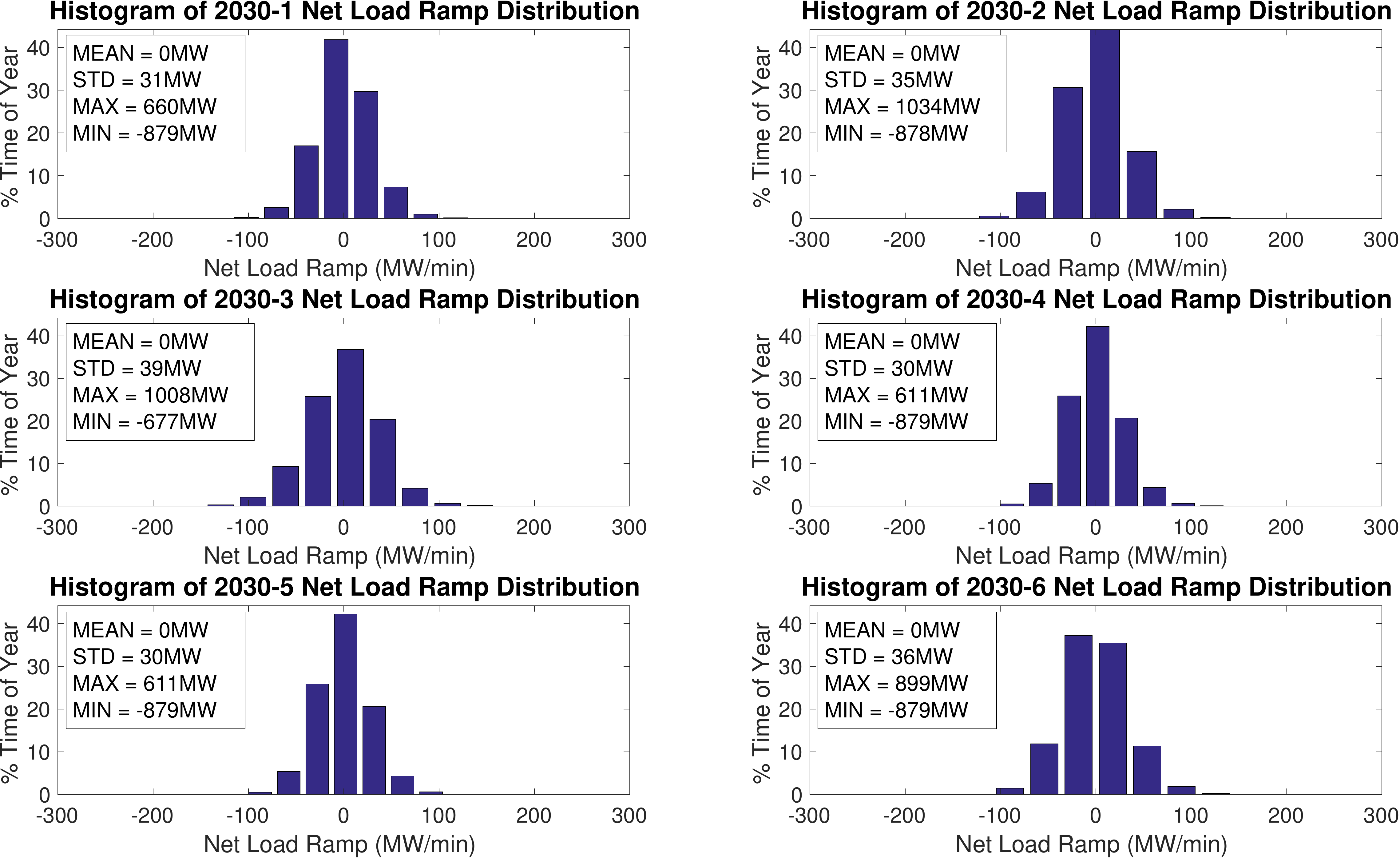}
\caption{Net load ramping histograms for 2030 scenarios}
\label{fig:2030NetLoadRampDistributions}
\end{figure}

Similarly, the ramping rate distributions for the year 2030 are plotted in Figure~\ref{fig:2030NetLoadRampDistributions}.
The statistical characteristics in Table~\ref{tab:2030NetLoadRampStatistics} show that scenarios 2030-2, 2030-3 and 2030-6 exhibit the greatest net load ramp up at 1-minute, 10 minute, 1 hour and 4 hour resolutions. 
\begin{table}[!t]
\caption{Net load ramping statistics for 2030 scenarios}
\vspace{-0.1in}
\begin{footnotesize}
\begin{center}
\begin{tabular}{lrrrrrr}\toprule
 & \textbf{2030-1}& \textbf{2030-2}& \textbf{2030-3}& \textbf{2030-4}& \textbf{2030-5}& \textbf{2030-6}\\\toprule
\textbf{Max 1-Min-Up$^1$} (MW/min)&   660 &  1034 &  1008 &   611 &   611 &   899 \\\midrule
\textbf{Max 1-Min-Down$^1$} (MW/min)&   879 &   878 &   677 &   879 &   879 &   879 \\\bottomrule
\textbf{Max 10-Min-Up$^2$} (MW/min)&   228 &   748 &   383 &   126 &   126 &   672 \\\midrule
\textbf{Max 10-Min-Down$^2$} (MW/min)&   109 &   108 &   115 &   109 &   109 &   161 \\\bottomrule
\textbf{Max 1h-Up$^2$} (MW/min)&    53 &   103 &    95 &    52 &    52 &    99 \\\midrule
\textbf{Max 1h-Down$^2$} (MW/min)&    45 &    76 &    94 &    40 &    40 &    67 \\\bottomrule
\textbf{Max 4h-Up$^3$} (MW/min)&    33 &    61 &    67 &    32 &    32 &    69 \\\midrule
\textbf{Max 4h-Down$^3$} (MW/min)&    39 &    49 &    63 &    36 &    36 &    51 \\\bottomrule
\end{tabular}
\end{center}
\end{footnotesize}

\label{tab:2030NetLoadRampStatistics}
\begin{footnotesize}
$^1$-- Inter 1-minute ramps are calculated as the difference between consecutive points on the net load profile with 1-minute resolution.  $^2$-- Inter 10 minute and Inter 1h ramps are calculated as the difference between consecutive points on the net load profile after it has been averaged into 10 minute or 1h blocks respectively.  $^3$  -- Intra 4 hour ramps are calculated as the average \emph{sustained} ramp within a four hour window that covers the minimum and maximum net load values of that time period.  
\end{footnotesize}
\end{table}
Also, the maximum 1 minute down ramp is similar for all 2030 scenarios except for 2030-3.  The maximum 10 minute ramps down are similar for all 2030 scenarios. Scenarios 2030-2, 2030-3, and 2030-6 exhibit the greatest net load ramp down at the 1-hour and 4-hour resolutions.

\clearpage
\subsection{Load, Solar, \& Wind Forecast Errors in the Net Load}
The forecast errors for each type of resource used in this study are defined in Table~\ref{tab:ForecastErrorStatistics}. The SCUC, RTUC and SCED optimization programs uses different forecasts of the net load and, thus, have different forecast errors respectively.  
\begin{table}[!h]
\caption{Forecast error statistics}
\begin{footnotesize}
\begin{center}
\begin{tabular}{p{1.25cm}rrr}\toprule
 & \textbf{Load}& \textbf{Wind}& \textbf{Solar}\\\toprule
\textbf{SCUC}&     1.65\% &     12\% &     7\%  \\\midrule
\textbf{RTUC} &    1.5\% &    3\% &    3\%  \\\midrule
\textbf{SCED}&  0.15\% &  3\% &  3\% \\\bottomrule
\end{tabular}
\end{center}
\end{footnotesize}

\label{tab:ForecastErrorStatistics}
\end{table}
The trend shows that the smaller the time horizon of the optimization program is, the more accurate the forecast becomes because it is easier to forecast closer time intervals. 

Also, there is a clear distinction between forecast errors for load and renewable energy sources such as wind and solar. Load forecast technologies have been developed and refined during long decades of power system operations, starting from its inception. In contrast, wind and solar powers are comparably recent phenomena for power system operations and their forecasts are not as accurate. This also has to do with different technologies used to predict the system load and renewable energy generation. This fact demonstrates another challenge associated with renewable energy integration;  significantly increased uncertainty in system resource scheduling and procurement.

\clearpage
\section{Case Study Results}\label{sec:results}
This section presents the case study results in terms of key performance characteristics of the power system including:  load following and ramping reserves, curtailment of renewables, interface \& tie-line performances, regulation reserves and the system balancing performance. Each of these metrics are analyzed in the following subsections.

\subsection{Load Following Reserves}
Upward and downward load following reserves are procured during the day-ahead resource scheduling and are dispatched in the real-time balancing in response to net load variations. Traditionally, the procurement of sufficient upward load following reserves has been of the primary concern, while the ability of the system to provide downward load following service by reducing the generation output was assumed generally unconstrained. However, for the power system configuration scenarios considered in this study, both upward and downward load following reserves are equally important, as demonstrated by the results below.

\begin{figure}[!h]
\centering
\includegraphics[width=6.5in]{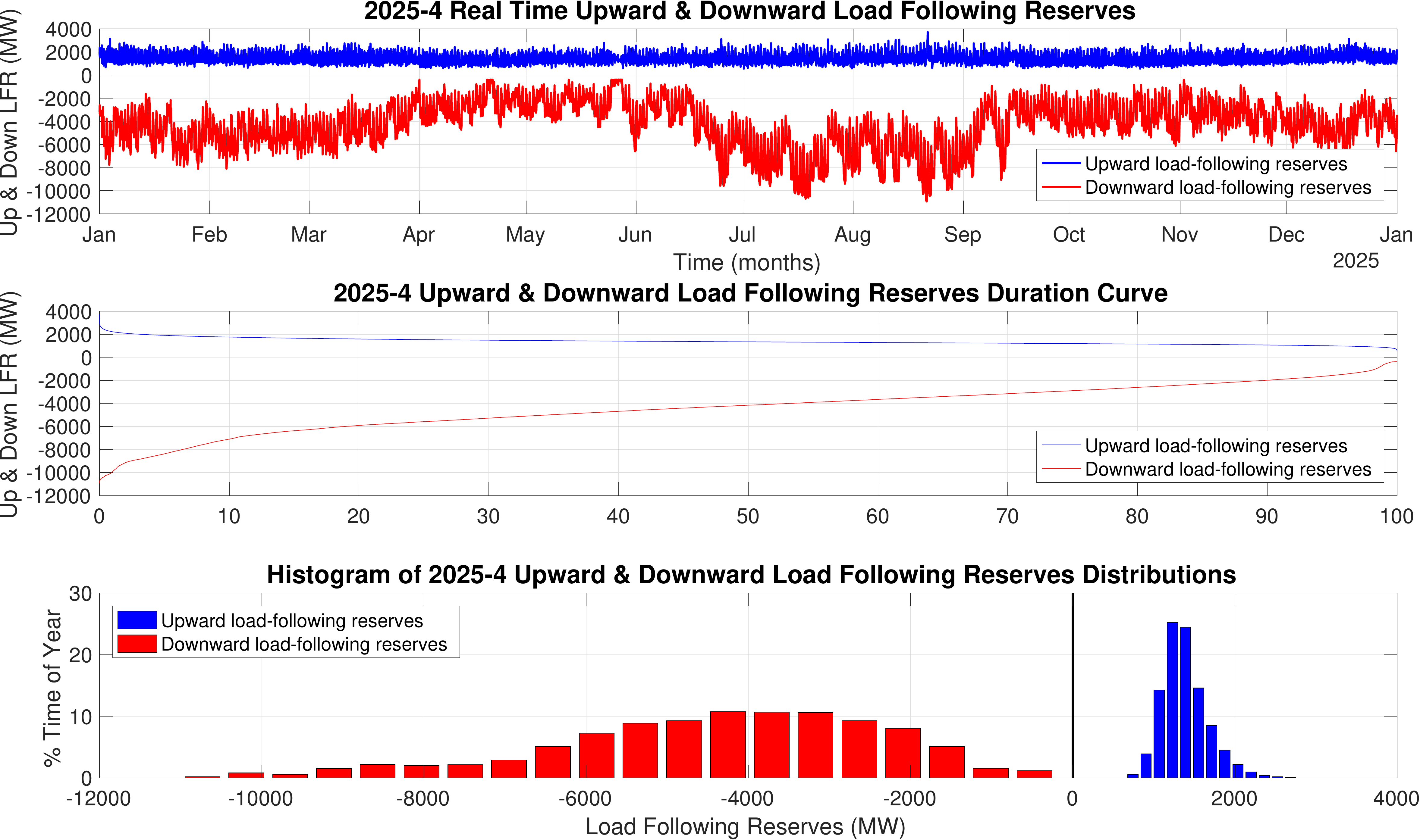}
\caption{Load following reserve profiles for Scenario 2025-4}
\label{fig:2025-4LFRProfile}
\end{figure}

As an example, the performance of load following reserves for Scenario 2025-4 is shown in Figure~\ref{fig:2025-4LFRProfile}. The amounts of upward and downward load following reserves fluctuate over time but are never completely exhausted (approach the zero black line). The closest the system gets to exhausting its downward load following reserves is during low-load spring and fall periods. Thus, when the system adheres to Scenario 2025-4, it is able to operate reliably without the need for more load following reserves. 

\begin{figure}[!h]
\centering
\includegraphics[width=6.5in]{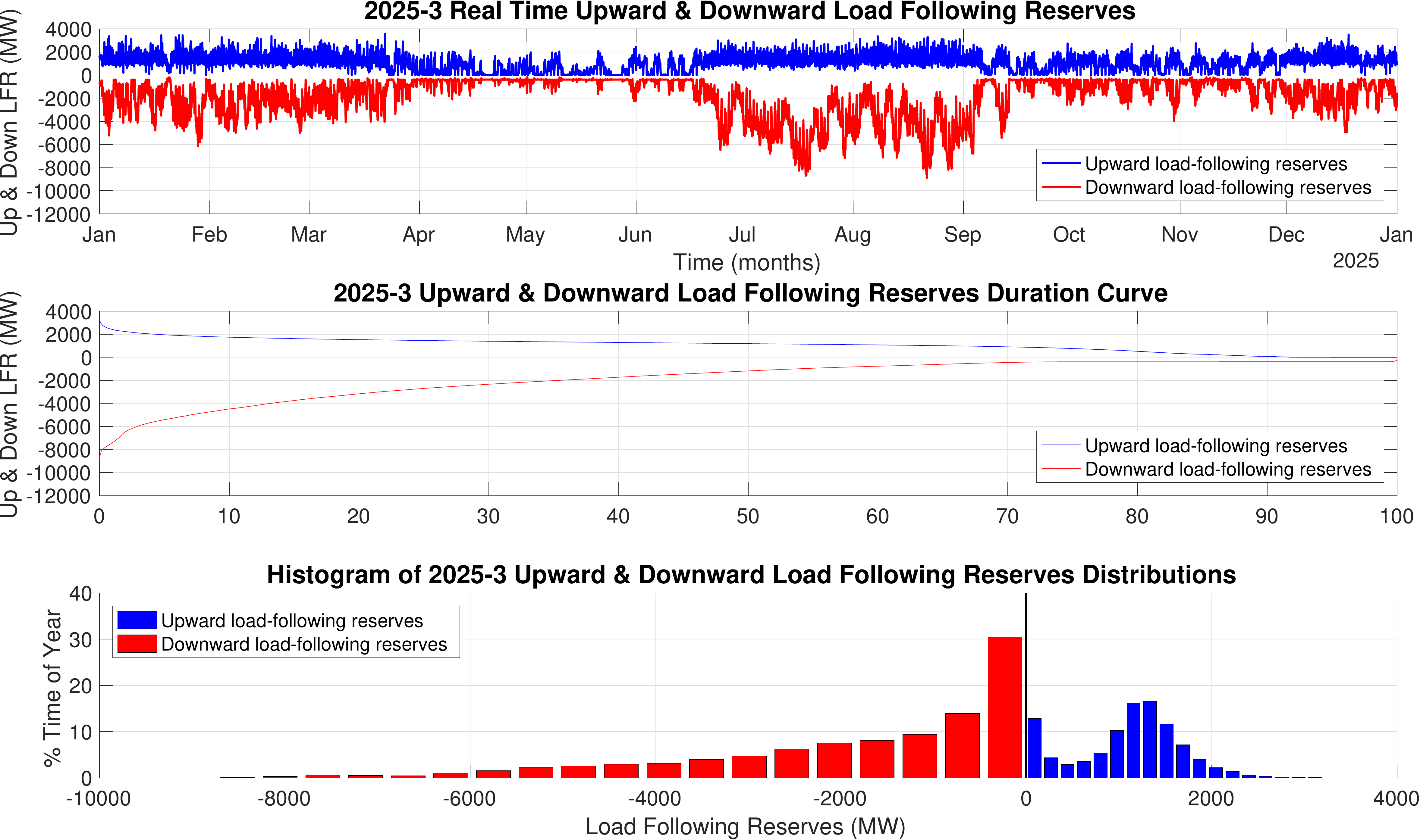}
\caption{Load following reserve profiles for Scenario 2025-3}
\label{fig:2025-3LFRProfile}
\end{figure}

In contrast to Scenario 2025-4, the results for Scenario 2025-3 in Figure~\ref{fig:2025-3LFRProfile} tell a different story. Both upward and downward load following reserves are often exhausted or nearly so.  The system is often unable to respond to the net load fluctuations. The integration of massive amounts of renewable energy for this scenario reduces the system net load significantly. This, coupled with the ``must-run" nuclear units, fosters situations with nearly no download load following reserves and an excess of upward load following reserves.  In the meantime, as the system net load rises, upward load following reserves can become constrained before additional units can be committed.  Figure~\ref{fig:2025-3LFRProfile} shows that in the Spring and the Fall, the ability to track such low net load conditions is particularly constrained.
\begin{figure}[!h]
\centering
\includegraphics[width=6.5in]{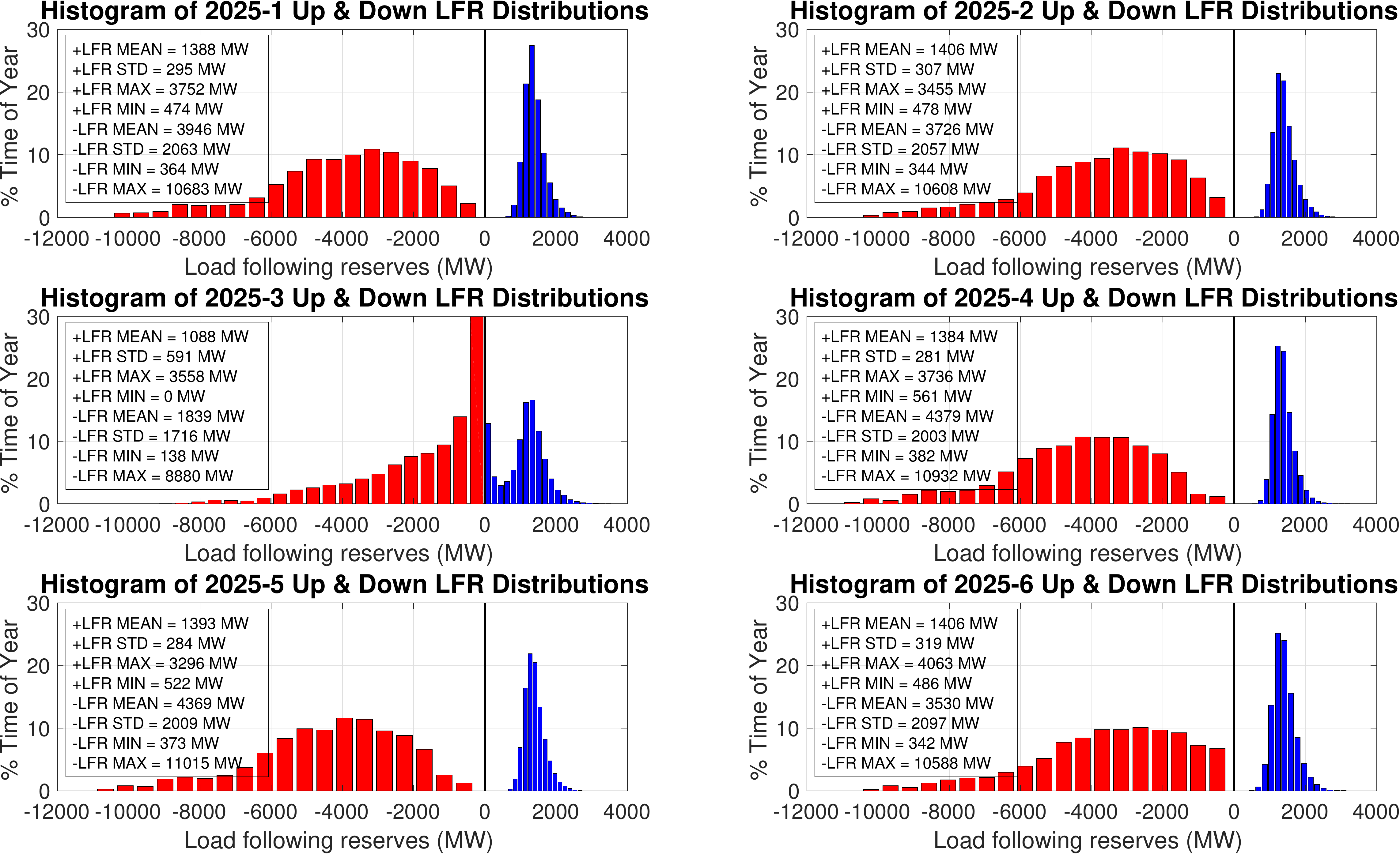}
\caption{Load following reserve distributions for 2025}
\label{fig:2025LFRDistributions}
\end{figure}
\begin{table}[!h]
\caption{Upward and downward load following reserve statistics for 2025 scenarios}
\begin{center}
\begin{tabular}{lrrrrrr}\toprule
 & \textbf{2025-1}& \textbf{2025-2}& \textbf{2025-3}& \textbf{2025-4}& \textbf{2025-5}& \textbf{2025-6}\\\toprule
\textbf{Up LFR Mean} (MW)&  1,376 &  1,385 &  1,160 &  1,377 &  1,380 &  1,392 \\\midrule
\textbf{Up LFR STD} (MW) &   302 &   307 &   558 &   286 &   285 &   321 \\\midrule
\textbf{Up LFR Min} (MW)&    10 &    28 &     0 &   277 &   142 &    81 \\\midrule
\textbf{Up LFR 95 percentile} (MW)&   958 &   957 &     1 &   977 &   976 &   937 \\\midrule
\textbf{Down LFR Mean} (MW)&  4,096 &  3,850 &  1,937 &  4,498 &  4,501 &  3,729 \\\midrule
\textbf{Down LFR STD} (MW) &  1,860 &  1,848 &  1,656 &  1,798 &  1,816 &  1,936 \\\midrule
\textbf{Down LFR Min} (MW)&   339 &   342 &    97 &   383 &   382 &   340 \\\midrule
\textbf{Down LFR 95 percentile} (MW)&  1,318 &  1,180 &   342 &  1,784 &  1,788 &   786 \\\bottomrule
\end{tabular}
\end{center}

\label{tab:2025LFRStatistics}
\end{table}

Figure~\ref{fig:2025LFRDistributions} shows the upward and downward load following reserve performances for all scenarios of 2025.  The key performance statistics for each scenario are extracted into Table~\ref{tab:2025LFRStatistics}. Here, the 95$^{th}$ percentile indicates that the system has more than this quantity of upward/downward load following reserves for 95\% of the time. The results show that for most of the scenarios the system exhibit sufficient downward load following reserves throughout the year. The lowest level of the downward load following reserves is only 97MW for Scenario 2025-3. However, judging by the corresponding 95$^{th}$ percentile value, such low-value occurrences are rare.  On the other hand, Scenario 2025-3 completely exhausts its upward load following reserves at least 5\% of the time, which represents a bigger issue. Thus, it can be concluded that despite the addition of significant renewable energy sources for Scenario 2025-3, the system is able to maintain adequate amount of downward load following reserves. However, such an increase of renewables energy capacity may require more upward load following reserves (in the form of newly committed units) to maintain the system's ability to follow upward net load trends as renewable energy generation diminishes.  

\begin{figure}[!h]
\centering
\includegraphics[width=5.5in]{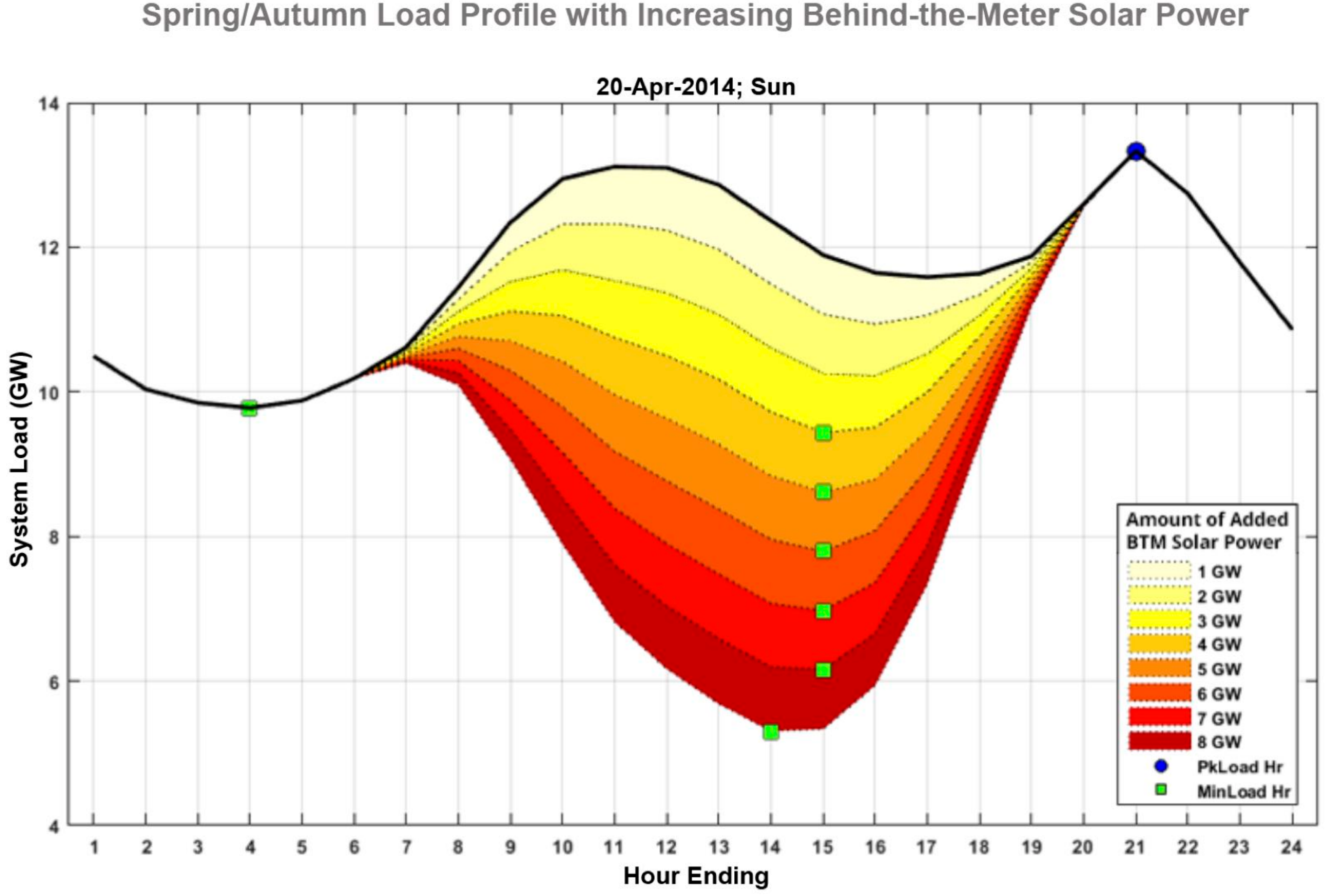}
\caption{The ``Duck Curve" in ISO New England\cite{iso-ne:2018:00}}
\label{fig:duckCurve}
\end{figure}
Such a result is consonant with the often discussed ``duck curve" shown in Figure \ref{fig:duckCurve} and first discussed by California ISO in the context of its solar integration planning studies.  The duck curve presents three important operational challenges.  During the midday hours, solar generation causes low net load conditions that will test a power system's ability to track downward using downward load following reserves.  Hours later, as solar generation wanes, net load conditions rise to their daily peak testing the power system's ability to track upward with upward load following reserves.  Finally, in the meantime, and as discussed in the next section, the transition hours between trough and peak conditions exhibits a sharp system ramp.  

\begin{figure}[!h]
\centering
\includegraphics[width=6.5in]{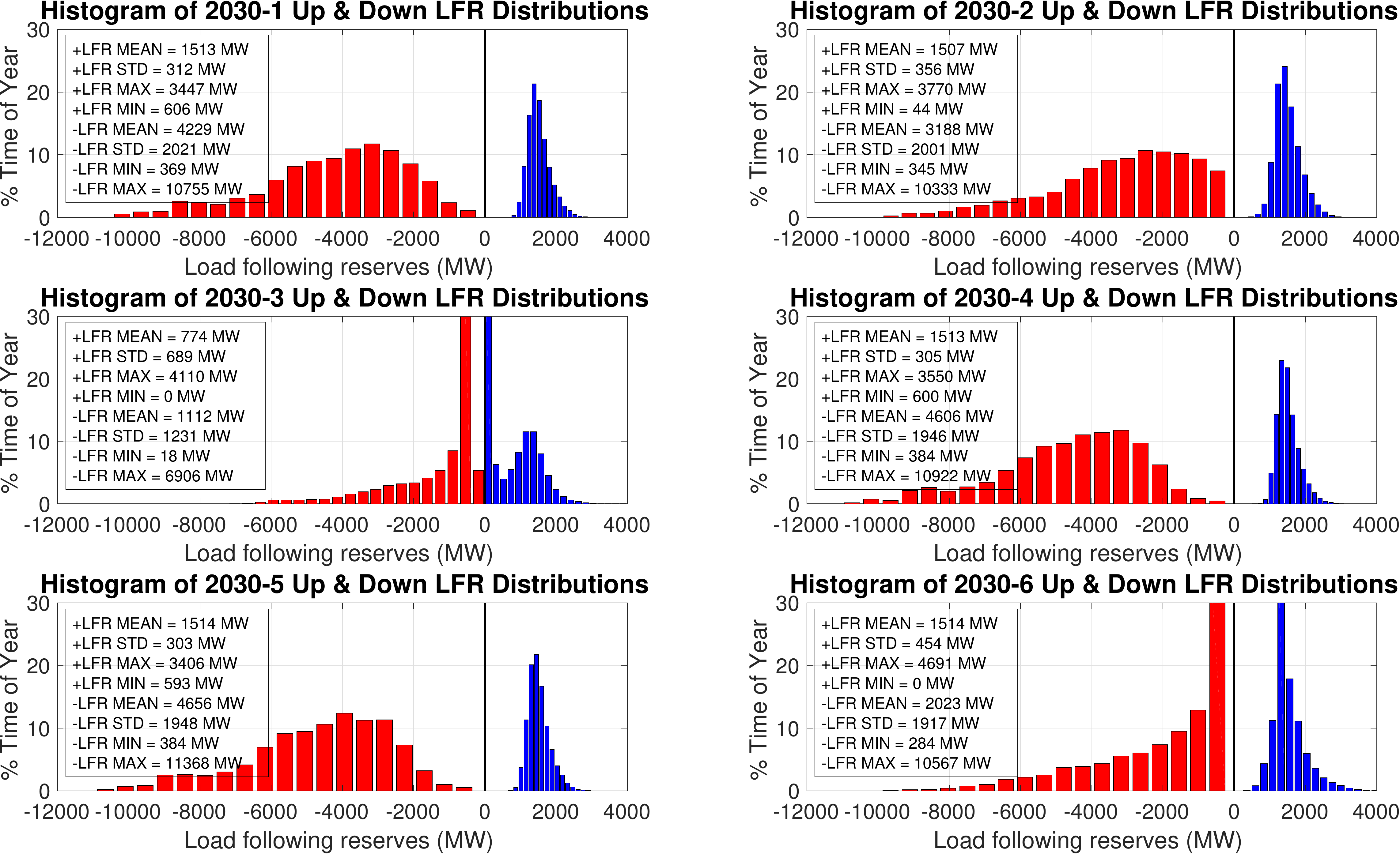}
\caption{Load following reserve distributions for 2030}
\label{fig:2030LFRDistributions}
\end{figure}
Similar to Figure \ref{fig:2025LFRDistributions}, Figure~\ref{fig:2030LFRDistributions} shows upward and downward load following reserve performances for all 2030 scenarios, and the key performance statistics for each scenario are extracted into Table~\ref{tab:2030LFRStatistics}. The results here are significantly different from 2025 scenarios and reveal several important aspects of systems overloaded with renewable energy sources. All scenarios, except for Scenario 2030-4, completely exhaust their upward load following reserves during some periods of the year. While such events occur in less than 5\% of time for all scenarios, except Scenario 2030-3, this is an overall negative trend compared to the system performance for 2025. Scenario 2030-3 experiences upward load following reserve shortages more often and requires an increase of such resources to maintain the balance of the system. The statistics in Table~\ref{tab:2030LFRStatistics} show that Scenarios 2030-3 and 2030-6 entirely exhaust their downward load following reserves; albeit for a fairly short part of the year. Despite such rare occurrences, the depletion of a resource that was assumed to be adequately available in the system for following the net load fluctuations shows the need for procurement of both upward and downward load following reserves in the day-ahead unit commitment. That said, the commitment of dispatchable resources and their associated quantities of commitment of load following and ramping reserves has a complex, difficult to predict, non-linear dependence on the amount of variable resources and the load profile statistics. Here, despite the similarities between Scenario 2030-4 and 2030-5, their associated quantities of load following reserves is quite different as a result of the differences in the resource characteristics between the two scenarios. 

\begin{table}[!h]
\caption{Upward and downward load following reserve statistics for 2030 scenarios}
\begin{center}
\begin{tabular}{lrrrrrr}\toprule
 & \textbf{2030-1}& \textbf{2030-2}& \textbf{2030-3}& \textbf{2030-4}& \textbf{2030-5}& \textbf{2030-6}\\\toprule
\textbf{Up LFR Mean} (MW)&  1,507 &  1,506 &   818 &  1,512 &  1,496 &  1,525 \\\midrule
\textbf{Up LFR STD} (MW) &   324 &   355 &   683 &   304 &   314 &   478 \\\midrule
\textbf{Up LFR Min} (MW)&     0 &     0 &     0 &   356 &     0 &     0 \\\midrule
\textbf{Up LFR 95 percentile} (MW)&  1,072 &  1,022 &     0 &  1,104 &  1,067 &   935 \\\midrule
\textbf{Down LFR Mean} (MW)&  4,374 &  3,333 &  1,145 &  4,730 &  4,805 &  2,125 \\\midrule
\textbf{Down LFR STD} (MW) &  1,805 &  1,827 &  1,212 &  1,738 &  1,714 &  1,865 \\\midrule
\textbf{Down LFR Min} (MW)&   351 &   340 &     0 &   425 &   389 &     0 \\\midrule
\textbf{Down LFR 95 percentile} (MW)&  1,728 &   714 &   335 &  2,167 &  2,285 &   342 \\\bottomrule
\end{tabular}
\end{center}

\label{tab:2030LFRStatistics}
\end{table}

\begin{figure}[!h]
\centering
\includegraphics[width=5.5in]{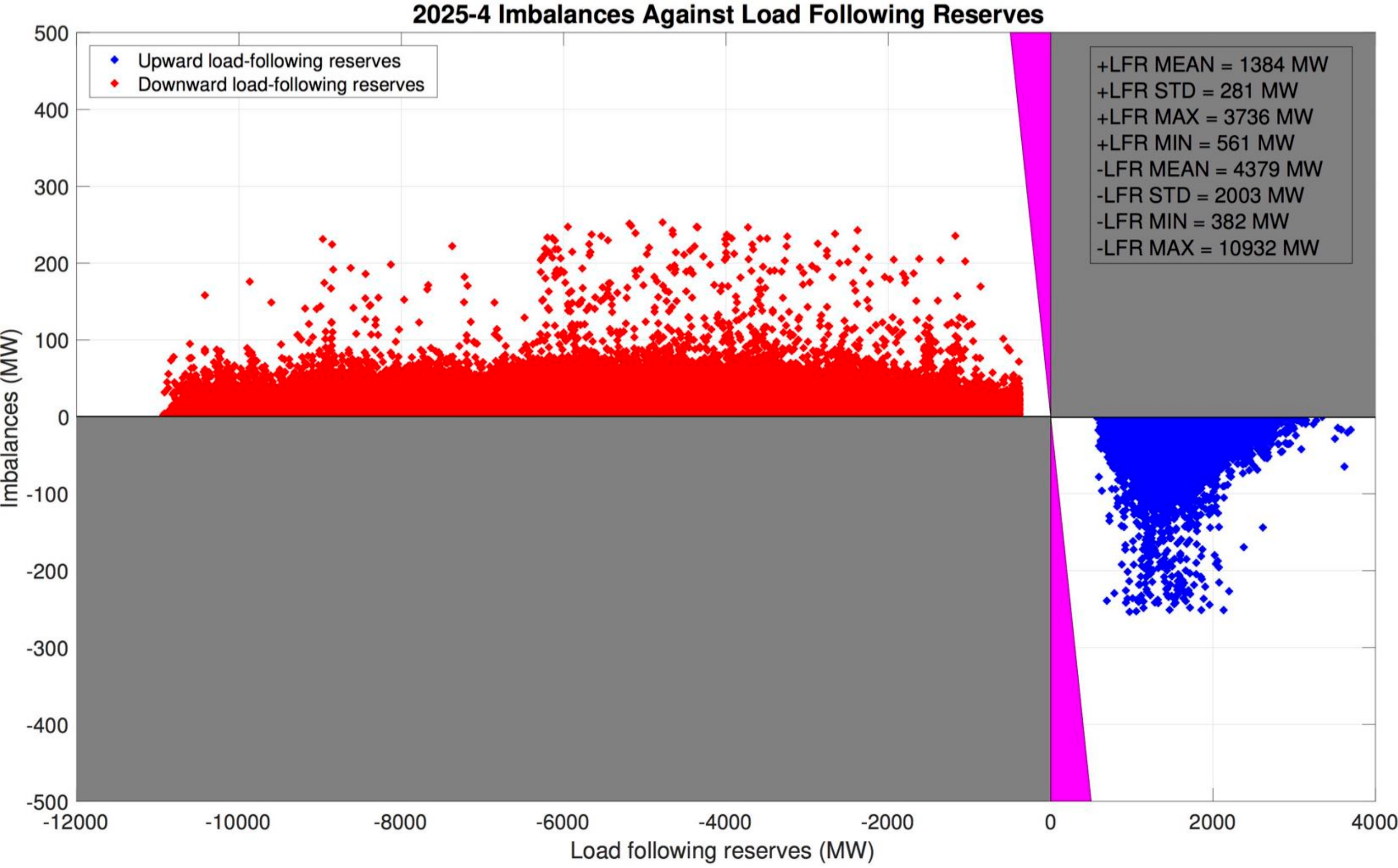}
\caption{Imbalances against load following reserves for Scenario 2025-4}
\label{fig:2025-4LFRImbalances}
\end{figure}

While the primary purpose of load following reserves is to mitigate the system imbalances induced by the net load variability and day-ahead forecast errors, increasing the quantity of load following reserves in the system is not always a comprehensive solution for imbalance mitigation. Certain portions of imbalances may be due to inadequate ramping capabilities of the resources or topological limitations of the system. To demonstrate this phenomenon, Figure~\ref{fig:2025-4LFRImbalances} shows the relationship between load following reserves and the imbalances for Scenario 2025-4. In the grey regions, upward and downward load following reserves do not serve to mitigate positive and negative imbalances respectively. In the white regions, upward and downward load following reserves serve to mitigate positive and negative imbalances respectively. In the magenta regions, a 1MW increase of load following reserves leads to a 1MW reduction of imbalances. This region represents when there are insufficient amounts of load following reserves to serve the system imbalance. In Scenario 2025-4, imbalances do not coincide with low load following reserves; suggesting that imbalance mitigation requires use of another means.

\begin{figure}[!h]
\centering
\includegraphics[width=6.5in]{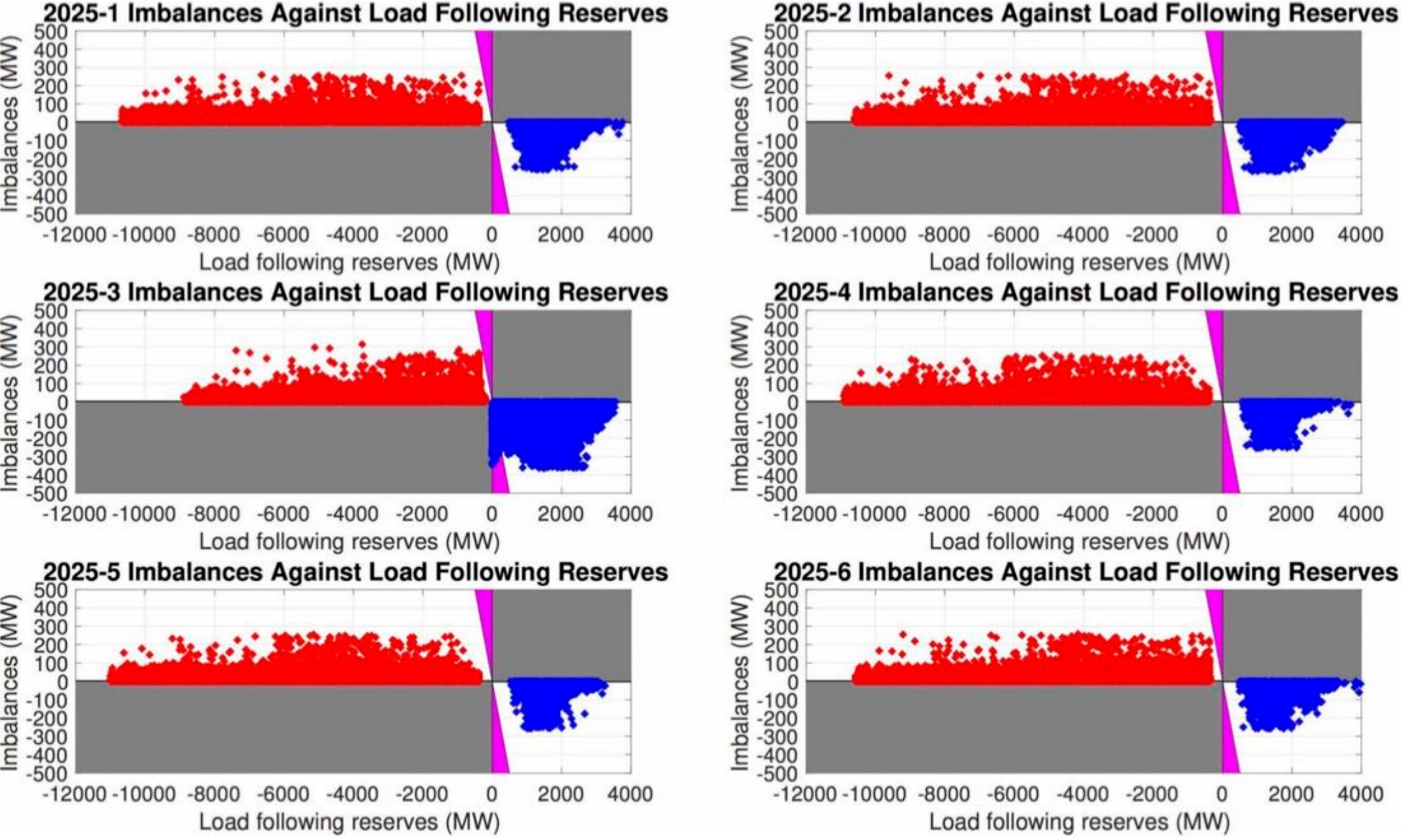}
\caption{Imbalances against load following reserves for 2025 scenarios}
\label{fig:2025ImbalanceLFR}
\end{figure}
\begin{figure}[!h]
\centering
\includegraphics[width=6.5in]{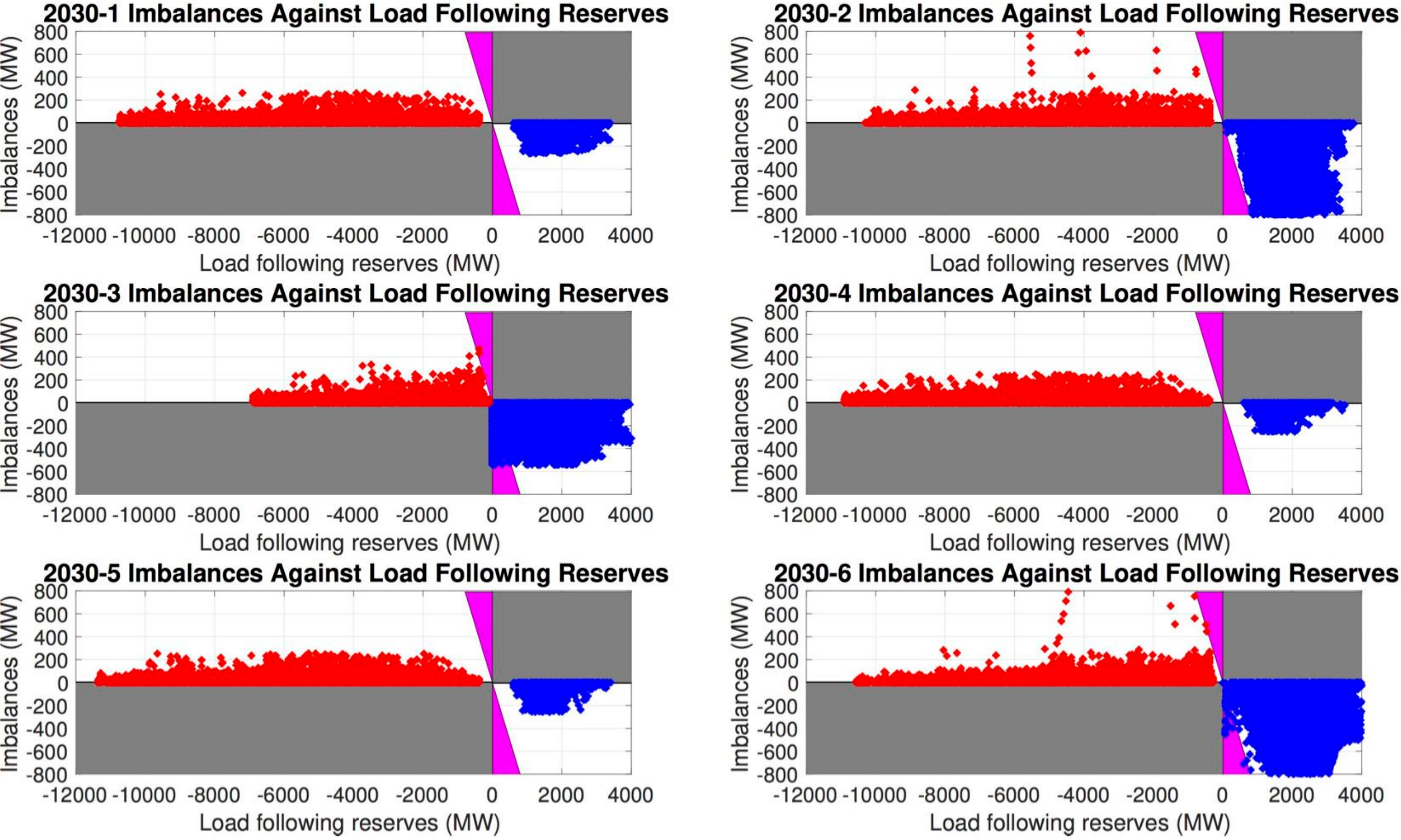}
\caption{Imbalances against load following reserves for 2030 scenarios}
\label{fig:2030ImbalanceLFR}
\end{figure}

Figure~\ref{fig:2025ImbalanceLFR} shows the relationship between load following reserves and the imbalances for the 2025 scenarios. These results confirm the conclusion reached above that all of the 2025 scenarios have sufficient downward load following reserves and, of the 2025 scenarios, only scenario Scenario 2025-3 would benefit from additional upward load following reserves. Similarly, Figure~\ref{fig:2030ImbalanceLFR} shows the relationship between load following reserve requirements and the imbalances for the 2030 scenarios. Here, Scenarios 2030-3 and 2030-6 show the coincidence of downward load following reserves and positive imbalances; suggesting a need for more of this type of reserve. Also, all 2030 scenarios, except for Scenarios 2030-4, would benefit from additional upward load following reserves. To summarize, most scenarios require varying degrees of additional load following reserves.

\subsection{Ramping Reserves}
Upward and downward ramping reserves are procured during the day-ahead resource scheduling to enhance the ramping cabilities of generation units when responding to net load variations in real-time. Traditionally, procurement of sufficient load following reserves has been of the primary concern, while the generator ramping constraints in the day-ahead scheduling were assumed to provide sufficient ramping capabilities to the system. However, for the power system configuration scenarios considered in this study, both load following and ramping reserves are equally important, as demonstrated by the results below.

\begin{figure}[!h]
\centering
\includegraphics[width=6.5in]{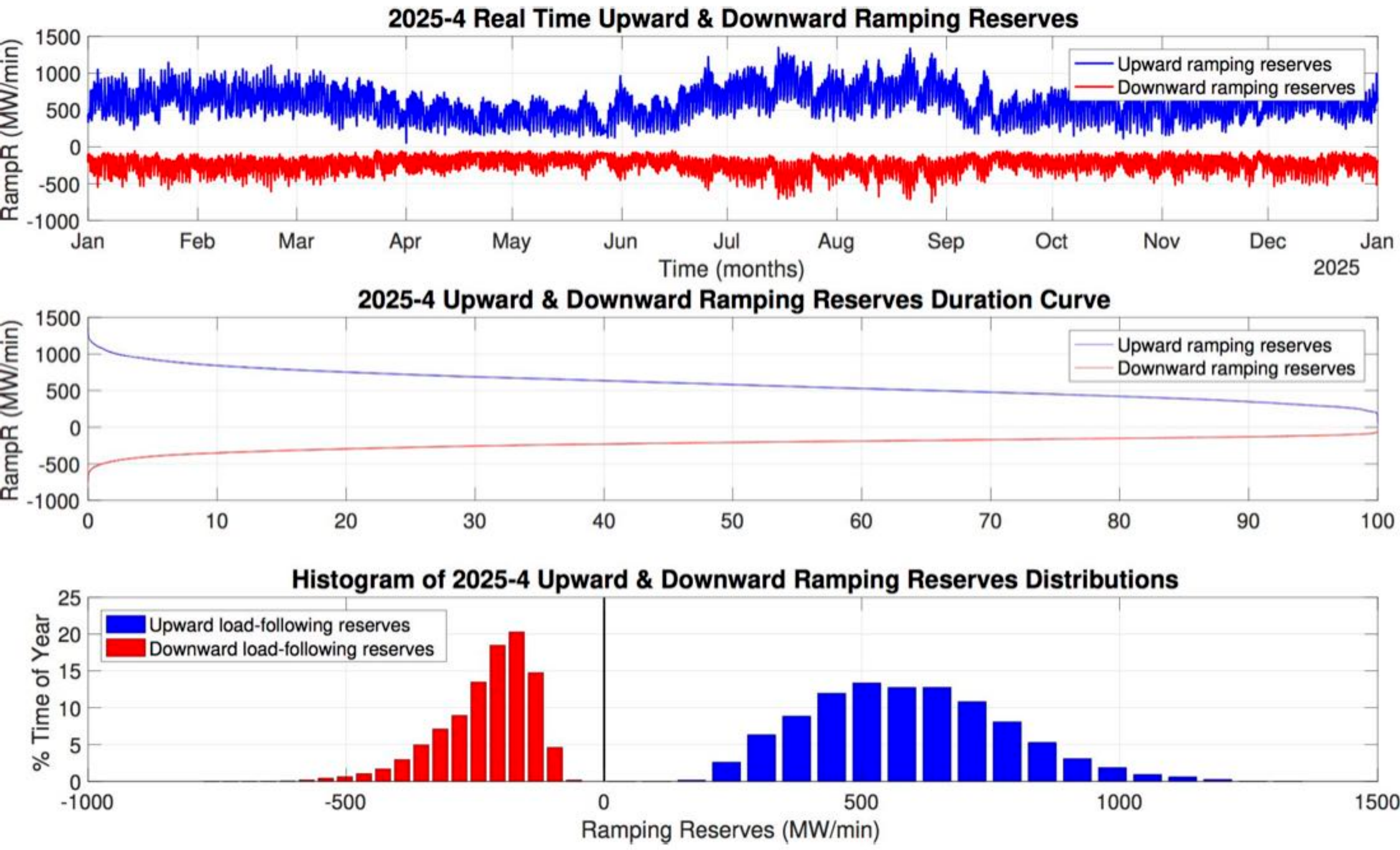}
\caption{Ramping reserves profile for Scenario 2025-4}
\label{fig:2025-4RampRProfile}
\end{figure}

As an example, the performance of ramping reserves for Scenario 2025-4 is shown in Figure~\ref{fig:2025-4RampRProfile}. The amount of upward ramping reserves fluctuate over time but is never completely exhausted (approach the zero black line). Downward ramping reserves hit the zero line for a few brief instances only. Similar to load following reserves, ramping reserves get the closest to depletion during low-load spring and fall periods. Thus, when the system follows the Scenario 2025-4, it is generally able to operate reliably without the need for more ramping capabilities. 

\begin{figure}[!t]
\centering
\includegraphics[width=6.5in]{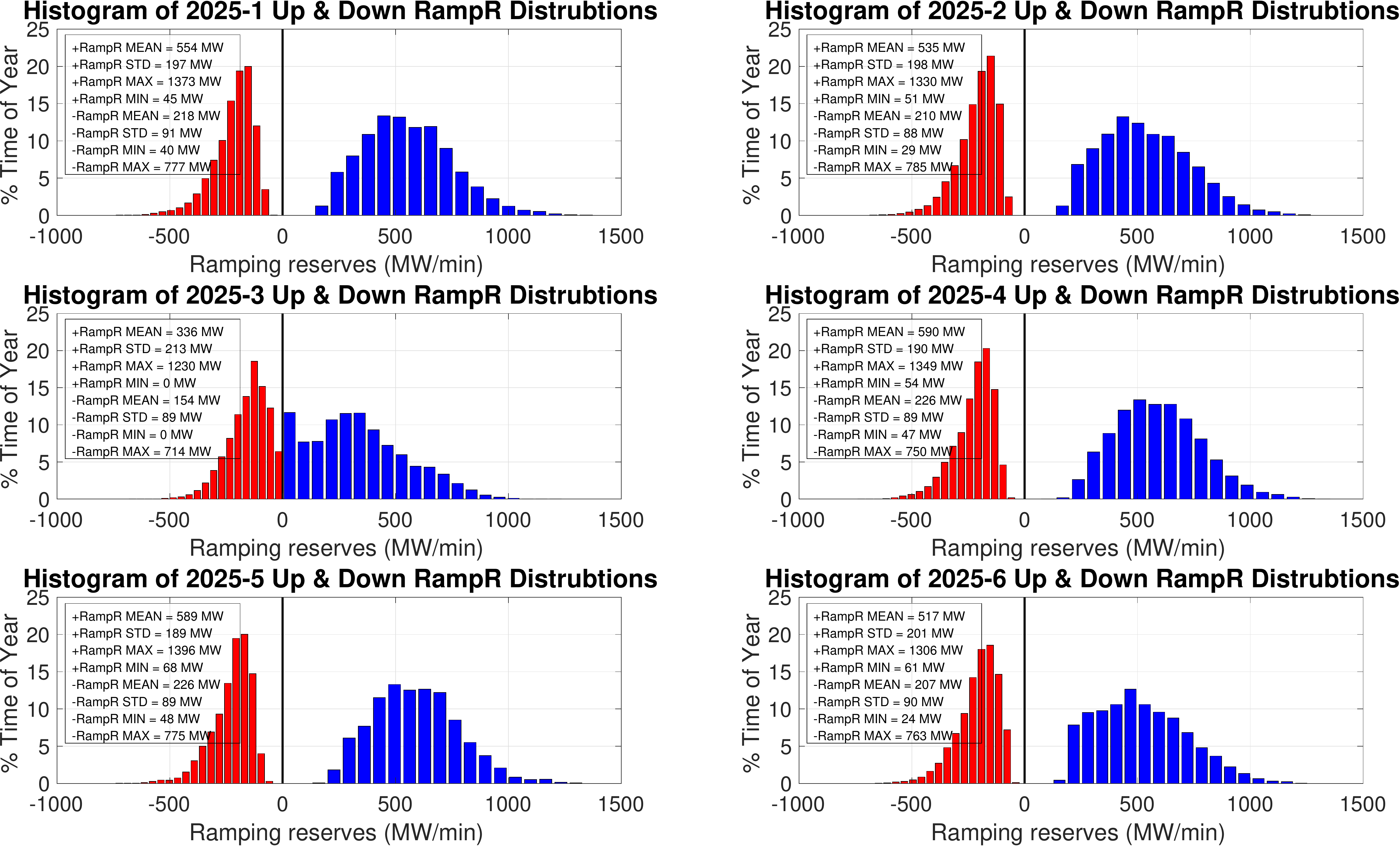}
\caption{Ramping reserves distributions for 2025 scenarios}
\label{fig:2025RampProfiles}
\end{figure}

Figure~\ref{fig:2025RampProfiles} shows the upward and downward ramping reserve performances for all scenarios of 2025, and the key performance statistics for each scenario are extracted into Table~\ref{tab:2025RampRStatistics}. Here, the 95$^{th}$ percentile indicates that the system has more than this quantity of upward/downward ramping reserves for 95\% of the time. The results show that downward ramping reserves for all scenarios hit the zero value at some point during the year. However, except for Scenario 2025-3, such occurrences are brief.  For Scenario 2025-3, on the other hand, both upward and downward ramping reserves have zero values far more often. Furthermore, their distributions are shifted closer to the zero black line, which also explains the low 95$^{th}$ percentile values. Thus, it can be concluded that the addition of significant renewable energy sources for Scenario 2025-3 requires more upward and downward ramping reserves to maintain the system's ability to follow the net load fluctuations.
\begin{table}[!t]
\caption{Upward and downward ramping reserves statistics for 2025 scenarios}
\begin{center}
\begin{tabular}{lrrrrrr}\toprule
 & \textbf{2025-1}& \textbf{2025-2}& \textbf{2025-3}& \textbf{2025-4}& \textbf{2025-5}& \textbf{2025-6}\\\toprule
\textbf{Up RampR Mean} (MW/min)&   591 &   571 &   367 &   621 &   623 &   554 \\\midrule
\textbf{Up RampR STD} (MW/min) &   204 &   204 &   218 &   194 &   197 &   210 \\\midrule
\textbf{Up RampR Max} (MW/min)&  1,412 &  1,390 &  1,291 &  1,420 &  1,433 &  1,362 \\\midrule
\textbf{Up RampR Min} (MW/min)&    78 &    85 &     0 &    69 &    38 &    95 \\\midrule
\textbf{Up RampR 95 percentile} (MW/min)&   285 &   267 &    38 &   329 &   326 &   243 \\\midrule
\textbf{Down RampR Mean} (MW/min)&   235 &   226 &   167 &   238 &   243 &   220 \\\midrule
\textbf{Down RampR STD} (MW/min) &   102 &   100 &    94 &    98 &   100 &   100 \\\midrule
\textbf{Down RampR Min} (MW/min)&    0 &    0 &    0 &    0 &    0 &    0 \\\midrule
\textbf{Down RampR Max} (MW/min)&   805 &   782 &   766 &   802 &   819 &   780 \\\midrule
\textbf{Down RampR 95 percentile} (MW/min)&   112 &   105 &    36 &   120 &   123 &    93 \\\bottomrule
\end{tabular}
\end{center}

\label{tab:2025RampRStatistics}
\end{table}

\begin{figure}[!t]
\centering
\includegraphics[width=6.5in]{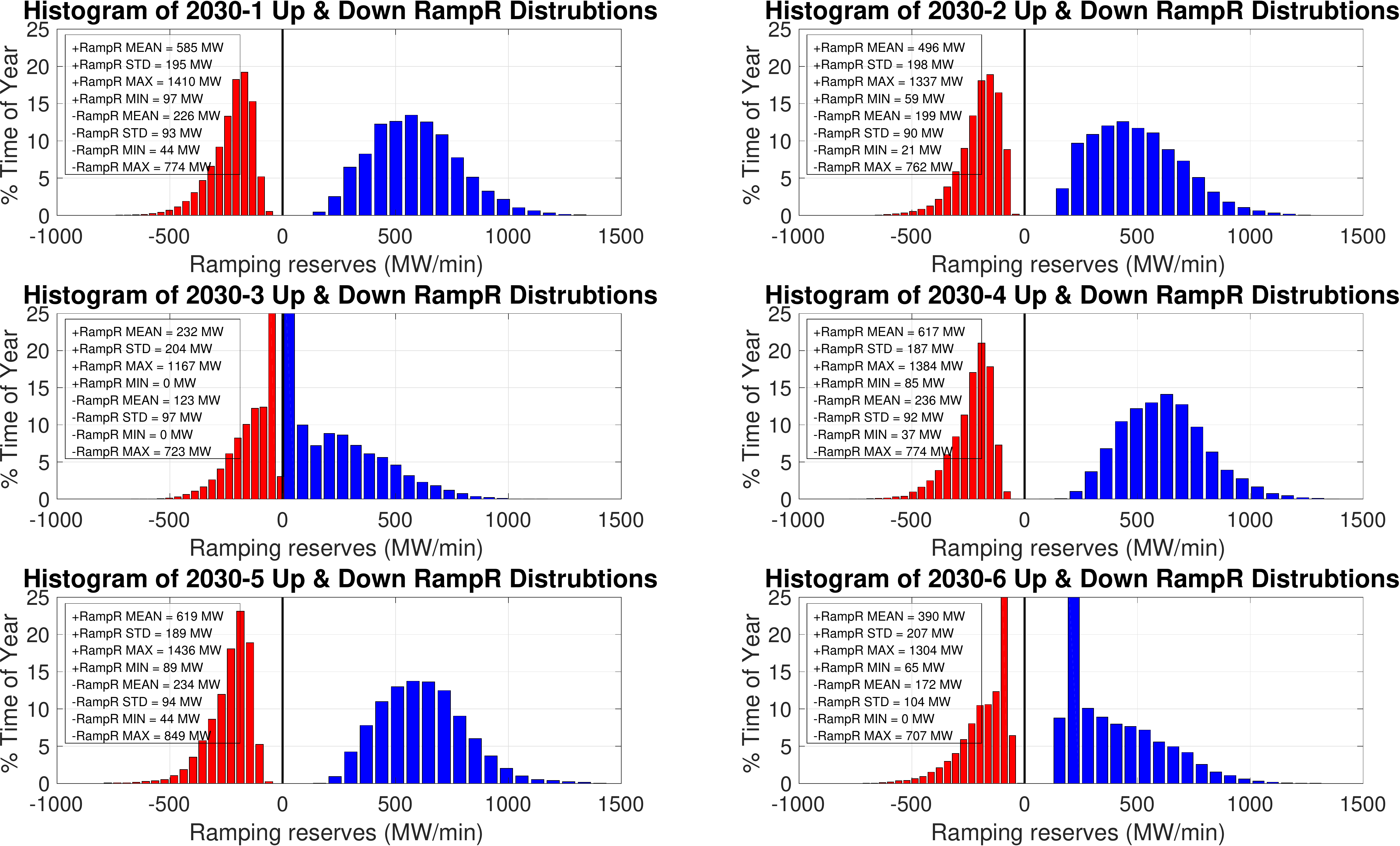}
\caption{Ramping reserves distributions for 2030 scenarios}
\label{fig:2030RampProfiles}
\end{figure}

Similarly, Figure~\ref{fig:2030RampProfiles} shows upward and downward ramping reserve performances for all scenarios of 2030, and the key performance statistics for each scenario are extracted into Table~\ref{tab:2030RampRStatistics}. The results here are similar to those for 2025 scenarios. Downward ramping reserves for all scenarios hit the zero value at some point during the year. However, except for Scenario 2030-3, such occurrences are brief.  Despite that, the depletion of a resource that was assumed to be adequately available in the system shows the need for the procurement of both upward and downward ramping reserves in the day-ahead unit commitment. For Scenario 2030-3, on the other hand, both upward and downward ramping reserves have zero values far more often. Furthermore, their distributions are shifted closer to the zero black line, which also explains the low 95$^{th}$ percentile values. Thus, it can be concluded that the addition of significant renewable energy sources for Scenarios 2025-3 and 2030-3 requires more upward and downward ramping reserves to maintain the system's ability to follow the net load fluctuations.  These results are consonant wit the ``duck curve" discussion found in the previous section on load following reserves.  
\begin{table}[!t]
\caption{Upward and downward ramping reserves statistics for 2030 scenarios}
\begin{center}
\begin{tabular}{lrrrrrr}\toprule
 & \textbf{2030-1}& \textbf{2030-2}& \textbf{2030-3}& \textbf{2030-4}& \textbf{2030-5}& \textbf{2030-6}\\\toprule
\textbf{Up RampR Mean} (MW/min)&   623 &   531 &   254 &   656 &   659 &   414 \\\midrule
\textbf{Up RampR STD} (MW/min) &   206 &   209 &   216 &   190 &   200 &   220 \\\midrule
\textbf{Up RampR Max} (MW/min)&  1,458 &  1,420 &  1,239 &  1,424 &  1,459 &  1,388 \\\midrule
\textbf{Up RampR Min} (MW/min)&    87 &    59 &     0 &    95 &    86 &    52 \\\midrule
\textbf{Up RampR 95 percentile} (MW/min)&   316 &   228 &    33 &   370 &   362 &   177 \\\midrule
\textbf{Down RampR Mean} (MW/min)&   242 &   213 &   134 &   251 &   250 &   182 \\\midrule
\textbf{Down RampR STD} (MW/min) &   109 &   101 &   105 &   102 &   112 &   111 \\\midrule
\textbf{Down RampR Min} (MW/min)&    0 &    0 &    0 &    0 &    0 &    0 \\\midrule
\textbf{Down RampR Max} (MW/min)&   850 &   801 &   771 &   845 &   836 &   791 \\\midrule
\textbf{Down RampR 95 percentile} (MW/min)&   118 &    91 &    31 &   129 &   123 &    70 \\\bottomrule
\end{tabular}
\end{center}
\label{tab:2030RampRStatistics}
\end{table}

\begin{figure}[!t]
\centering
\includegraphics[width=6.5in]{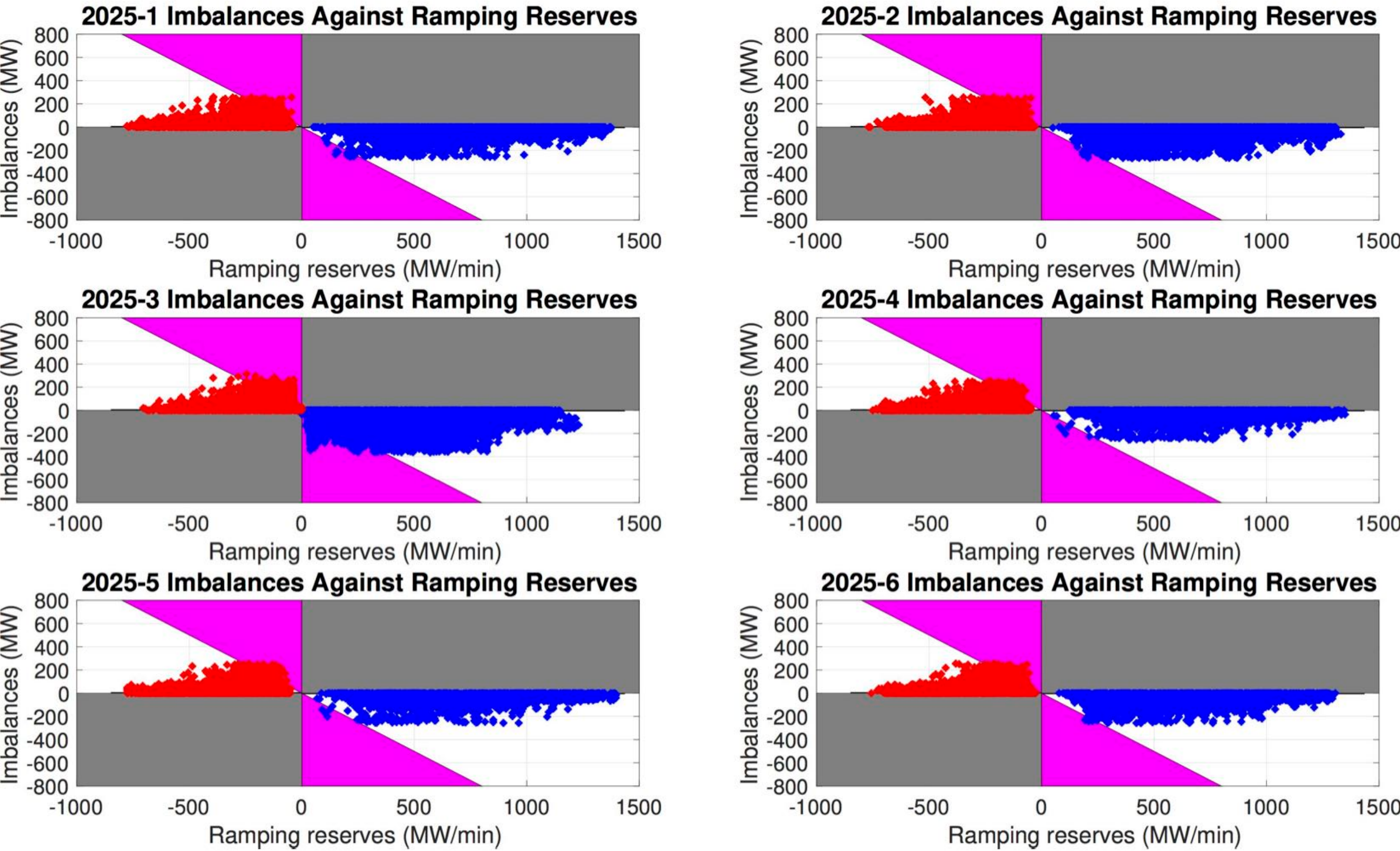}
\caption{Imbalances against ramping reserves for 2025 scenarios}
\label{fig:2025ImbalanceRampR}
\end{figure}

\begin{figure}[!t]
\centering
\includegraphics[width=6.5in]{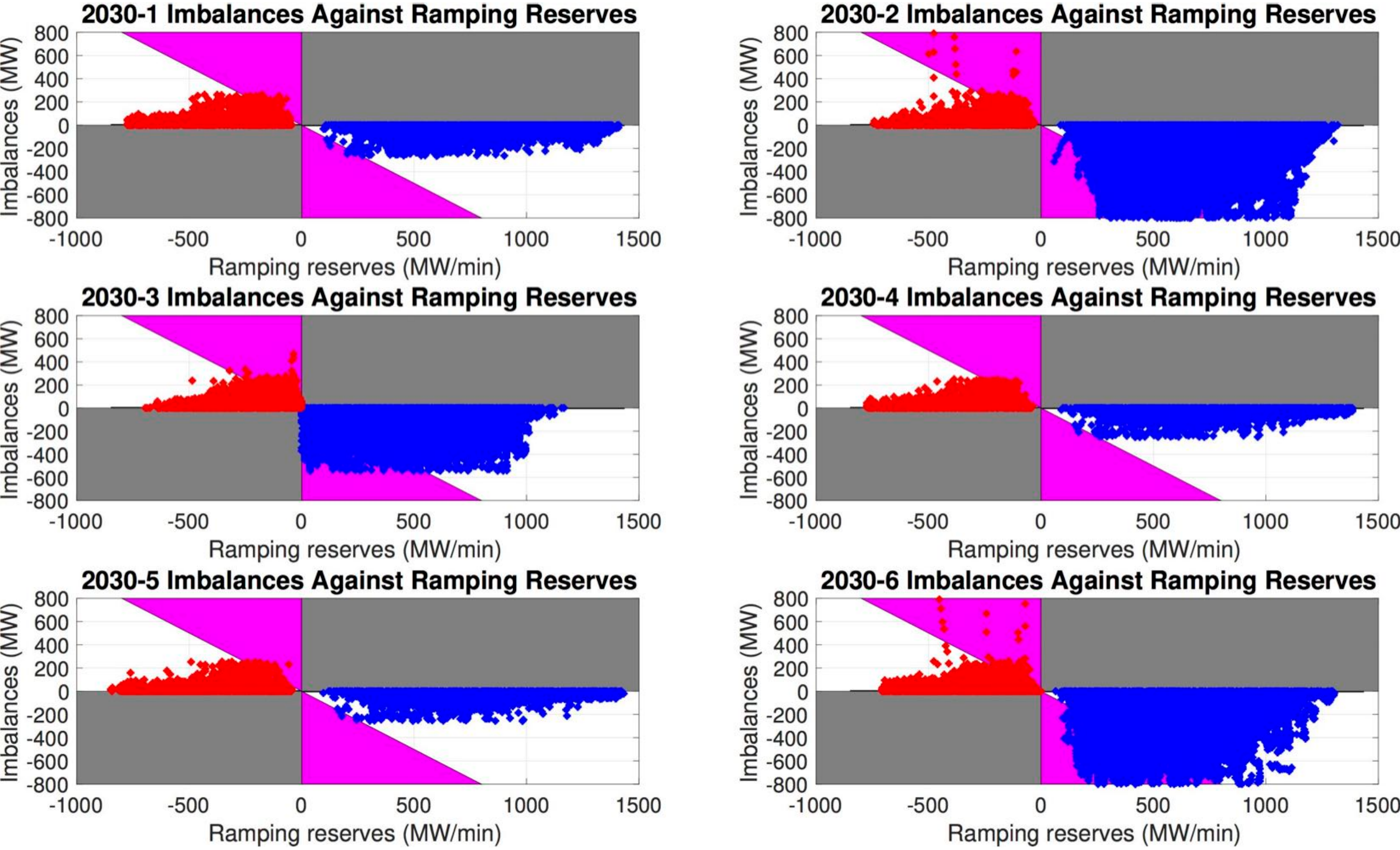}
\caption{Imbalances against ramping reserves for 2030 scenarios}
\label{fig:2030ImbalanceRampR}
\end{figure}

Along with load following reserves, the primary purpose of ramping reserves is to mitigate the system imbalances induced by the net load variability and day-ahead forecast errors. Figure~\ref{fig:2025ImbalanceRampR} and Figure~\ref{fig:2030ImbalanceRampR} show the relationship between ramping reserves and the imbalances for 2025 and 2030 scenarios respectively. In the grey regions, upward and downward ramping reserves do not serve to mitigate positive and negative imbalances respectively. In the white regions, upward and downward ramping reserves serve to mitigate positive and negative imbalances respectively. In the magenta regions, a 1MW/min increase of ramping reserves leads to a 1MW reduction of imbalances. This region represents when there are insufficient amounts of ramping reserves to serve the system imbalance. The results in Figure~\ref{fig:2025ImbalanceRampR} and Figure~\ref{fig:2030ImbalanceRampR} confirm the conclusions reached above that while for Scenarios 2025-3 and 2030-3 both shortages of upward and downward ramping reserves experience far more often, all scenarios would benefit from varying degrees of additional upward and downward ramping reserves.

\clearpage
\subsection{Curtailment of Semi-Dispatchable Resources}
\begin{figure}[!h]
\centering
\includegraphics[width=6.5in]{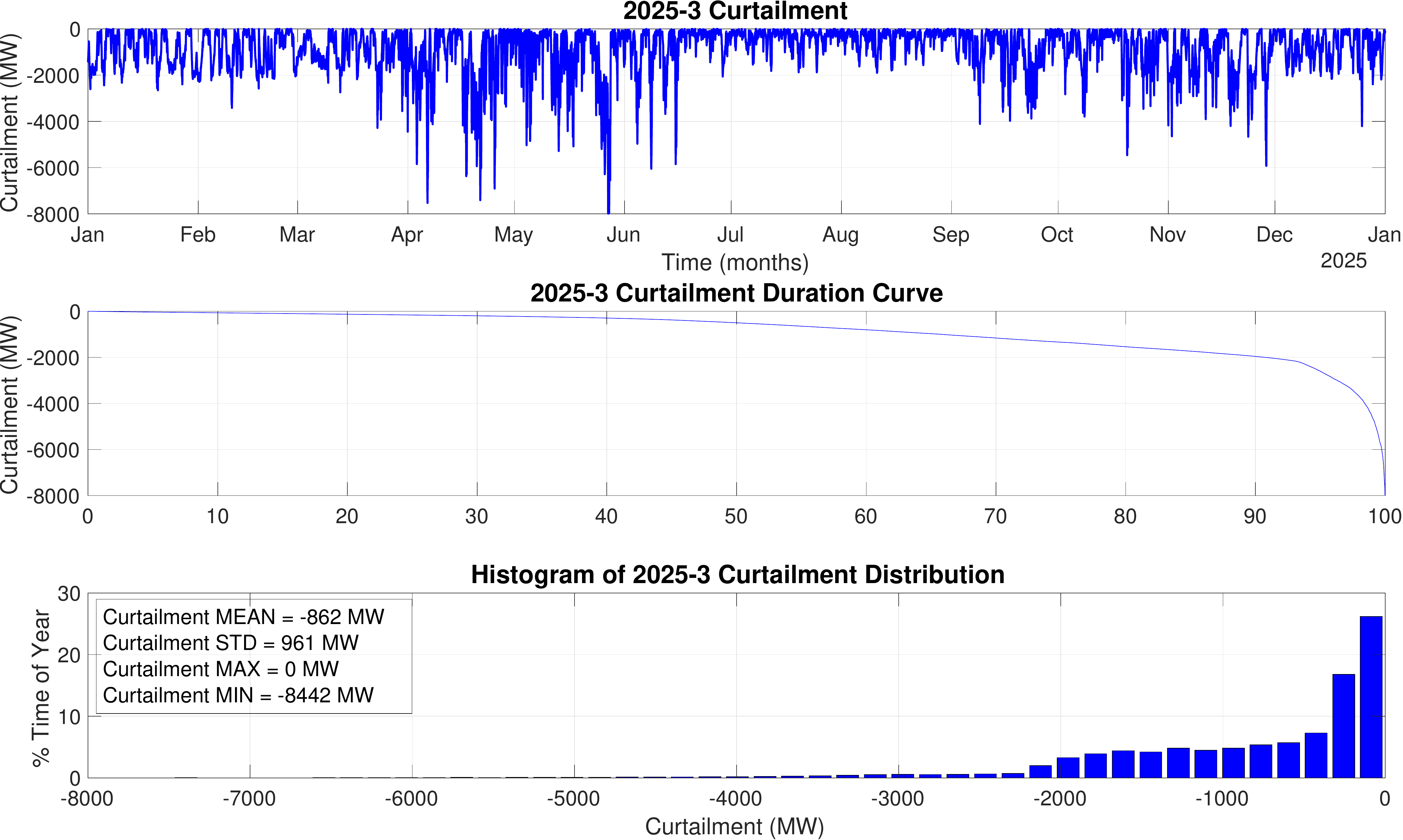}
\caption{Curtailments for Scenario 2025-3}
\label{fig:2025-4CurtailProfile}
\end{figure}

In the absence of adequate load following and ramping reserves, the curtailment of production from renewable energy sources, serves a vital balancing function. To emphasize the ability to reduce their power outputs, this study also refers to renewable energy sources as ``semi-dispatchable resources''. While curtailment of semi-dispatchable resources wastes generally cheaper and greener energy and, therefore, is a less desirable balancing method, it allows more flexibility and can help overcome topological limitations of the system where load following and ramping reserves might be ineffective. It is also important to emphasize that some of these topological limitations are due to the integration of semi-dispatchable resources in remote areas that replace the traditional generation units located close to the main consumption centers. Thus, semi-dispatchable resources might have a self-limiting feature which also defines the ability of the system to accommodate them. As an example, Figure~\ref{fig:2025-4CurtailProfile} shows the curtailment profile for Scenario 2025-3. The graph shows that some form of curtailment occurs for all but 0.1\% of the year.  Also, the largest curtailments occur during spring and fall when the system load is at its lowest.

\begin{figure}[!t]
\centering
\includegraphics[width=6.5in]{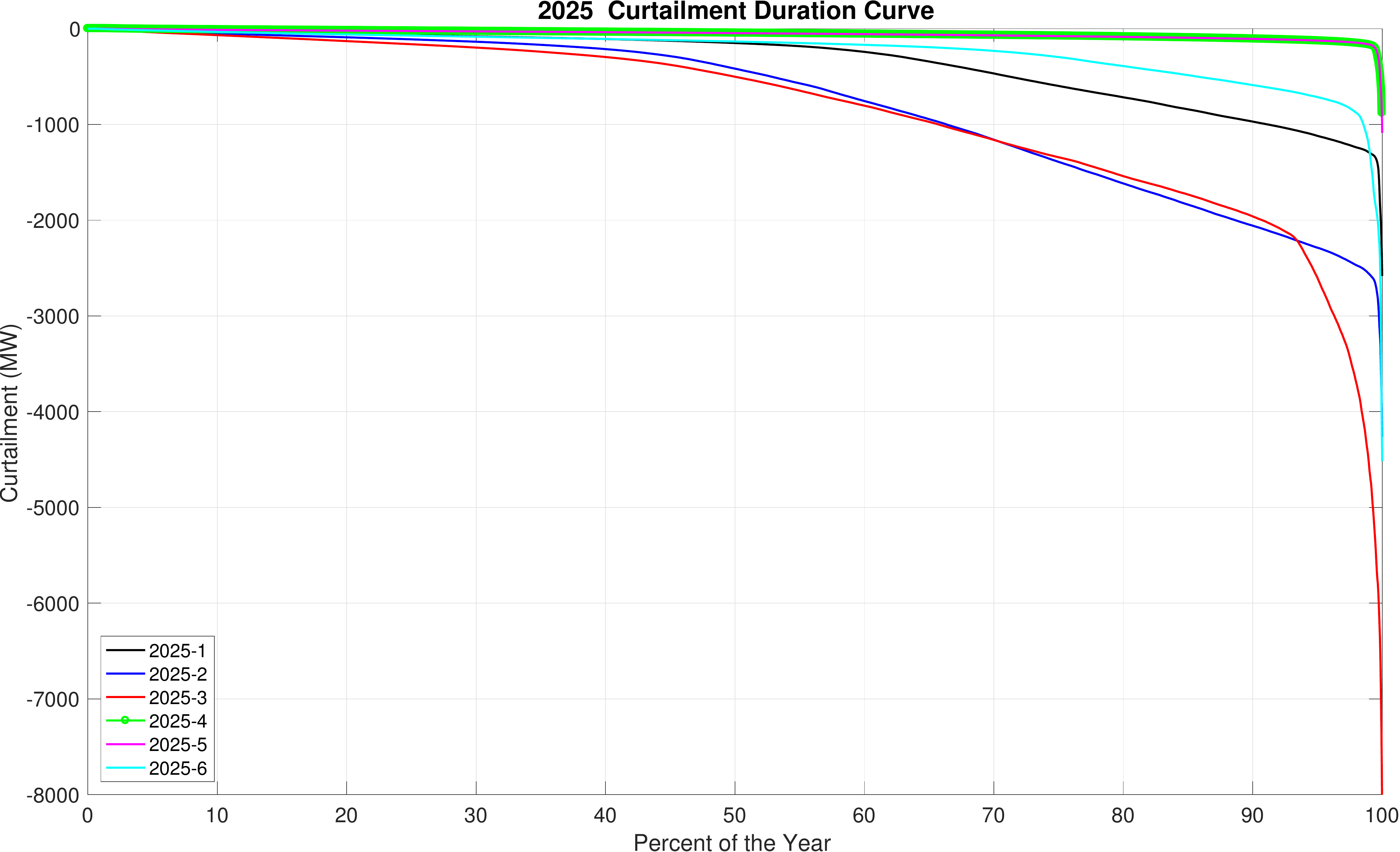}
\caption{Curtailment duration curves for 2025 scenarios}
\label{fig:2025CurtailProfile}
\end{figure}

Curtailment duration curves in Figure~\ref{fig:2025CurtailProfile} show that curtailment becomes an integral part of balancing operations for all 2025 Scenarios except 2025-4 and 2025-5. The results show that the largest curtailments occur for Scenarios 2025-2 and 2025-3. This is explained by the fact that Scenario 2025-3 is defined by the integration of large amounts of semi-dispatchable resources, and in Scenario 2025-2 the retired oil and coal units are replaced by semi-dispatchable resources instead of NGCC. This supports the statement above that the curtailment of semi-dispatchable resources is often times a way to mitigate topological limitations of the system amplified by the integration of the same semi-dispatchable resources. The curtailment statistics for 2025 scenarios in Table~\ref{tab:2025CurtailmentStats} show that they are used almost for the whole duration of the year for all scenarios. However, their magnitudes vary significantly for different scenarios, reaching the maximum of 8,442MW for Scenario 2025-3.
\begin{table}[!t]
\caption{Curtailment statistics for 2025 scenarios}
\begin{center}
\begin{tabular}{lrrrrrr}\toprule
 & \textbf{2025-1}& \textbf{2025-2}& \textbf{2025-3}& \textbf{2025-4}& \textbf{2025-5}& \textbf{2025-6}\\\toprule
\textbf{Tot. Semi-Disp. Res.} (GWh)& 48,674 & 55,215 & 63,850 & 41,532 & 41,532 & 53,118 \\\midrule
\textbf{Tot. Curtailed Semi-Disp.  Energy} (GWh) &  3,604 &  7,333 &  7,600 &  1,130 &  1,123 &  2,585 \\\midrule
\textbf{\% Semi-Disp. Energy Curtailed}&  7.41 & 13.28 & 11.90 &  2.72 &  2.70 &  4.87 \\\midrule
\textbf{\% Time Curtailed}& 99.61 & 99.79 & 99.90 & 98.89 & 98.83 & 99.63 \\\midrule
\textbf{Max Curtailment Level} (MW)&  2,880 &  4,115 &  8,442 &  1,605 &  1,701 &  4,748 \\\bottomrule
\end{tabular}
\end{center}
\label{tab:2025CurtailmentStats}
\end{table}

\begin{figure}[!t]
\centering
\includegraphics[width=6.5in]{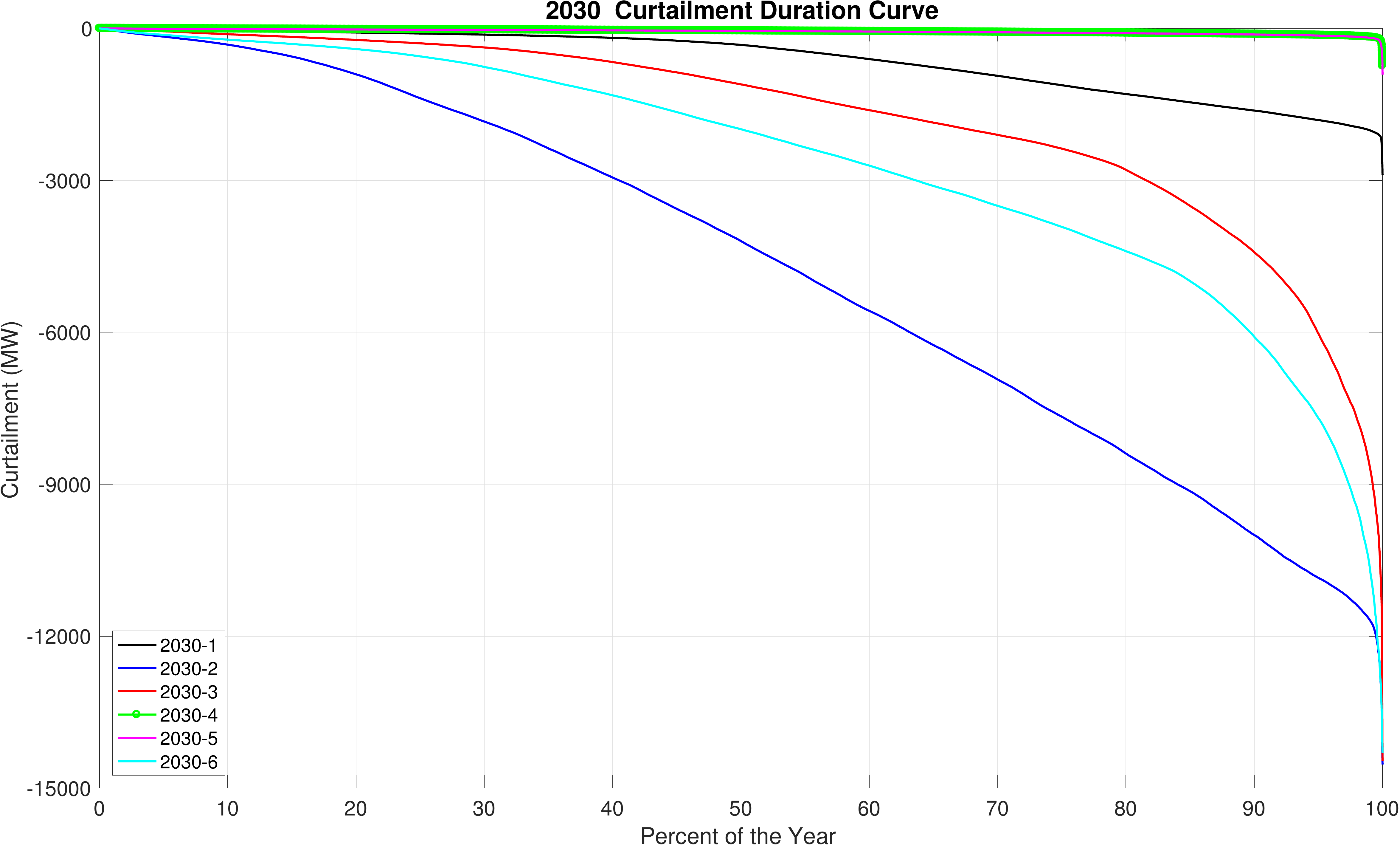}
\caption{Curtailment duration curves for 2030 scenarios}
\label{fig:2030CurtailProfile}
\end{figure}

Curtailment duration curves for 2030 scenarios are shown in Figure~\ref{fig:2030CurtailProfile}. Here too, curtailment plays an integral part of balancing operations for all Scenarios except 2030-4 and 2030-5. The results show that the largest curtailments occur for Scenarios 2030-2, 2030-3 and 2030-6. The emergence of Scenario 2030-6 as the case with the second largest curtailment is due to integration of large amounts of offshore wind units in 2030. Again, it can be observed that the curtailment of semi-dispatchable resources is often times a way to mitigate topological limitations of the system amplified by the integration of the same semi-dispatchable resources. The curtailment statistics for 2030 scenarios in Table~\ref{tab:2030CurtailmentStats} show that they are used almost for the whole duration of the year for all scenarios. However, their magnitudes vary significantly for different scenarios, reaching the maximum of 14,534MW for Scenario 2030-2.
\begin{table}[!t]
\caption{Curtailment statistics for 2030 scenarios}
\begin{center}
\begin{tabular}{lrrrrrr}\toprule
 & \textbf{2030-1}& \textbf{2030-2}& \textbf{2030-3}& \textbf{2030-4}& \textbf{2030-5}& \textbf{2030-6}\\\toprule
\textbf{Tot. Semi-Disp. Res.} (GWh)& 52,748 & 100,786 & 76,606 & 42,662 & 42,662 & 97,115 \\\midrule
\textbf{Tot. Curtailed Semi-Disp.   Energy} (GWh) &  5,993 & 41,517 & 14,495 &  1,149 &  1,162 & 22,531 \\\midrule
\textbf{\% Semi-Disp. Energy Curtailed} & 11.36 & 41.19 & 18.92 &  2.69 &  2.72 & 23.20 \\\midrule
\textbf{\% Time Curtailed} & 99.85 & 99.95 & 99.88 & 98.84 & 98.91 & 99.95 \\\midrule
\textbf{Max Curtailment Level} (MW)&  3,378 & 14,534 & 14,468 &  1,640 &  1,637 & 14,234 \\\bottomrule
\end{tabular}
\end{center}
\label{tab:2030CurtailmentStats}
\end{table}

\clearpage
\subsection{Interface and Tie-Line Performances}
As mentioned in the previous section, one of the main reasons for curtailment of semi-dispatchable resources are topological limitations of the system. These limitations are primarily due to the enforcement of several interface flow limits depicted in Figure~\ref{fig:topology}, which may cause congestion in the system and require curtailment. The performance of the following four key interfaces are discussed here:
\begin{itemize}
\item Orrington-South
\item Surowiec-South
\item North-South
\item SEMA-RI Import
\end{itemize}
The other interfaces and tie-lines in their respective scenarios exhibit negligible or no congestion.

\begin{figure}[!b]
\centering
\includegraphics[width=6.5in]{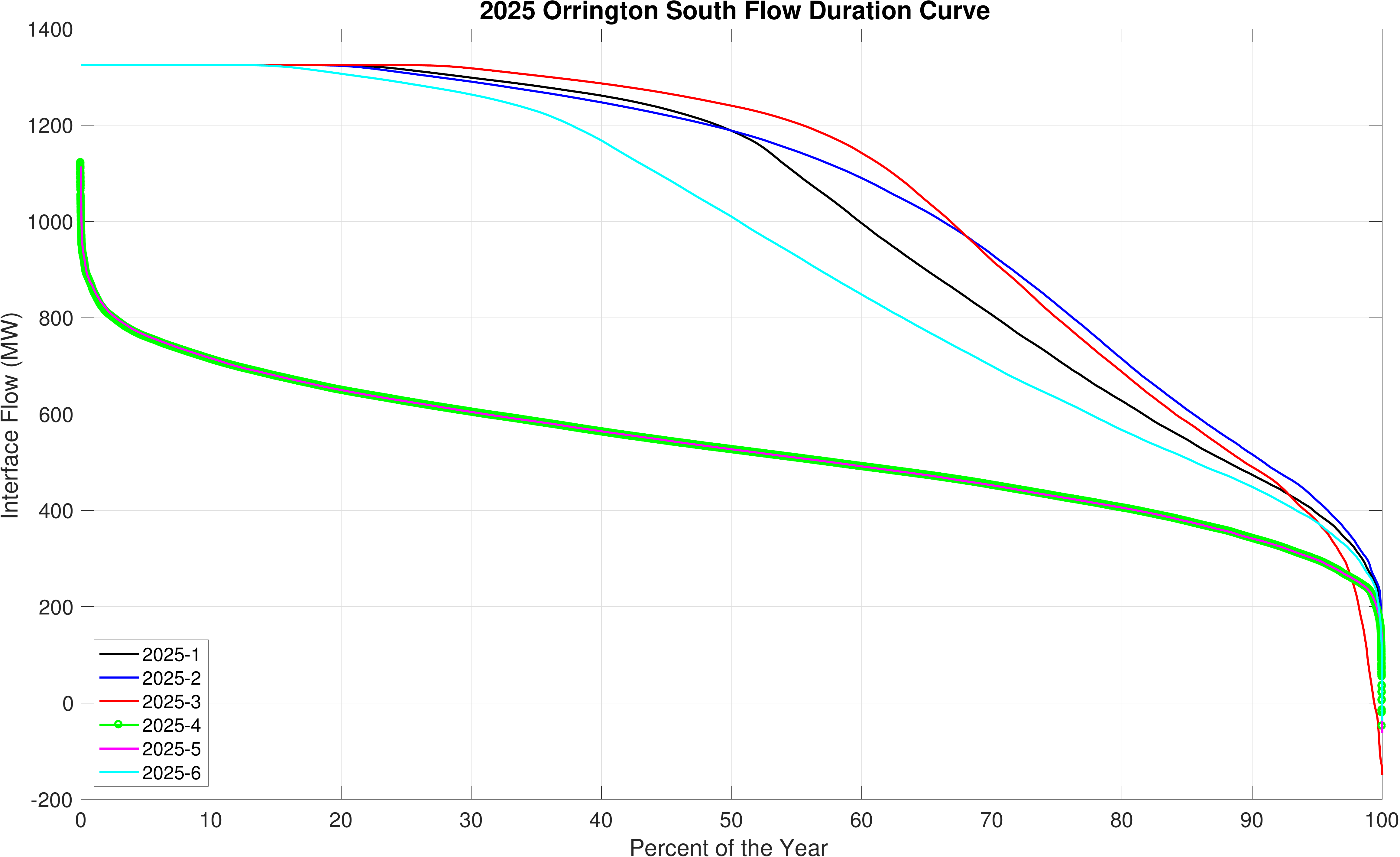}
\caption{Orrington-South flow duration curves for 2025 scenarios}
\label{fig:2025Interface1Duration}
\end{figure}

\begin{figure}[!h]
\centering
\includegraphics[width=6.5in]{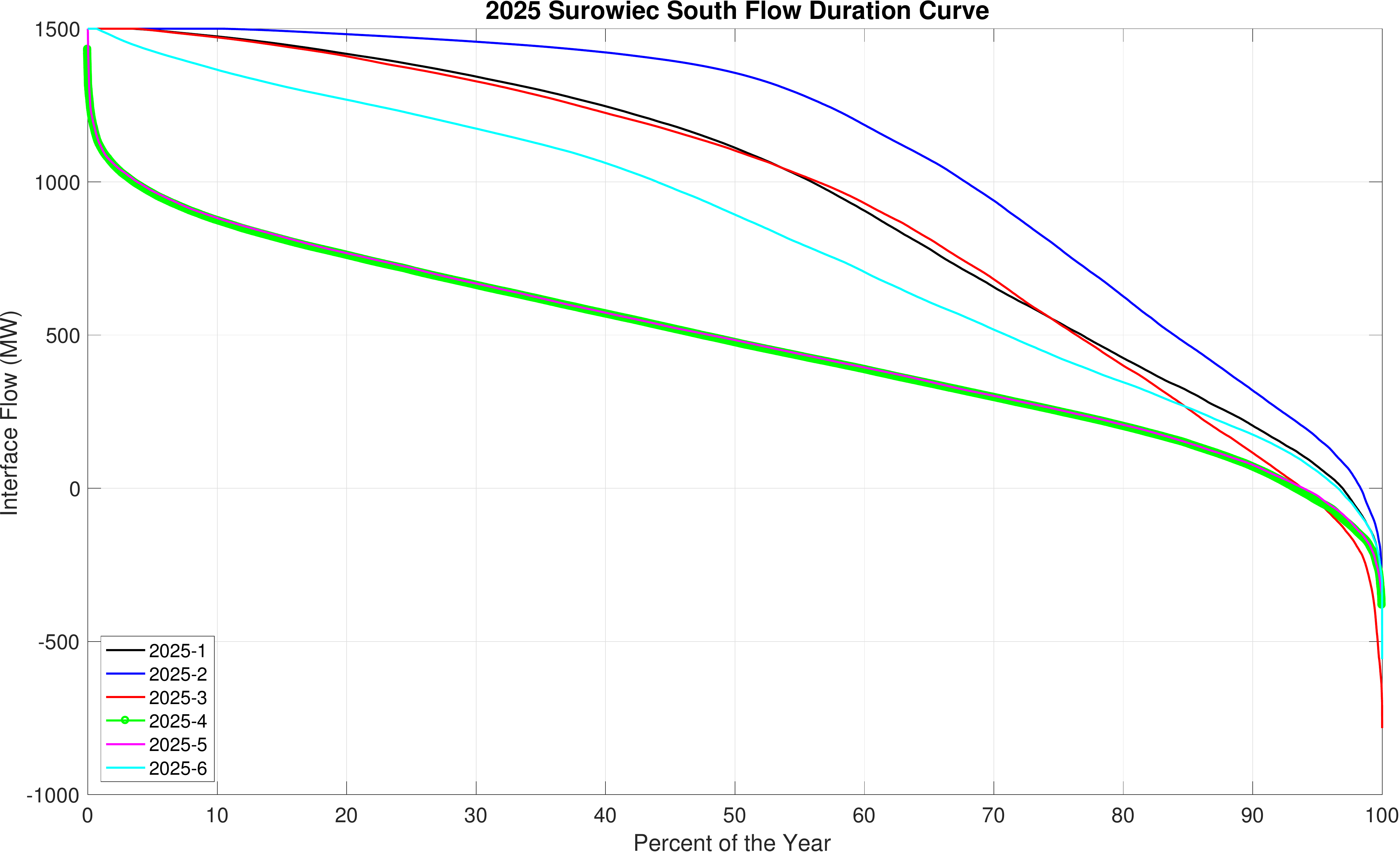}
\caption{Surowiec-South flow duration curves for 2025 scenarios}
\label{fig:2025Interface2Duration}
\end{figure}

\begin{figure}[!h]
\centering
\includegraphics[width=6.5in]{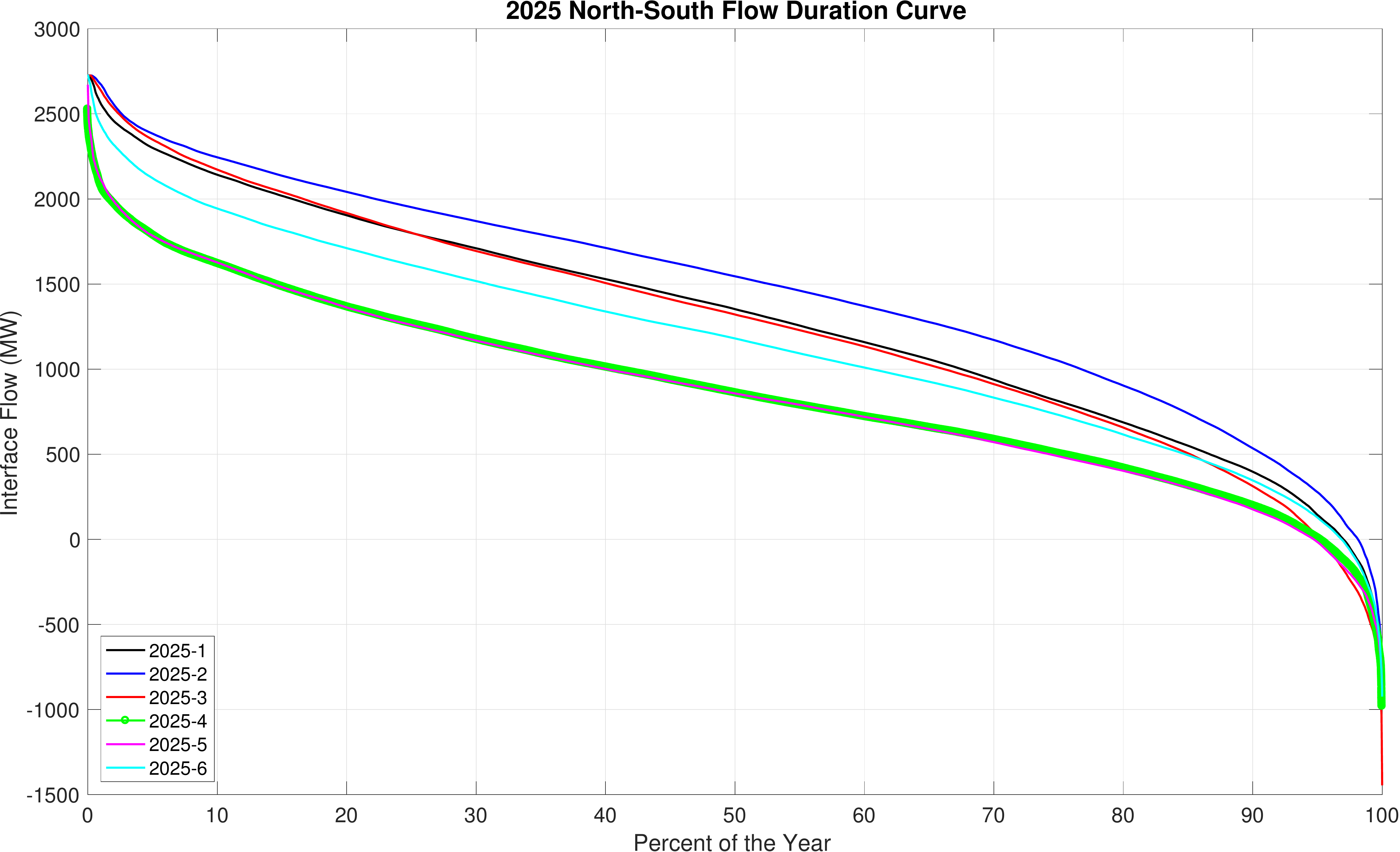}
\caption{North-South flow duration curves for 2025 scenarios}
\label{fig:2025Interface4Duration}
\end{figure}

\begin{figure}[!h]
\centering
\includegraphics[width=6.5in]{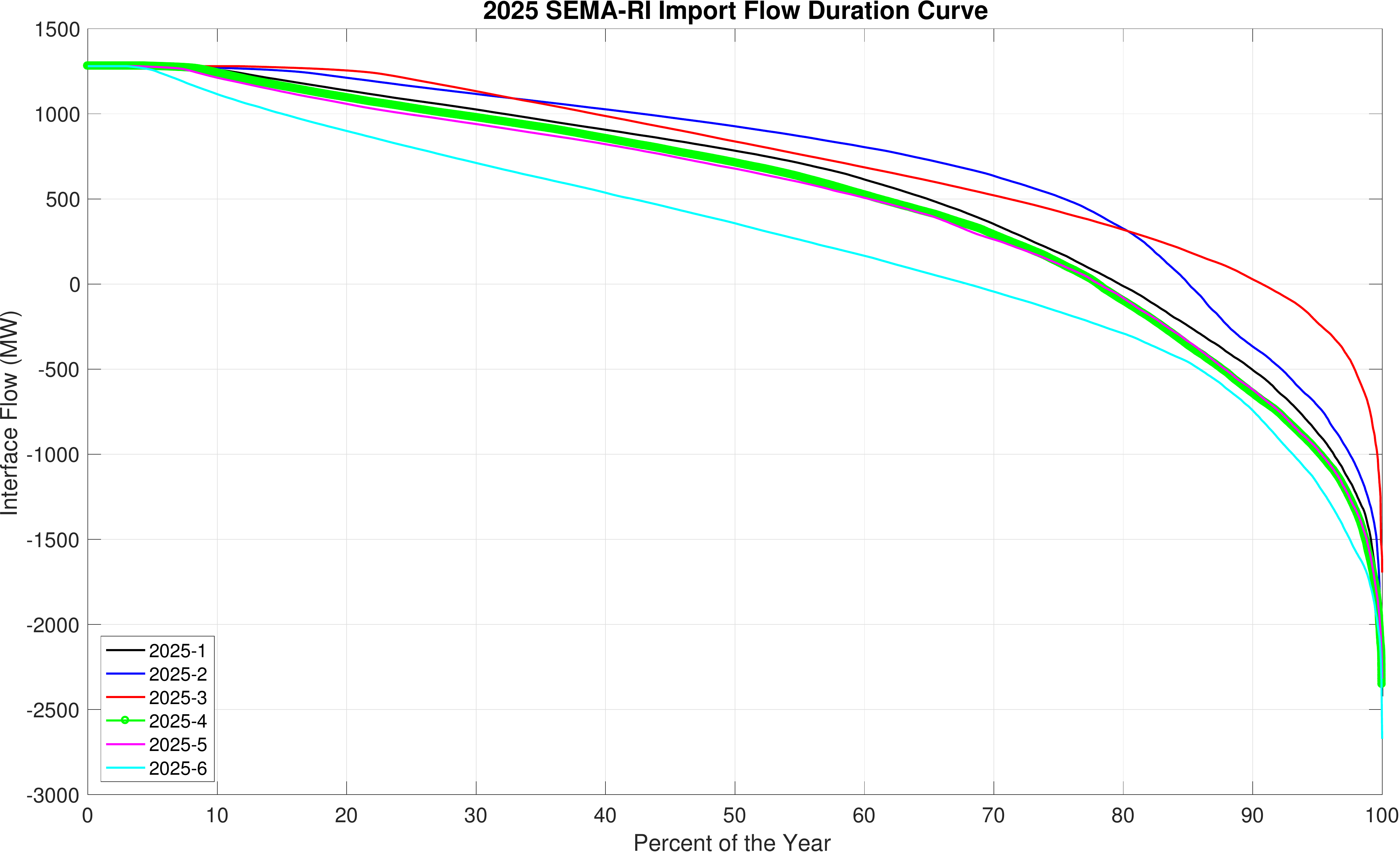}
\caption{SEMA-RI Import flow duration curves for 2025 scenarios}
\label{fig:2025Interface8Duration}
\end{figure}

Figure~\ref{fig:2025Interface1Duration} shows the duration curve for flows across the Orrington South interface.  It shows that the system experiences significant congestion on the Orrington-South interface for Scenarios 2025-1, 2025-2, 2025-3, and 2025-6 compared to Scenarios 2025-4 and 2025-5 that have no congestions at all. A similar pattern, but to a lesser degree, is observed on the Surowiec-South interface shown in Figure~\ref{fig:2025Interface2Duration}. The important observation here is that these scenarios are defined by a significant increase of renewable energy resources in the system. The power generated by these resources needs to flow down from remote areas of Maine towards the main consumption centers, such as Massachusetts, which causes the congestion on these two interfaces. On the other hand, the North-South interface shown in Figure~\ref{fig:2025Interface4Duration} exhibits congestion only in rare cases. This is due to the fact that the North-South interface has a much higher interface limit of 2,725MW, and is able to pass the additional renewable energy generation coming through the Orrington-South and the Surowiec-South interfaces without being congested. Finally, the SEMA-RI import interface in Figure~\ref{fig:2025Interface8Duration} exhibits some congestion for all 2025 scenarios. The interface congestion statistics for 2025 are summarized in Table~\ref{tab:2025InterfaceCongStats}. 
\begin{table}[!h]
\caption{Interface congestion statistics for 2025 scenarios}
\begin{center}
\begin{tabular}{lrrrrrr}\toprule
 & \textbf{2025-1}& \textbf{2025-2}& \textbf{2025-3}& \textbf{2025-4}& \textbf{2025-5}& \textbf{2025-6}\\\toprule
\textbf{Orrington South  \% Time Congested}& 20.49 & 19.05 & 27.06 &  0.00 &  0.00 & 13.91 \\\midrule
\textbf{Surowiec South  \% Time Congested}&  4.39 & 11.82 &  4.41 &  0.00 &  0.00 &  0.90 \\\midrule
\textbf{North-South  \% Time Congested}&  0.15 &  0.38 &  0.51 &  0.00 &  0.00 &  0.04 \\\midrule
\textbf{SEMA-RI Import  \% Time Congested}&  3.09 &  3.61 &  9.88 &  3.22 &  3.07 &  2.00 \\\bottomrule
\end{tabular}
\end{center}
\label{tab:2025InterfaceCongStats}
\end{table}

\begin{figure}[!h]
\centering
\includegraphics[width=6.5in]{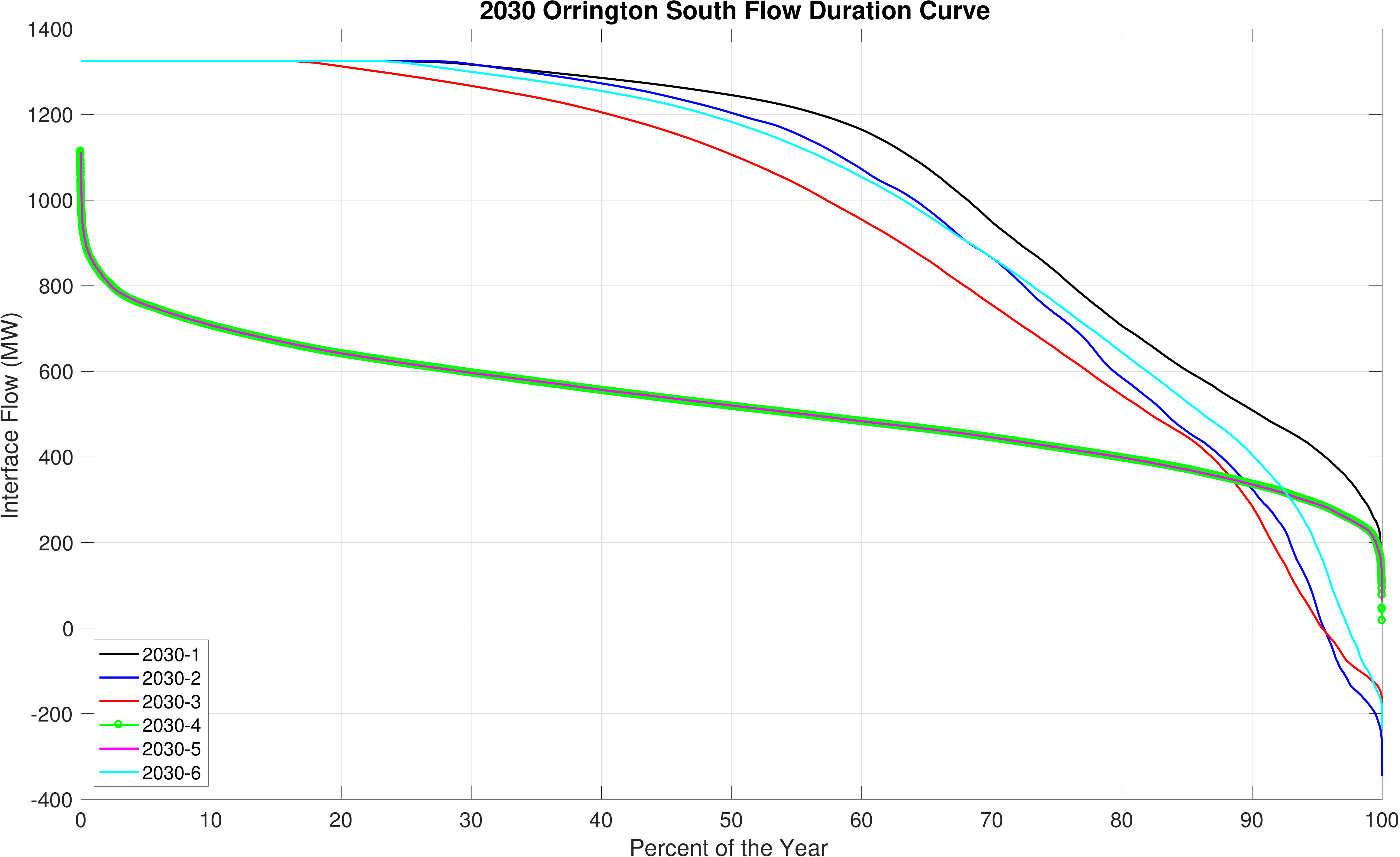}
\caption{Orrington-South flow duration curves for 2030 scenarios}
\label{fig:2030Interface1Duration}
\end{figure}

\begin{figure}[!h]
\centering
\includegraphics[width=6.5in]{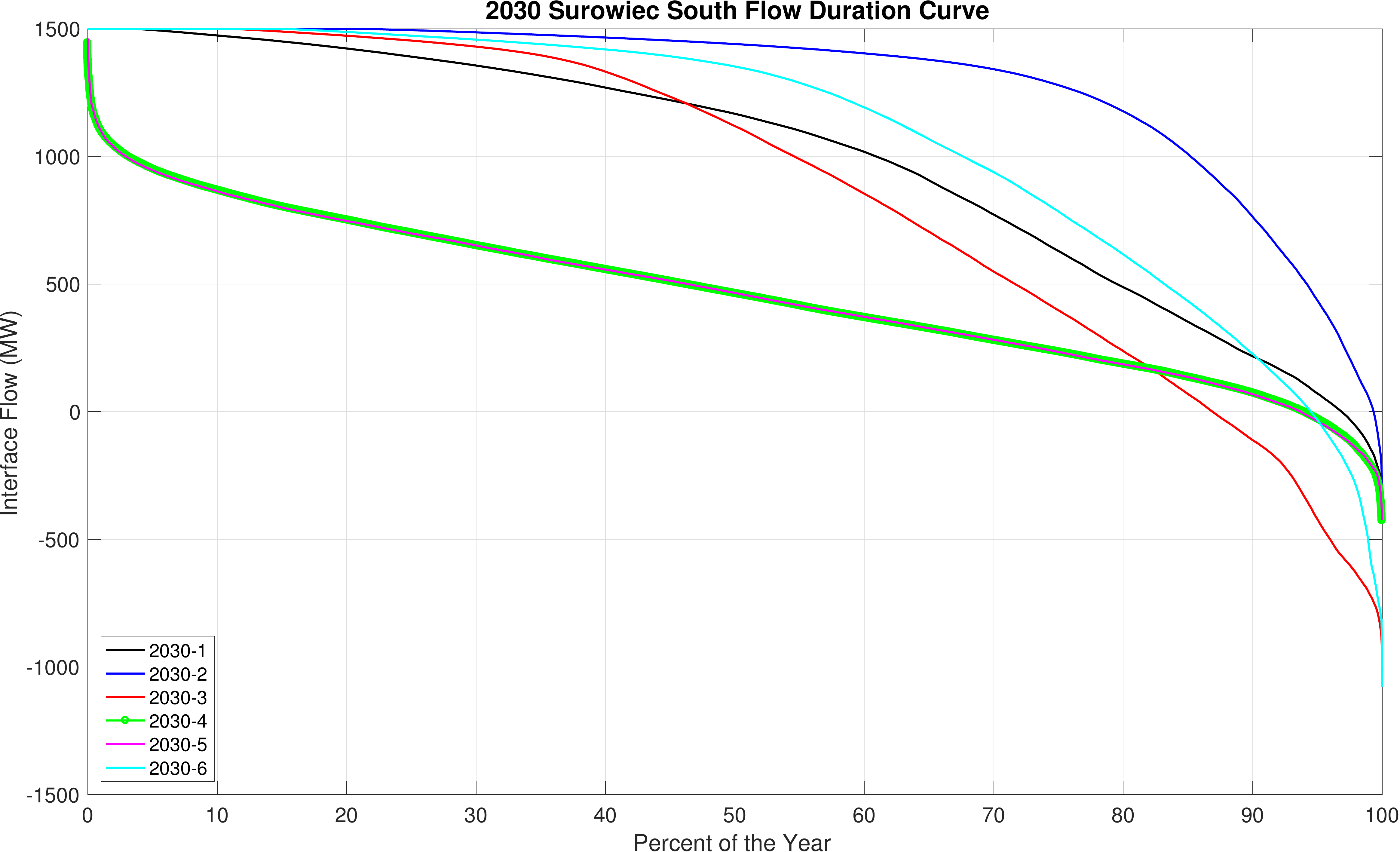}
\caption{Surowiec-South flow duration curves for 2030 scenarios}
\label{fig:2030Interface2Duration}
\end{figure}

\begin{figure}[!h]
\centering
\includegraphics[width=6.5in]{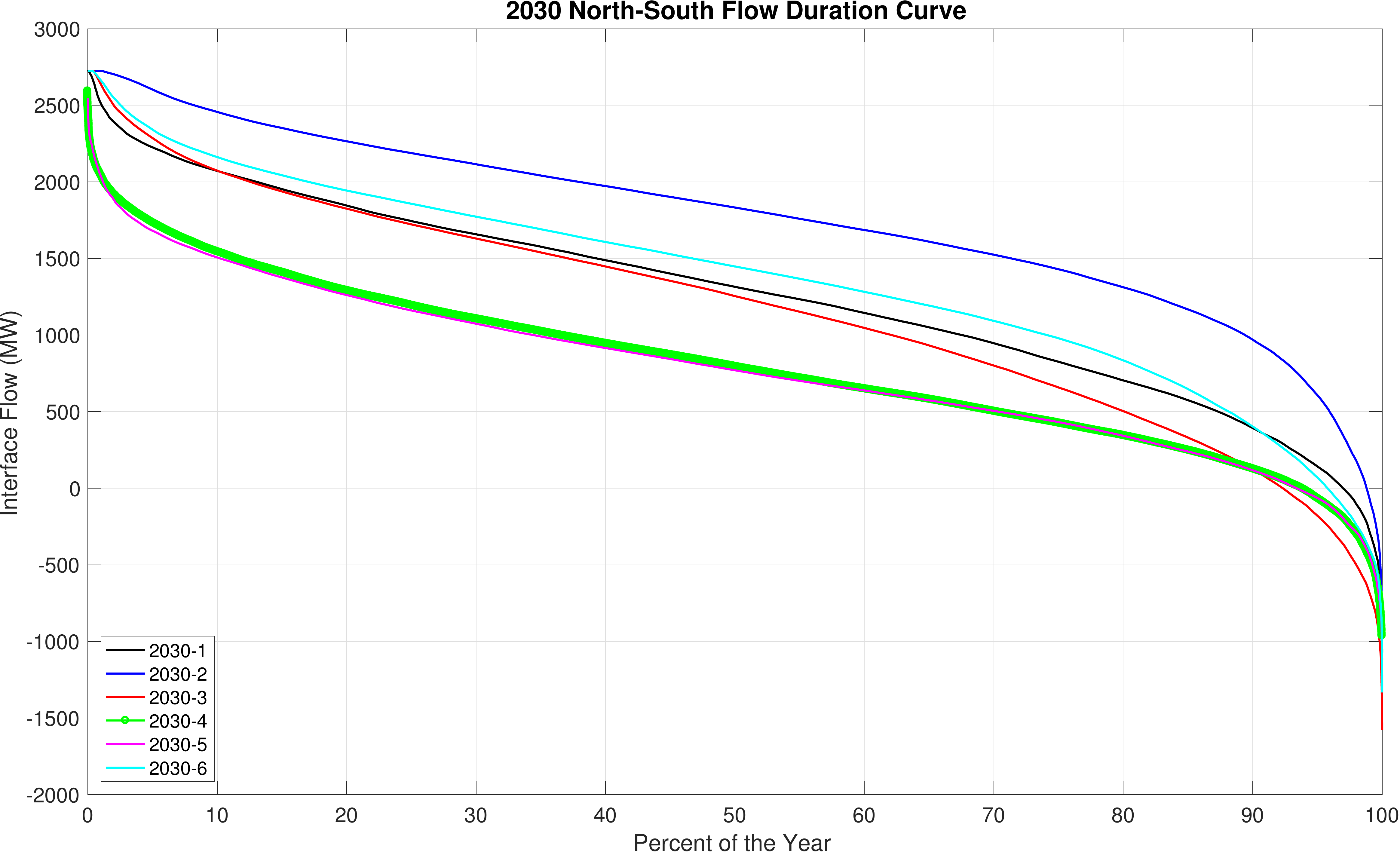}
\caption{North-South flow duration curves for 2030 scenarios}
\label{fig:2030Interface4Duration}
\end{figure}

\begin{figure}[!h]
\centering
\includegraphics[width=6.5in]{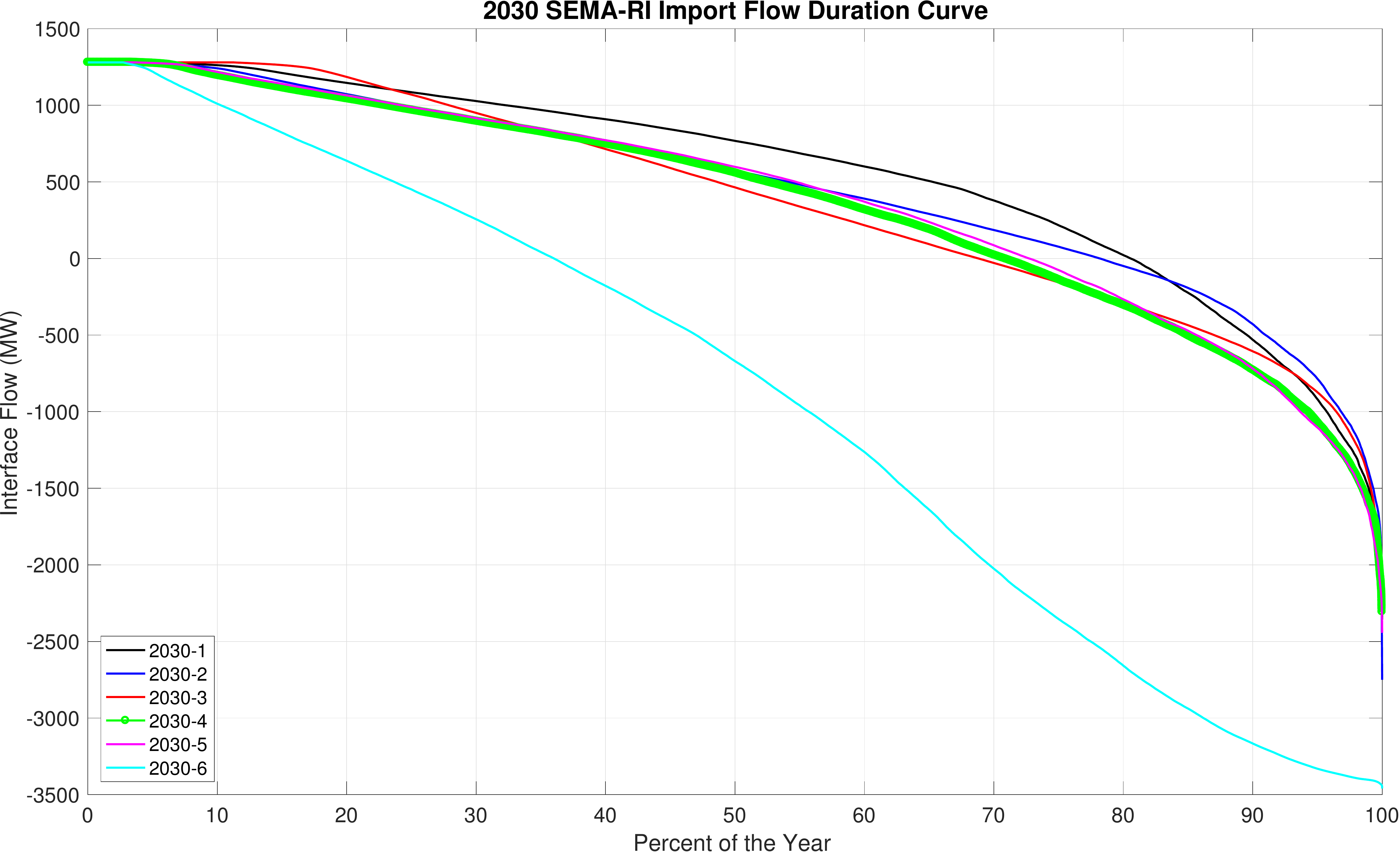}
\caption{SEMA-RI Import flow duration curves for 2025 scenarios}
\label{fig:2030Interface8Duration}
\end{figure}

The results for 2030 are fairly similar to those for 2025. Figure~\ref{fig:2030Interface1Duration} shows that the system experiences significant congestions on the Orrington-South interface for Scenarios 2030-1, 2030-2, 2030-3, and 2030-6 compared to Scenarios 2030-4 and 2030-5 that have no congestions at all. A similar pattern, but to a lesser degree, is observed on the Surowiec-South interface shown in Figure~\ref{fig:2030Interface2Duration}. On the other hand, the North-South interface shown in Figure~\ref{fig:2030Interface4Duration} exhibits congestion only in rare cases. This is due to the fact that the North-South interface has much higher interface limit of 2,725MW, and is able to pass the additional renewable energy generation coming through the Orrington-South and the Surowiec-South interfaces without being congested. Finally, the SEMA-RI import interface in Figure~\ref{fig:2030Interface8Duration} exhibits some congestion for all 2030 scenarios. The SEMA area has high penetrations of PV.  During the midday, dispatchable units are turned off.  As the sun sets, dispatchable units can not be turned on and ramped fast enough to meet the demand and consequently the imported power exceeds the import limit.  The interface congestion statistics for 2030 are summarized in Table~\ref{tab:2030InterfaceCongStats}. 
\begin{table}[!h]
\caption{Interface congestion statistics for 2030 scenarios}
\begin{center}
\begin{tabular}{lrrrrrr}\toprule
 & \textbf{2030-1}& \textbf{2030-2}& \textbf{2030-3}& \textbf{2030-4}& \textbf{2030-5}& \textbf{2030-6}\\\toprule
\textbf{Orrington South  \% Time Congested}& 25.80 & 27.84 & 17.14 &  0.00 &  0.00 & 24.05 \\\midrule
\textbf{Surowiec South  \% Time Congested}&  4.17 & 21.83 & 12.00 &  0.00 &  0.00 & 16.30 \\\midrule
\textbf{North-South  \% Time Congested}&  0.15 &  1.13 &  0.48 &  0.00 &  0.00 &  0.54 \\\midrule
\textbf{SEMA-RI Import  \% Time Congested}&  3.45 &  2.92 &  9.91 &  2.65 &  3.07 &  1.63 \\\bottomrule
\end{tabular}
\end{center}
\label{tab:2030InterfaceCongStats}
\end{table}

Matching the congestion results of this section to the curtailment results in the previous section, the following conclusion can be drawn; the integration of significant renewable energy resources increases the potential of congestion on several key interfaces, such as Orrington-South and the Surowiec-South, and, therefore, require heavy curtailments of these resources. Thus, the ability of the system to accommodate more renewables is limited by its topology, particularly, some of the key interfaces discussed here.

\clearpage
\subsection{Regulation Reserves}
\begin{figure}[!h]
\centering
\includegraphics[width=6.5in]{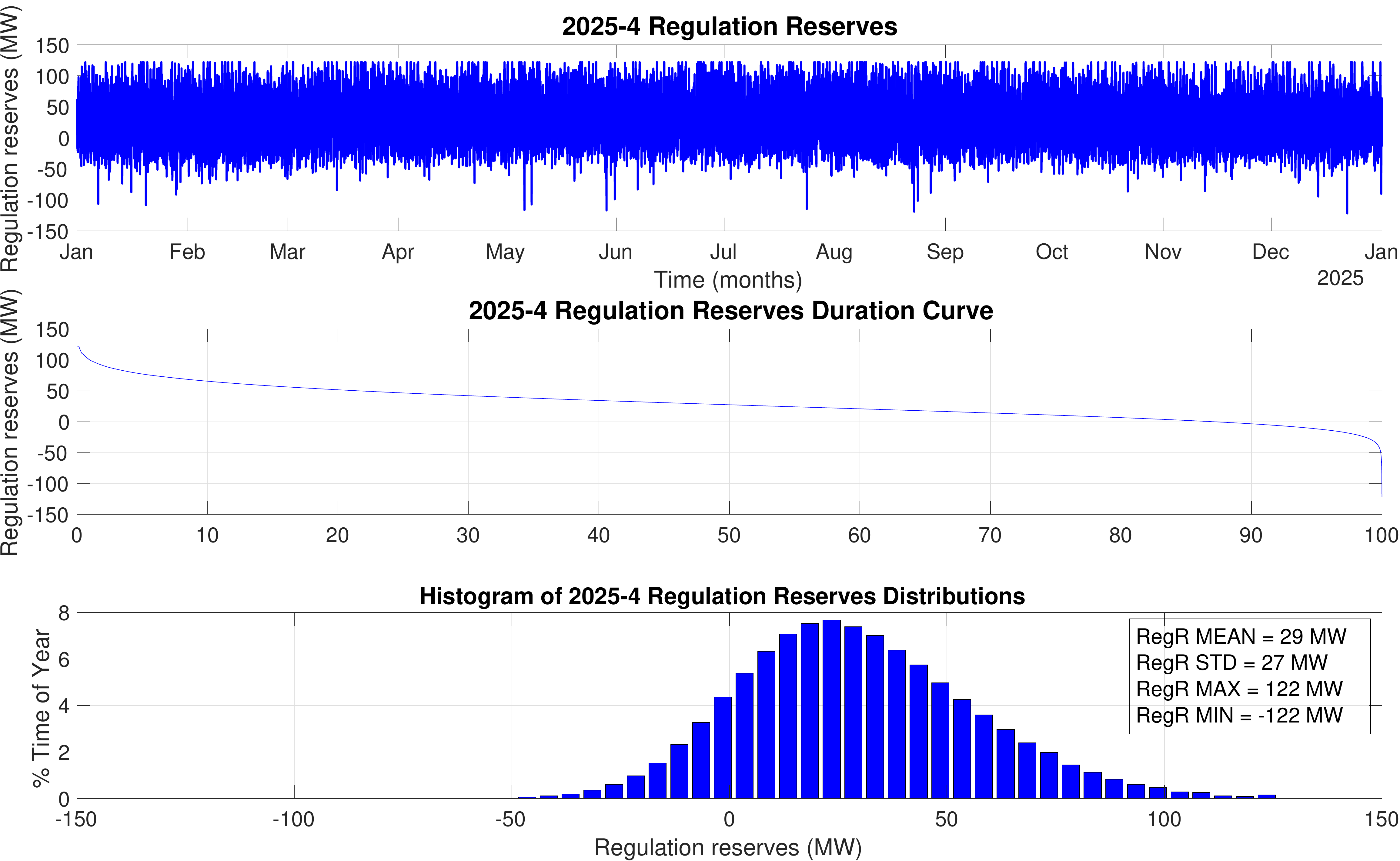}
\caption{Regulation reserve performance for Scenario 2025-4}
\label{fig:2025-4RegRProfile}
\end{figure}

Regulation service is the fastest resource that responds to the residual imbalances after the deployment of load following and ramping reserves and performing the necessary curtailment of semi-dispatchable resources. Therefore, sufficient regulation reserves are instrumental for effective mitigation of imbalances. Figure~\ref{fig:2025-4RegRProfile} shows the regulation reserve performance for Scenario 2025-4. None of three graphs indicate saturation, and, therefore, the system has sufficient regulation reserves. The distribution has a smooth bell-like shape which indicates an efficient use of these resources.  Similarly, the associated duration curve does not indicate any clipping at either end of the regulation range.  

\begin{figure}[!h]
\centering
\includegraphics[width=6.5in]{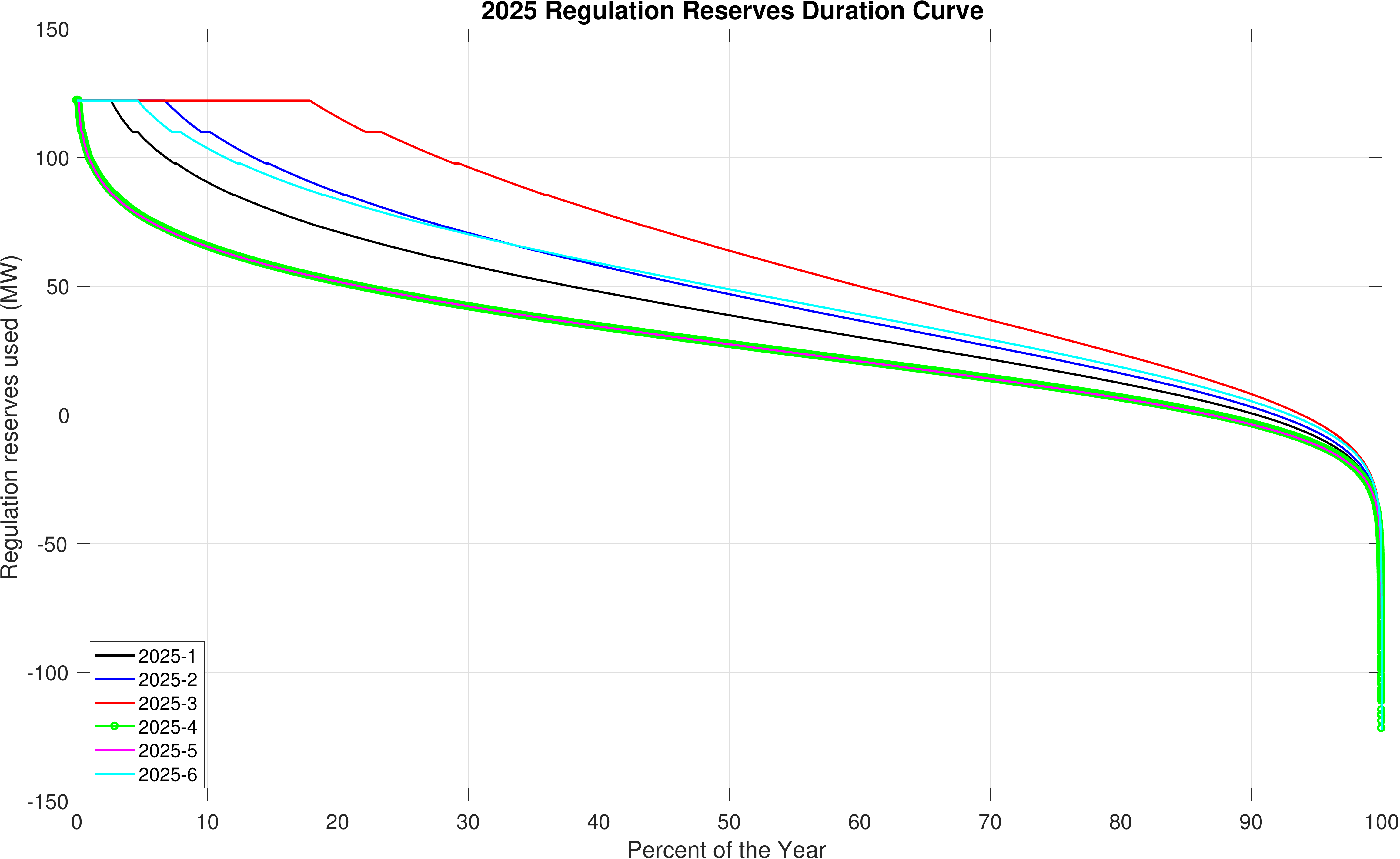}
\caption{Regulation reserve performances for 2025 scenarios}
\label{fig:2025RegResDuration}
\end{figure}

\begin{figure}[!b]
\centering
\includegraphics[width=6.5in]{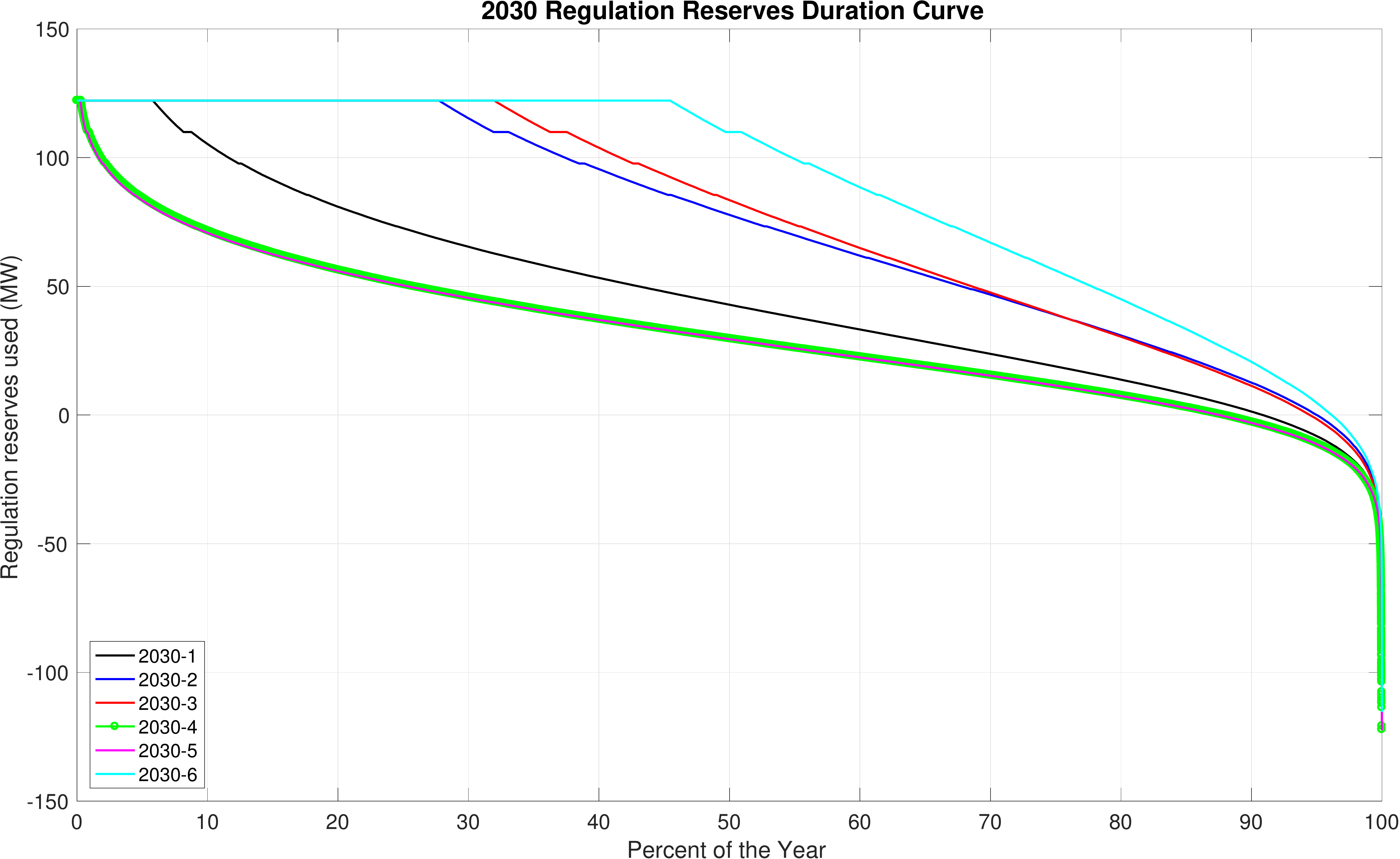}
\caption{Regulation reserve performances for 2030 scenarios}
\label{fig:2030RegResDuration}
\end{figure}

Figure~\ref{fig:2025RegResDuration} shows regulation reserve performances for all 2025 scenarios. Scenarios 2025-1, 2025-2, 2025-3, 2025-6 show heavy saturation of regulation reserves, in contrast to Scenarios 2025-4 and 2025-5. Is should be noted that these scenarios are defined by a significant increase of renewable energy resources in the system. Thus, the increased renewable energy sources in the system will also require additional regulation reserves to effectively respond to the residual imbalances. Regulation reserve performances for 2030 scenarios in Figure~\ref{fig:2030RegResDuration} show a similar effect. However, regulation reserve saturation occur more often compared to 2025 since 2030 has more renewable energy sources in the system. The regulation reserve performance statistics are summarized in Table~\ref{tab:RegResStatistics}.
\begin{table}[!t]
\caption{Regulation reserve statistics}
\begin{center}
\begin{tabular}{lrrrrrr}\toprule
 & \textbf{2025-1}& \textbf{2025-2}& \textbf{2025-3}& \textbf{2025-4}& \textbf{2025-5}& \textbf{2025-6}\\\toprule
\textbf{\% Time Reg. Res  Exhausted}&  2.74 &  6.98 & 18.32 &  0.17 &  0.14 &  4.87 \\\midrule
\textbf{Reg. Res.  Mileage} (GWh)& 389.53 & 461.72 & 582.15 & 283.49 & 283.73 & 462.53 \\\bottomrule
 & \textbf{2030-1}& \textbf{2030-2}& \textbf{2030-3}& \textbf{2030-4}& \textbf{2030-5}& \textbf{2030-6}\\\toprule
\textbf{\% Time Reg. Res  Exhausted}&  6.07 & 28.15 & 33.03 &  0.37 &  0.43 & 46.20 \\\midrule
\textbf{Reg. Res.  Mileage} (GWh)& 433.23 & 659.09 & 684.21 & 307.50 & 305.54 & 778.99 \\\bottomrule
\end{tabular}
\end{center}
\label{tab:RegResStatistics}
\end{table}

\clearpage
\subsection{Balancing Performance}
\begin{figure}[!h]
\centering
\includegraphics[width=6.5in]{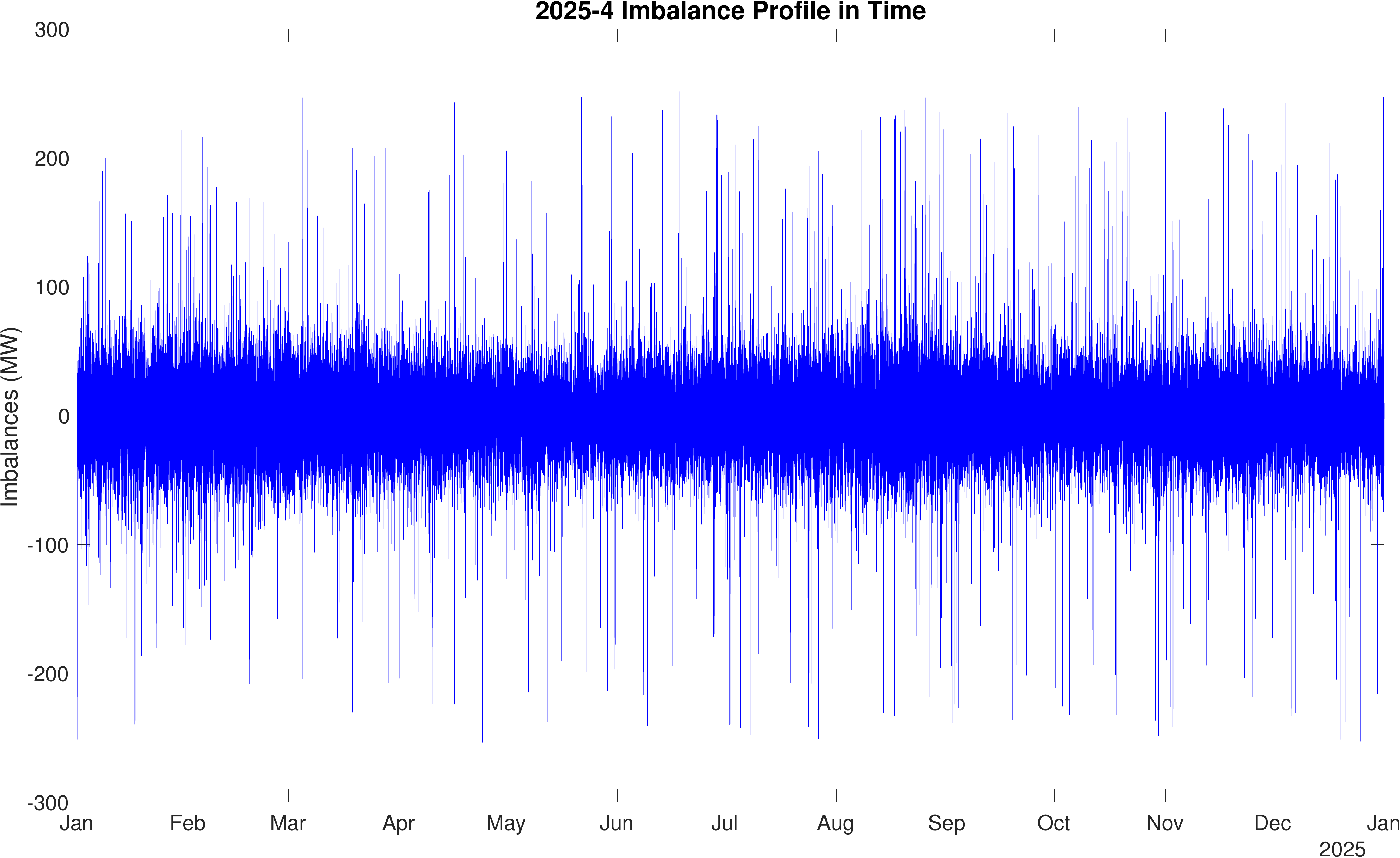}
\caption{Imbalance profile for Scenario 2025-4}
\label{fig:2025-4ImbalanceProfile}
\end{figure}

The balancing performance of the system can be assessed from the residual imbalances after the regulation service was deployed. As shown in the previous section, all scenarios exhibit regulation service saturation to varying degrees. For that reason, all scenarios are expected to have residual imbalances. Figure~\ref{fig:2025-4ImbalanceProfile} shows that the imbalances for Scenario 2025-4 are well-controlled with zero mean and moderate variability on the order of 75MW for the overwhelming majority of the year. Such a low value indicates that the system is well-equipped to mitigate the imbalances effectively.

\begin{figure}[!h]
\centering
\includegraphics[width=6.5in]{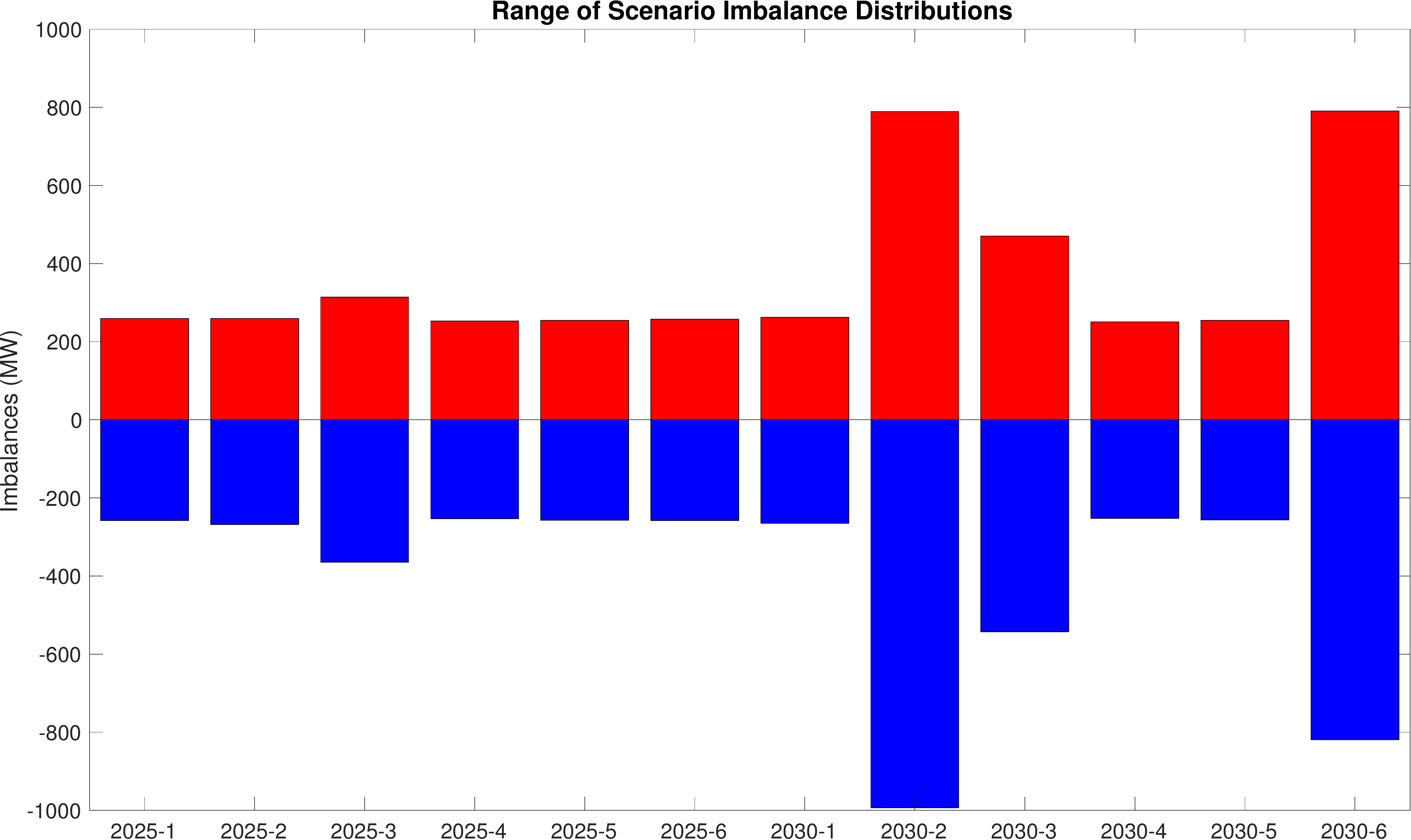}
\caption{Imbalance ranges for 2025 and 2030 scenarios}
\label{fig:ImbalanceDistRange}
\end{figure}

\begin{figure}[!h]
\centering
\includegraphics[width=6.5in]{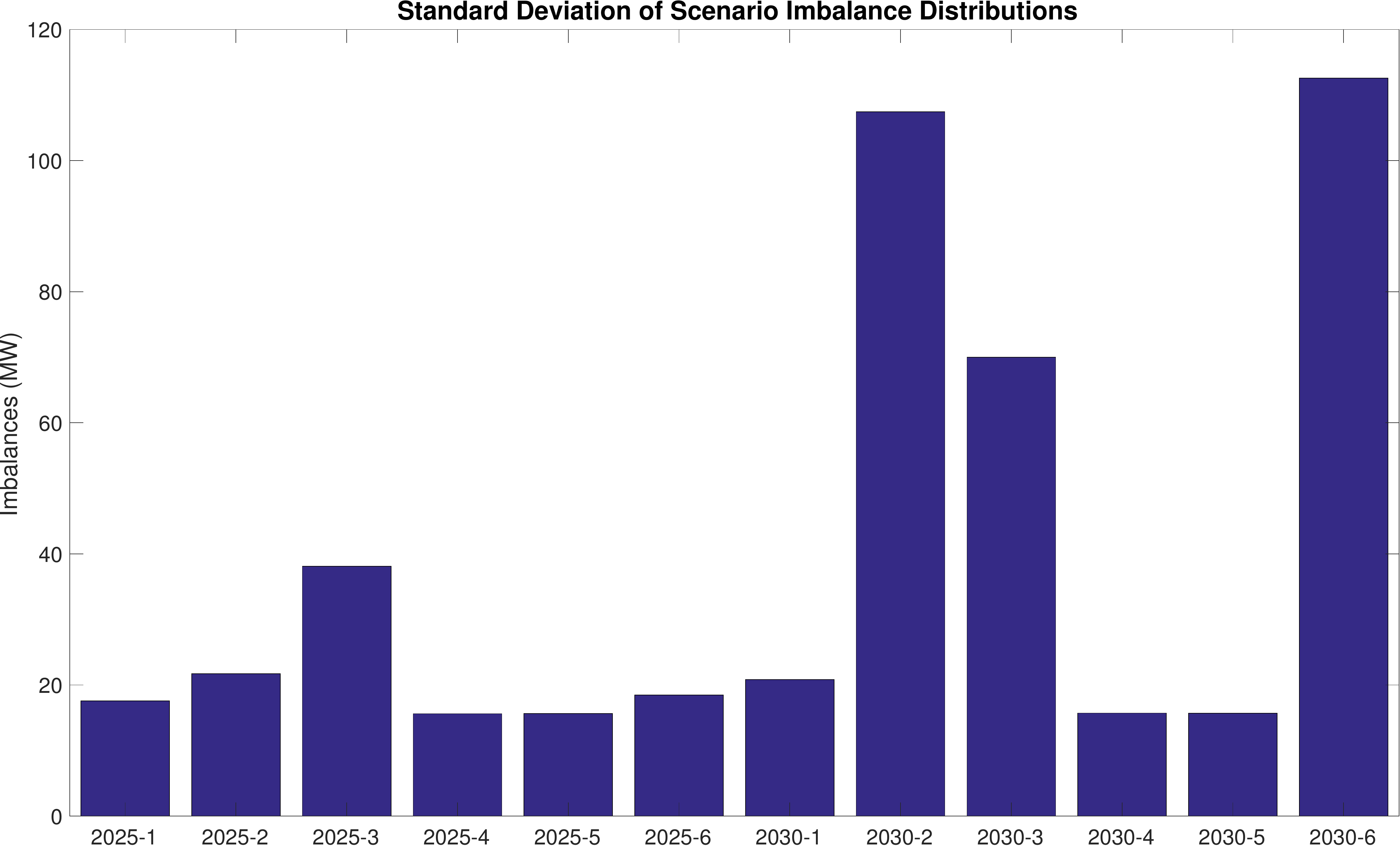}
\caption{Imbalance standard deviations for 2025 and 2030 scenarios}
\label{fig:ImbalanceDistSTD}
\end{figure}

Figure~\ref{fig:ImbalanceDistRange} and Figure~\ref{fig:ImbalanceDistSTD} show the imbalance ranges and standard deviations for all scenarios respectively. Scenarios 2030-2 and 2030-6 and to a lesser extent Scenario 2030-3 have a wider range between the maximum and minimum imbalance values, which can be described as a measure of the intensity of improbable or extreme events. The use of imbalance \emph{range} as a statistic emphasizes single points in time in which brief imbalance excursions can occur.  Upon further investigation, these three scenarios also demonstrate significantly higher values of the standard deviation of imbalances as a measure of the continual balancing ``stress" on the enterprise control of the power system.  From these two complementary results, one can conclude that the 2030-2, 2030-3, and 2030-6 scenarios have a balancing performance that is significantly degraded relative to the other scenarios and a complementary set of measures would be required to achieve the performance of the other scenarios.  That said, it would be premature to conclude that these scenarios would result in degraded overall system reliability in real life because it is not clear at which (absolute) imbalance levels disruptive events might occur.  Imbalance excursions of several hundred megawatts are found within the historical data and \emph{do not} immediately correspond to a disruptive reliability event.   Finally, all scenarios except these three maintain imbalance variability of less than 50MW, despite the saturation of regulations reserves observed in the previous section.

\clearpage
\section{Conclusion \& Final Insights}\label{sec:conclusion}
\subsection{Summary}
This report describes the methodology and the key findings of the 2017 ISO New England System Operational Analysis and Renewable Energy Integration Study (SOARES). It argues that with respect to operations, the integration of variable energy resources should not be considered as ``business-as-usual," and instead a more holistic approach is required. It lays out the requirements for such a rigorous assessment. That discussion contextualizes a review of the methodological adequacy of existing renewable energy integration studies. 

The heart of the project's methodology is a novel, but now extensively published, holistic assessment approach called the Electric Power Enterprise Control System (EPECS) simulator. Most fundamentally, the EPECS methodology is \emph{integrated} and \emph{techno-economic}. It characterizes a power system in terms of the physical power grid and its multiple layers of control including commitment decisions, economic dispatch, and regulation services. The report provides precise definitions of how variable energy resources and operating reserves are modeled. It also includes detailed models of day-ahead resource scheduling, same-day resource scheduling, real-time balancing operations and regulation service. The report also includes the zonal-network (i.e. pipe \& bubble) model of the physical power grid. 

\subsection{Key Findings}
The key findings of this study can be summarized in the following points: 
\begin{enumerate}
\item The commitment of dispatchable resources and their associated quantities of committed load following and ramping reserves has a complex, difficult to predict, non-linear dependence on the amount of VERs and the load profile statistics. High and low levels of VERs do not necessarily correspond to high or low quantities of operating reserves respectively.  For example, during the midday hours, solar generation causes low net load conditions that will test a power system's ability to track downward using downward load following reserves.  Hours later, as solar generation wanes, net load conditions rise to their daily peak testing the power system's ability to track upward with upward load following reserves.  In the meantime, the transition hours between trough and peak conditions exhibits a sharp system ramp.  
\item For the scenarios with significant presence of VERs (2025-3, 2030-3 and 2030-6), the system may require additional amounts of upward load following reserves to effectively mitigate imbalances and maintain its reliable operations. Furthermore, these scenarios entirely exhaust their downward load following reserves; albeit for a fairly short part of the year. Despite such occurrences being rare, depletion of a resource that was assumed to be adequately available in the system for following the net load fluctuations shows the need for the procurement of both upward and downward load following reserves in the day-ahead unit commitment.
\item For the scenarios with significant presence of VERs (2025-3, 2030-3 and 2030-6), the system entirely exhausts its upward and downward ramping capabilities. Such moments coincide with power system imbalances.  These results indicate that the assumption that the generator ramping constraints in the day-ahead scheduling provide sufficient ramping capabilities to the system is inadequate. Therefore, both load following and ramping reserves should be procured in the day-ahead unit commitment.
\item Along with the load following and ramping reserves provided by dispatchable resources, the curtailment of semi-dispatchable resources becomes an integral part of balancing performance; in part to complement operating reserves and in part to mitigate the topological limitations of the system. Every scenario uses curtailment in some way at least 98.6\% of the time.  The maximum level of curtailment for all scenarios ranges from 1,605MW (in Scenario 2025-4) to 14,534MW (in Scenario 2030-2).  In all, these curtailments correspond to a loss of between 2.72\% (in Scenario 2030-4) and 41.19\% (in Scenario 2030-2) of the total semi-dispatchable energy available.    It is also important to emphasize that some of the associated topological limitations only start affecting the system performance after the integration of VERs in remote areas that replace the traditional generation units located close to the main consumption centers. Thus, VERs might have a self-limiting feature which also defines the ability of the system to accommodate them.
\item The integration of significant amounts of VERs increases the potential of congestion on several key interfaces (Orrington-South and Surowiec-South), and, therefore, require heavy curtailments of these resources. Thus, the ability of the system to accommodate more renewables is limited by its topology. A longer-term solution to accommodating large amounts of VERs while avoiding such congestions would be the construction of new transmission lines from remote areas of VER installation to the main consumption centers. 
\item For the scenarios with significant presence of VERs (Scenarios 2025-3, 2030-2, 2030-3, and 2030-6), the system experiences heavy saturations of regulation reserves and, therefore, requires additional regulation reserves to effectively respond to the residual imbalances.  Scenarios 2025-1, 2025-2, 2025-6, 2030-1 also experience moderate saturations of regulation reserves indicating the need for their increase in 8 out of the 12 scenarios studied.  
\item The scenarios with significant presence of VERs (Scenarios 2030-2, 2030-3 and 2030-6) have significantly degraded balancing performance relative to the other scenarios studied and a complementary set of new measures would be required to achieve similar performance.  It would be premature to conclude that these scenarios would result in degraded overall system reliability in real life because it is not clear at which absolute imbalance levels disruptive events might occur.  The simulated imbalance excursions in all scenarios are comparable to the historical normal operation data.  
\end{enumerate}

\subsection{Final Insights}
The above key findings indicate that Scenarios 2030-2, 2030-3 and 2030-6 and to a lesser extent 2025-3 have qualitatively and quantitatively different behavior in large part due to the integration of variable energy resources.  Consequently, it is important to ask what makes the integration of variable energy resources so different in New England and what actions could potentially serve to bring the balancing performance back to a level similar to that of other scenarios.  

The presence of congestion caused by the integration of VERs in remote areas geographically isolates operating reserves in other parts of New England.  Simply speaking, because the variable energy resource generation occurs ``behind" a congested interface, operating reserves are unable to respond accordingly.  Higher interface limits in the form of additional transmission lines would help accommodate these VERs because: 
\begin{enumerate}
\item it would provide access to load centers that can absorb their power injections; 
\item it would provide access to dispatchable resources with additional operating reserves
\item It would reduce the need for and reliance upon curtailment as a mitigating measure.  
\end{enumerate}
\noindent Removing system transmission constraints would improve the overall system balancing performance.  The reduced curtailment levels may also result in a systemic shift in generated energy from dispatchable resources to relatively cheaper semi-dispatchable resources.  

These scenarios also prominently show the emergence of curtailment as a balancing performance control lever.  Indeed, in the presence of congestion and the absence of other (generation or load) energy resources in remote areas, curtailment often becomes the only control lever.  The prominence of curtailment in these scenarios should also inspire deeper reflection.  In many cases, curtailment levels are commensurate with the available load following reserves from dispatchable energy resources, and in other cases greatly surpass these values.  Furthermore, that the mathematical form of curtailment is nearly equivalent to that of load following reserves should raise the question as to why they are treated differently.  The only difference between dispathable and semi-dispatchable resources is that the former has a fixed upper capacity limit while the latter has a variable upper capacity limited by forecast.  Similarly, curtailment directly contributes to a system's ramping capability.   The \emph{time rate of change} of a curtailment signal is mathematically equivalent to ramping reserves.  The values of operating reserves reported in this study should be interpreted in this context.  In keeping with the ISO New England mission, a reconciliation of operating reserve and curtailment definitions could serve to clarify our understanding of overall system reliability and provide for an equitable administration of how these values are maintained on the system.  

The increased penetration of variable energy resources is likely to require additional regulation reserves.  While historically, this type of reserve has come exclusively from dispatchable resources, there is little to suggest that they could not come from semi-dispatchable resources in the future.  Indeed, the curtailment signal used in this study did effectively move both upward and downward at the relatively time scale of the 10-minute real-time balancing optimization.  If this signal were to become more temporally granular (perhaps with greater telemetry) to a 1-minute level, then it could serve the role of regulation.  

In this study, many of the most challenging operational periods occurred during low-load spring and fall months.  During these times, nuclear generation units were assumed to ``must-run" at full capacity.  Balancing performance is likely to improve if this assumption were to be relaxed.  In other parts of the world, nuclear generation facilities have been shown to run at less than full capacity and to exhibit modest load following capability.  Alternatively, and perhaps more imminently, it is possible to coordinate the scheduled maintenance of these units during these low load periods.

Finally, within these scenarios, energy storage and demand response played relatively minor roles either due to the associated penetration rates and due to the choice of their associated threshold prices.   Investigating alternative scenarios in which these types of resources have a more prominent role can serve to rebalance the balancing burden away from curtailment as a primary lever to a broader diversity of energy resources across the system.  Again, a deep reflection of how such resources offer load following, ramping, and regulation reserve capability is needed.  

In all, if these considerations are taken into account it is likely that the above scenarios could have a balancing performance that is equal or better than the other eight scenarios studied here.
\section*{Acknowledgement}
The authors are grateful to ISO New England for the funding required to complete this work.  The authors would also like to thank the ISO New England technical staff for its technical review of the work throughout the duration of the research project.  

\clearpage
\bibliographystyle{IEEEtran}
\bibliography{LIINESLibrary,LIINESPublications,ISONE}

\begin{thebibliography}{10}
\providecommand{\url}[1]{#1}
\csname url@samestyle\endcsname
\providecommand{\newblock}{\relax}
\providecommand{\bibinfo}[2]{#2}
\providecommand{\BIBentrySTDinterwordspacing}{\spaceskip=0pt\relax}
\providecommand{\BIBentryALTinterwordstretchfactor}{4}
\providecommand{\BIBentryALTinterwordspacing}{\spaceskip=\fontdimen2\font plus
\BIBentryALTinterwordstretchfactor\fontdimen3\font minus
  \fontdimen4\font\relax}
\providecommand{\BIBforeignlanguage}[2]{{%
\expandafter\ifx\csname l@#1\endcsname\relax
\typeout{** WARNING: IEEEtran.bst: No hyphenation pattern has been}%
\typeout{** loaded for the language `#1'. Using the pattern for}%
\typeout{** the default language instead.}%
\else
\language=\csname l@#1\endcsname
\fi
#2}}
\providecommand{\BIBdecl}{\relax}
\BIBdecl

\bibitem{Welie:2018:00}
\BIBentryALTinterwordspacing
G.~van Welie. (2018, May) {ISO New England Identifies Fuel-Security Risk as the
  Power System Undergoes Rapid Transformation}. [Online]. Available:
  \url{https://www.iso-ne.com/static-assets/documents/2018/05/ma_joint_tue_committee_briefing_gvw_may_23_2018_final.pdf}
\BIBentrySTDinterwordspacing

\bibitem{ISO-NE:2017:00}
\BIBentryALTinterwordspacing
{ISO New England}, ``{2015 ISO New England Electric Generator Air Emissions
  Report},'' ISO New England Inc. System Planning, Tech. Rep., January 2017.
  [Online]. Available:
  \url{https://www.iso-ne.com/static-assets/documents/2017/01/2015\_emissions\_report.pdf}
\BIBentrySTDinterwordspacing

\bibitem{UVIG:2017:00}
UVIG, ``2017 spring technical workshop,'' Utility Variable-Generation
  Integration Group, Tech. Rep., {March} 2017.

\bibitem{IEA:2017:00}
\BIBentryALTinterwordspacing
IEA. (2017, January 13) Natural gas prices in 2016 were the lowest in nearly 20
  years. [Online]. Available:
  \url{https://www.eia.gov/todayinenergy/detail.php?id=29552#}
\BIBentrySTDinterwordspacing

\bibitem{ISO-NE:2016:00}
\BIBentryALTinterwordspacing
{ISO New England}, ``{ISO New Englands Internal Market Monitor -- 2015 Annual
  Markets Report},'' ISO New England, Holyoke, MA, Tech. Rep., May 2016.
  [Online]. Available:
  \url{https://www.iso-ne.com/static-assets/documents/2016/05/2015\_imm\_amr\_final\_5\_25\_2016.pdf}
\BIBentrySTDinterwordspacing

\bibitem{ISO-NE:2017:02}
\BIBentryALTinterwordspacing
------. (2017) {Resource Mix-ISO New England}. [Online]. Available:
  \url{https://www.iso-ne.com/about/key-stats/resource-mix}
\BIBentrySTDinterwordspacing

\bibitem{Rourke:2015:00}
S.~J. Rourke. (2015, June) {New England's Energy Resource Mix Is Changing
  Rapidly}.
  \url{https://www.iso-ne.com/static-assets/documents/2015/06/eia_june_15_presentation_final.pdf}.

\bibitem{Farid:2014:SPG-J26}
\BIBentryALTinterwordspacing
A.~M. Farid, B.~Jiang, A.~Muzhikyan, and K.~Youcef-Toumi, ``{The Need for
  Holistic Enterprise Control Assessment Methods for the Future Electricity
  Grid},'' \emph{Renewable \& Sustainable Energy Reviews}, vol.~56, no.~1, pp.
  669--685, 2015. [Online]. Available:
  \url{http://dx.doi.org/10.1016/j.rser.2015.11.007}
\BIBentrySTDinterwordspacing

\bibitem{Meier:2006:00}
\BIBentryALTinterwordspacing
A.~von Meier, \emph{{Electric power systems : a conceptual
  introduction}}.\hskip 1em plus 0.5em minus 0.4em\relax Hoboken, N.J.: IEEE
  Press : Wiley-Interscience, 2006. [Online]. Available:
  \url{http://ieeexplore.ieee.org/xpl/bkabstractplus.jsp?bkn=5238205}
\BIBentrySTDinterwordspacing

\bibitem{Schavemaker:2008:00}
\BIBentryALTinterwordspacing
P.~Schavemaker, L.~{Van der Sluis}, and {Books24x7 Inc.}, \emph{{Electrical
  power system essentials}}.\hskip 1em plus 0.5em minus 0.4em\relax Chichester,
  England ; Hoboken, NJ: Wiley, 2008. [Online]. Available:
  \url{http://www.loc.gov/catdir/enhancements/fy0810/2008007359-d.html
  http://www.loc.gov/catdir/enhancements/fy0810/2008007359-t.html}
\BIBentrySTDinterwordspacing

\bibitem{Gellings:1985:00}
C.~W. Gellings, ``The concept of demand-side management for electric
  utilities,'' \emph{Proceedings of the IEEE}, vol.~73, no.~10, pp. 1468--1470,
  1985.

\bibitem{Wood:2014:00}
A.~J. Wood and B.~F. Wollenberg, \emph{{Power generation, operation, and
  control}}, 3rd~ed.\hskip 1em plus 0.5em minus 0.4em\relax Hoboken, NJ, USA:
  John Wiley \& Sons, 2014.

\bibitem{Gomez-Exposito:2008:00}
A.~Gomez-Exposito, A.~J. Conejo, and C.~Canizares, \emph{Electric Energy
  Systems: Analysis and Operation}.\hskip 1em plus 0.5em minus 0.4em\relax Boca
  Raton, FL: CRC Press, 2008.

\bibitem{Farid:2013:SPG-C13}
\BIBentryALTinterwordspacing
A.~M. Farid and A.~Muzhikyan, ``{The Need for Holistic Assessment Methods for
  the Future Electricity Grid (Best Applied Research Paper Award)},'' in
  \emph{GCC CIGRE Power 2013}, Abu Dhabi, UAE, 2013, pp. 1--12. [Online].
  Available:
  \url{http://amfarid.scripts.mit.edu/resources/Conferences/SPG-C13.pdf}
\BIBentrySTDinterwordspacing

\bibitem{Diaz-Gonzalez:2014:00}
F.~Diaz-Gonzalez, M.~Hau, A.~Sumper, O.~Gomis-Bellmunt, and
  F.~D{\'\i}az-gonz{\'a}lez, ``Participation of wind power plants in system
  frequency control : Review of grid code requirements and control methods,''
  \emph{Renewable and Sustainable Energy Reviews}, vol.~34, pp. 551--564, 2014.

\bibitem{Mohseni:2012:00}
M.~Mohseni and S.~M. Islam, ``Review of international grid codes for wind power
  integration: Diversity, technology and a case for global standard,''
  \emph{Renewable and Sustainable Energy Reviews}, vol.~16, no.~6, pp.
  3876--3890, Aug 2012.

\bibitem{Anonymous:2012:05}
Anonymous, N.~Grid, E.~Transmission, E.~Act, G.~Britain, and P.~Act, ``The grid
  code,'' National Grid Electricity Transmission plc, Warwick, UK, Tech.
  Rep.~5, 2012.

\bibitem{Kassakian:2011:SPG-B01}
\BIBentryALTinterwordspacing
J.~G. Kassakian, R.~Schmalensee, G.~Desgroseilliers, T.~D. Heidel, K.~Afridi,
  A.~M. Farid, J.~M. Grochow, W.~W. Hogan, H.~D. Jacoby, J.~L. Kirtley, H.~G.
  Michaels, I.~Perez-Arriaga, D.~J. Perreault, N.~L. Rose, G.~L. Wilson,
  N.~Abudaldah, M.~Chen, P.~E. Donohoo, S.~J. Gunter, P.~J. Kwok, V.~A.
  Sakhrani, J.~Wang, A.~Whitaker, X.~L. Yap, R.~Y. Zhang, and M.~I.
  of~Technology, \emph{{The Future of the Electric Grid: An Interdisciplinary
  MIT Study}}.\hskip 1em plus 0.5em minus 0.4em\relax Cambridge, MA: MIT Press,
  2011. [Online]. Available:
  \url{http://web.mit.edu/mitei/research/studies/documents/electric-grid-2011/Electric\_Grid\_Full\_Report.pdf}
\BIBentrySTDinterwordspacing

\bibitem{Giebel:2011:00}
G.~Giebel, R.~Brownsword, G.~Kariniotakis, M.~Denhard, and C.~Draxl, \emph{{The
  State-Of-The-Art in Short-Term Prediction of Wind Power: A Literature
  Overview}}, 2nd~ed.\hskip 1em plus 0.5em minus 0.4em\relax Roskilde, Denmark:
  ANEMOS.plus, 2011.

\bibitem{Gellings:2011:00}
C.~Gellings, F.~Functioning, and S.~Grid, ``Estimating the costs and benefits
  of the smart grid,'' EPRI, Palo Alto, CA, USA, Tech. Rep., 2011.

\bibitem{Easton:2012:01}
B.~Easton, K.~House, and J.~Byars, ``Smart grid : a race worth winning ? a
  report on the economic benefits of smart grid,'' Ernst \& Young, London, UK,
  Tech. Rep. April, 2012.

\bibitem{Martin:2012:01}
P.~G. Martin, ``The need for enterprise control,'' \emph{InTech}, vol. Nov/Dec,
  pp. 1--5, 2012.

\bibitem{ANSI-ISA:2005:00}
ANSI-ISA, ``{Enterprise Control System Integration Part 3: Activity Models of
  Manufacturing Operations Management},'' The International Society of
  Automation, Tech. Rep., 2005.

\bibitem{Sanchez:2001:00}
L.~M. Sanchez and R.~Nagi, ``A review of agile manufacturing systems,''
  \emph{International Journal of Production Research}, vol.~39, no.~16, pp.
  3561--3600, Jan 2001.

\bibitem{Gunasekaran:1998:00}
A.~Gunasekaran, ``Agile manufacturing: enablers and an implementation
  framework,'' \emph{International Journal of Production Research}, vol.~36,
  no.~5, pp. 1223--1247, 1998.

\bibitem{Beach:2000:02}
\BIBentryALTinterwordspacing
R.~Beach, A.~P. Muhlemann, D.~H.~R. Price, A.~Paterson, and J.~A. Sharp,
  ``{Review of manufacturing flexibility},'' \emph{European Journal of
  Operational Research}, vol. 122, no.~1, pp. 41--57, 2000. [Online].
  Available: \url{http://dx.doi.org/10.1016/S0377-2217(99)00062-4}
\BIBentrySTDinterwordspacing

\bibitem{De-Toni:1998:00}
A.~{De Toni} and S.~Tonchia, ``{Manufacturing Flexibility: a literature
  review},'' \emph{International Journal of Production Research}, vol.~36,
  no.~6, pp. 1587--1617, 1998.

\bibitem{Pels:1997:00}
H.~J. Pels, J.~C. Wortmann, and A.~J.~R. Zwegers, ``Flexibility in
  manufacturing: an architectural point of view,'' \emph{Computers in
  Industry}, vol.~33, no. 2-3, pp. 271--283, 1997.

\bibitem{Lapalus:1995:00}
E.~Lapalus, S.~G. Fang, C.~Rang, and R.~J. van Gerwen, ``{Manufacturing
  integration},'' \emph{Computers in Industry}, vol.~27, no.~2, pp. 155--165,
  1995.

\bibitem{Williams:2001:00}
T.~J. Williams, G.~A. Rathwell, and H.~Li, \emph{{A Handbook on Master Planning
  and Implementation for Enterprise Integration Programs}}.\hskip 1em plus
  0.5em minus 0.4em\relax Purdue University Institute for Interdisciplinary
  Engineering Studies, 2001.

\bibitem{Kosanke:1999:01}
K.~Kosanke, F.~Vernadat, and M.~Zelm, ``Cimosa: Enterprise engineering and
  integration,'' \emph{Computers in Industry}, vol.~40, no. 2-3, pp. 83--87,
  1999.

\bibitem{ANSI-ISA:2000:00}
ANSI-ISA, \emph{{Enterprise-Control System Integration Part 1: Models and
  Terminology}}.\hskip 1em plus 0.5em minus 0.4em\relax Instrument Society of
  America, 2000, no. July.

\bibitem{Wu:2005:00}
F.~F. Wu, K.~Moslehi, and A.~Bose, ``{Power System Control Centers: Past,
  Present, and Future},'' \emph{Proceedings of the IEEE}, vol.~93, no.~11, pp.
  1890--1908, 2005.

\bibitem{Yan:2013:00}
Y.~Yan, Y.~Qian, H.~Sharif, and D.~Tipper, ``A survey on smart grid
  communication infrastructures: Motivations, requirements and challenges,''
  \emph{IEEE Communications Surveys {\&} Tutorials}, vol.~15, no.~1, pp. 5--20,
  2013.

\bibitem{Anonymous:2010:01}
Anonymous, ``{NIST Framework and Roadmap for Smart Grid Interoperability
  Standards Release 1.0: NIST Special Publication 1108},'' Office of the
  National Coordinator for Smart Grid Interoperability, National Institute of
  Standards and Technology, United States Department of Commerce, Washington
  D.C., Tech. Rep. NIST Special Special Publication 1108, 2010.

\bibitem{Amin:2001:01}
M.~Amin, ``{Toward self-healing energy infrastructure systems},''
  \emph{Computer Applications in Power, IEEE}, vol.~14, no.~1, pp. 20--28,
  2001.

\bibitem{Amin:2006:00}
------, ``Toward a self-healing energy infrastructure,'' in \emph{Power
  Engineering Society General Meeting}, ser. 2006 IEEE Power Engineering
  Society General Meeting.\hskip 1em plus 0.5em minus 0.4em\relax Montreal,
  Canada: IEEE, 2006, pp. 7 pp. BN ---- 1 4244 0493 2.

\bibitem{Amin:2008:00}
------, ``Challenges in reliability, security, efficiency, and resilience of
  energy infrastructure: Toward smart self-healing electric power grid,'' in
  \emph{BT - 2008 IEEE Power and Energy Society General Meeting, 20-24 July
  2008}, ser. 2008 IEEE Power Energy Society General Meeting.\hskip 1em plus
  0.5em minus 0.4em\relax Pittsburgh, PA, USA: IEEE, Jul 2008, pp. 1--5.

\bibitem{Amin:2011:00}
S.~M. Amin, ``Smart grid overview issues and opportunities. advances and
  challenges in sensing modeling simulation optimization and control,''
  \emph{European Journal of Control}, vol.~17, no. 5-6, pp. 547--567, 2011.

\bibitem{McArthur:2012:00}
S.~D.~J. McArthur, P.~C. Taylor, G.~W. Ault, J.~E. King, D.~Athanasiadis, V.~D.
  Alimisis, and M.~Czaplewski, ``The autonomic power system - network operation
  and control beyond smart grids,'' \emph{2012 3rd IEEE PES Innovative Smart
  Grid Technologies Europe (ISGT Europe)}, pp. 1--7, Oct 2012.

\bibitem{Henderson:2016:00}
M.~I. Henderson, ``2016 economic studies:s2 sensitivity study draft results,''
  ISO New England Planning Advisory Committee, Tech. Rep., November 2016.

\bibitem{Coste:2016:00}
W.~Coste, ``{2016 Economic Studies Phase I Assumptions},'' ISO New England
  Planning Advisory Committee, Tech. Rep., June 2016.

\bibitem{Muzhikyan:2015:SPG-J16}
\BIBentryALTinterwordspacing
A.~Muzhikyan, A.~M. Farid, and K.~Youcef-Toumi, ``{An Enterprise Control
  Assessment Method for Variable Energy Resource Induced Power System
  Imbalances Part 2: Results},'' \emph{IEEE Transactions on Industrial
  Electronics}, vol.~62, no.~4, pp. 2459 -- 2467, 2015. [Online]. Available:
  \url{http://dx.doi.org/10.1109/TIE.2015.2395380}
\BIBentrySTDinterwordspacing

\bibitem{Muzhikyan:2015:SPG-J15}
\BIBentryALTinterwordspacing
------, ``{An Enterprise Control Assessment Method for Variable Energy Resource
  Induced Power System Imbalances Part 1: Methodology},'' \emph{IEEE
  Transactions on Industrial Electronics}, vol.~62, no.~4, pp. 2448--2458,
  2015. [Online]. Available: \url{http://dx.doi.org/10.1109/TIE.2015.2395391}
\BIBentrySTDinterwordspacing

\bibitem{Muzhikyan:2015:SPG-J22}
\BIBentryALTinterwordspacing
------, ``{Relative Merits of Load Following Reserves and Energy Storage Market
  Integration Towards Power System Imbalances},'' \emph{International Journal
  of Electrical Power and Energy Systems}, vol.~74, no.~1, pp. 222--229, 2016.
  [Online]. Available: \url{http://dx.doi.org/10.1016/j.ijepes.2015.07.013}
\BIBentrySTDinterwordspacing

\bibitem{Muzhikyan:2014:SPG-C32}
\BIBentryALTinterwordspacing
------, ``{An Enhanced Method for the Determination of Load Following
  Reserves},'' in \emph{American Control Conference, 2014}, Portland, Oregon,
  2014, pp. 1--8. [Online]. Available:
  \url{http://dx.doi.org/10.1109/ACC.2014.6859254}
\BIBentrySTDinterwordspacing

\bibitem{Muzhikyan:2015:SPG-C47}
\BIBentryALTinterwordspacing
------, ``{An Enhanced Method for Determination of the Regulation Reserves},''
  in \emph{IEEE American Control Conference}, Los Angeles, CA, USA, 2015, pp.
  1--8. [Online]. Available: \url{http://dx.doi.org/10.1109/ACC.2015.7170866}
\BIBentrySTDinterwordspacing

\bibitem{Muzhikyan:2015:SPG-C46}
\BIBentryALTinterwordspacing
------, ``{An Enhanced Method for Determination of the Ramping Reserves},'' in
  \emph{IEEE American Control Conference}, Los Angeles, CA, USA, 2015, pp.
  1--8. [Online]. Available: \url{http://dx.doi.org/10.1109/ACC.2015.7170863}
\BIBentrySTDinterwordspacing

\bibitem{Muzhikyan:2016:SPG-J28}
\BIBentryALTinterwordspacing
------, ``{An A Priori Analytical Method for Determination of Operating
  Reserves Requirements},'' \emph{International Journal of Energy and Power
  Systems}, vol.~86, no.~3, pp. 1--11, 2016. [Online]. Available:
  \url{http://dx.doi.org/10.1016/j.ijepes.2016.09.005}
\BIBentrySTDinterwordspacing

\bibitem{Muzhikyan:2014:SPG-C43}
\BIBentryALTinterwordspacing
------, ``{A Power Grid Enterprise Control Method for Energy Storage System
  Integration},'' in \emph{IEEE Innovative Smart Grid Technologies Conference
  Europe}, Istanbul, Turkey, 2014, pp. 1--6. [Online]. Available:
  \url{http://dx.doi.org/10.1109/ISGTEurope.2014.7028898}
\BIBentrySTDinterwordspacing

\bibitem{ISO-New-England:2017:00}
{ISO New England}, ``2016 economic study: {NEPOOL} scenario analysis,'' ISO New
  England, Tech. Rep., 2017.

\bibitem{ABB:2018:00}
\BIBentryALTinterwordspacing
{\relax ABB Inc. Electric Systems Consulting}, ``Gridview -- an analytic tool
  for market simulation \& asset performance evaluations.'' [Online].
  Available:
  \url{https://library.e.abb.com/public/d25b0020b72d94eac1256fda00488560/GridView\%20Presentation.pdf}
\BIBentrySTDinterwordspacing

\bibitem{Ela:2009:00}
E.~Ela, M.~Milligan, B.~Parsons, D.~Lew, and D.~Corbus, ``The evolution of wind
  power integration studies: past, present, and future,'' in \emph{Power \&
  Energy Society General Meeting, 2009. PES'09. IEEE}.\hskip 1em plus 0.5em
  minus 0.4em\relax IEEE, 2009, pp. 1--8.

\bibitem{Holttinen:2012:01}
H.~Holttinen, M.~O. Malley, J.~Dillon, and D.~Flynn, ``Recommendations for wind
  integration studies -- {IEA} task 25,'' International Energy Agency,
  Helsinki, Tech. Rep., 2012.

\bibitem{Holttinen:2013:00}
H.~Holttinen, A.~Orths, H.~Abilgaard, F.~van Hulle, J.~Kiviluoma, B.~Lange,
  M.~OMalley, D.~Flynn, A.~Keane, J.~Dillon, E.~M. Carlini, J.~O. Tande,
  A.~Estanquiro, E.~G. Lazaro, L.~Soder, M.~Milligan, C.~Smith, and C.~Clark,
  ``Iea wind export group report on recommended practices wind integration
  studies,'' International Energy Agency, Paris, France, Tech. Rep., 2013.

\bibitem{Brouwer:2014:00}
A.~S. Brouwer, M.~van~den Broek, A.~Seebregts, and A.~Faaij, ``{Impacts of
  large-scale Intermittent Renewable Energy Sources on electricity systems ,
  and how these can be modeled},'' \emph{Renewable and Sustainable Energy
  Reviews}, vol.~33, pp. 443--466, 2014.

\bibitem{Brooks:2002:01}
D.~L. Brooks, E.~O. Lo, J.~W. Smith, J.~H. Pease, and M.~McGree, ``Assessing
  the impact of wind generation on system operations at xcel energy -- north
  and bonneville power administration,'' North and Bonneville Power
  Administraiton, Tech. Rep., 2002.

\bibitem{Ummels:2009:01}
B.~C. Ummels, ``Power system operation with large-scale wind power in
  liberalised environments,'' Ph.D. dissertation, Technical University of
  Delft, 2009.

\bibitem{Georgilakis:2008:00}
P.~S. Georgilakis, ``Technical challenges associated with the integration of
  wind power into power systems,'' \emph{Renewable and Sustainable Energy
  Reviews}, vol.~12, no.~3, pp. 852--863, Apr 2008.

\bibitem{Soder:2008:00}
L.~Soder, H.~Holttinen, and G.~E. Issues, ``On methodology for modelling wind
  power impact on power systems,'' \emph{International Journal of Global Energy
  Issues}, vol.~29, no. 1-2, pp. 181--198, 2008.

\bibitem{Bird:2012:00}
L.~Bird, M.~Milligan, and NREL, ``Lessons from large-scale renewable energy
  integration studies preprint,'' in \emph{2012 World Renewable Energy Forum},
  no. June, Denver, CO, United states, 2012, pp. ----8.

\bibitem{Holttinen:2008:00}
\BIBentryALTinterwordspacing
H.~Holttinen, M.~Milligan, B.~Kirby, T.~Acker, V.~Neimane, and T.~Molinski,
  ``{Using Standard Deviation as a Measure of Increased Operational Reserve
  Requirement for Wind Power},'' \emph{Wind Engineering}, vol.~32, no.~4, pp.
  355--377, 2008. [Online]. Available: \url{citeulike-article-id:8373954
  http://dx.doi.org/10.1260/0309-524X.32.4.355}
\BIBentrySTDinterwordspacing

\bibitem{Robitaille:2012:00}
A.~Robitaille, I.~Kamwa, A.~H. Oussedik, M.~de~Montigny, N.~Menemenlis,
  M.~Huneault, A.~Forcione, R.~Mailhot, J.~Bourret, and L.~Bernier,
  ``Preliminary impacts of wind power integration in the hydro-quebec system,''
  \emph{Wind Engineering}, vol.~36, no.~1, pp. 35--52, Feb 2012.

\bibitem{NERC:2012:00}
\BIBentryALTinterwordspacing
NERC, ``{Reliability Standards for the Bulk Electric Systems of North
  America},'' NERC--North American Electric Reliability Corporation, Tech.
  Rep., 2012. [Online]. Available:
  \url{http://www.nerc.com/files/Reliability\_Standards\_Complete\_Set\_1Dec08.pdf}
\BIBentrySTDinterwordspacing

\bibitem{Aigner:2012:00}
T.~Aigner, S.~Jaehnert, G.~L. Doorman, and T.~Gjengedal, ``The effect of
  large-scale wind power on system balancing in northern europe,'' \emph{IEEE
  Trans. Sustain. Energy}, vol.~3, no.~4, pp. 751--759, Oct 2012.

\bibitem{Ummels:2007:00}
\BIBentryALTinterwordspacing
B.~C. Ummels, M.~Gibescu, E.~Pelgrum, W.~L. Kling, and A.~J. Brand, ``{Impacts
  of Wind Power on Thermal Generation Unit Commitment and Dispatch},''
  \emph{IIEEE Transactions on Energy Conversion}, vol.~22, no.~1, pp. 44--51,
  2007. [Online]. Available: \url{http://dx.doi.org/10.1109/TEC.2006.889616
  http://ieeexplore.ieee.org/lpdocs/epic03/wrapper.htm?arnumber=4106021}
\BIBentrySTDinterwordspacing

\bibitem{Halamay:2011:00}
D.~A. Halamay, T.~K.~A. Brekken, A.~Simmons, and S.~McArthur, ``Reserve
  requirement impacts of large-scale integration of wind, solar, and ocean wave
  power generation,'' \emph{IEEE Trans. Sustain. Energy}, vol.~2, no.~3, pp.
  321--328, Jul 2011.

\bibitem{Hansen:2012:00}
C.~W. Hansen and A.~D. Papalexopoulos, ``Operational impact and cost analysis
  of increasing wind generation in the island of crete,'' \emph{IEEE Systems
  Journal}, vol.~6, no.~2, pp. 287--295, Jun 2012.

\bibitem{Luickx:2009:00}
P.~Luickx, E.~Delarue, and W.~D'haeseleer, ``Effect of the generation mix on
  wind power introduction,'' \emph{IET Renewable Power Generation}, vol.~3,
  no.~3, p. 267, 2009.

\bibitem{Albadi:2011:00}
M.~Albadi and E.~El-Saadany, ``Comparative study on impacts of wind profiles on
  thermal units scheduling costs,'' \emph{IET Renewable Power Generation},
  vol.~5, no.~1, p.~26, 2011.

\bibitem{Podmore:2010:00}
R.~Podmore and M.~R. Robinson, ``The role of simulators for smart grid
  development,'' \emph{IEEE Transactions on Smart Grid}, vol.~1, no.~2, pp.
  205--212, Sep 2010.

\bibitem{Muzhikyan:2013:SPG-C17}
\BIBentryALTinterwordspacing
A.~Muzhikyan, A.~M. Farid, and K.~Youcef-Toumi, ``{Variable Energy Resource
  Induced Power System Imbalances: A Generalized Assessment Approach},'' in
  \emph{IEEE Conference on Technologies for Sustainability}, Portland, Oregon,
  2013, pp. 1--8. [Online]. Available:
  \url{http://dx.doi.org.libproxy.mit.edu/10.1109/SusTech.2013.6617329}
\BIBentrySTDinterwordspacing

\bibitem{Muzhikyan:2013:SPG-C18}
\BIBentryALTinterwordspacing
------, ``{Variable Energy Resource Induced Power System Imbalances: Mitigation
  by Increased System Flexibility, Spinning Reserves and Regulation},'' in
  \emph{IEEE Conference on Technologies for Sustainability}, Portland, Oregon,
  2013, pp. 1--7. [Online]. Available:
  \url{http://dx.doi.org.libproxy.mit.edu/10.1109/SusTech.2013.6617292}
\BIBentrySTDinterwordspacing

\bibitem{Muzhikyan:2016:SPG-C55}
A.~Muzhikyan, A.~M. Farid, and T.~Mezher, ``{The Impact of Wind Power
  Geographical Smoothing on Operating Reserve Requirements},'' in \emph{IEEE
  American Control Conference}, Boston, MA, USA, 2016, pp. 1--6.

\bibitem{Jiang:2016:SPG-J30}
B.~Jiang, A.~Muzhikyan, A.~M. Farid, and K.~Youcef-Toumi, ``{Demand Side
  Management in Power Grid Enterprise Control -- A Comparison of Industrial and
  Social Welfare Approaches},'' \emph{Applied Energy (preprint in press)},
  vol.~1, no.~1, pp. 1--11, 2016.

\bibitem{Jiang:2015:SPG-J21}
\BIBentryALTinterwordspacing
B.~Jiang, A.~M. Farid, and K.~Youcef-Toumi, ``Demand side management in a
  day-ahead wholesale market a comparison of industrial and social welfare
  approaches,'' \emph{Applied Energy}, vol. 156, no.~1, pp. 642--654, 2015.
  [Online]. Available: \url{http://dx.doi.org/10.1016/j.apenergy.2015.07.014}
\BIBentrySTDinterwordspacing

\bibitem{Jiang:2015:SPG-C51}
\BIBentryALTinterwordspacing
B.~Jiang, A.~Muzhikyan, A.~M. Farid, and K.~Youcef-Toumi, ``Impacts of
  industrial baseline errors in demand side management enabled enterprise
  control,'' in \emph{IECON 2015 -- 41st Annual Conference of the IEEE
  Industrial Electronics Society}, Yokohama, Japan, 2015, pp. 1--6. [Online].
  Available: \url{http://dx.doi.org/10.1109/IECON.2015.7392637}
\BIBentrySTDinterwordspacing

\bibitem{Jiang:2015:SPG-C44}
\BIBentryALTinterwordspacing
B.~Jiang, A.~M. Farid, and K.~Youcef-Toumi, ``A comparison of day-ahead
  wholesale market: Social welfare vs industrial demand side management,'' in
  \emph{IEEE International Conference on Industrial Technology}, Sevilla,
  Spain, 2015, pp. 1--7. [Online]. Available:
  \url{http://dx.doi.org/10.1109/ICIT.2015.7125502}
\BIBentrySTDinterwordspacing

\bibitem{Jiang:2015:SPG-C45}
\BIBentryALTinterwordspacing
------, ``Impacts of industrial baseline errors on demand side management in
  day-ahead wholesale markets,'' in \emph{Proceedings of the ASME Power \&
  Energy 2015: Energy Solutions for Sustainable Future}, San Diego, CA, 2015,
  pp. 1--7. [Online]. Available:
  \url{http://engineering.dartmouth.edu/liines/resources/Conferences/SPG-C45.pdf}
\BIBentrySTDinterwordspacing

\bibitem{Wang:2012:00}
C.~Wang, Z.~Lu, and Y.~Qiao, ``{A Consideration of the Wind Power Benefits in
  Day-Ahead Scheduling of Wind-Coal Intensive Power Systems CaixiaWang, Student
  Member, IEEE, Zongxiang Lu, Member, IEEE,and},'' \emph{IEEE TRANSACTIONS ON
  POWER SYSTEMS}, pp. 1--10, 2012.

\bibitem{Apt:2007:01}
J.~Apt, ``The spectrum of power from wind turbines,'' \emph{Journal of Power
  Sources}, vol. 169, no.~2, pp. 369--374, jun 2007.

\bibitem{Curtright:2008:00}
A.~E. Curtright and J.~Apt, ``The character of power output from utility-scale
  photovoltaic systems,'' \emph{{Progress in Photovoltaics: Research and
  Applications}}, no. September 2007, pp. 241--247, 2008.

\bibitem{Monteiro:2009:00}
\BIBentryALTinterwordspacing
C.~Monteiro, R.~Bessa, V.~Miranda, A.~Botterud, J.~Wang, and G.~Conzelmann,
  ``{Wind Power Forecasting: State-of-the-Art 2009 Decision and Information
  Sciences Division},'' \emph{Argonne National Laboratory}, no. November 6, pp.
  1--216, 2009. [Online]. Available:
  \url{http://www.osti.gov/energycitations/product.biblio.jsp?osti\_id=968212}
\BIBentrySTDinterwordspacing

\bibitem{Moreno-Munoz:2008:00}
A.~Moreno-Munoz, J.~de~la Rosa, R.~Posadillo, and V.~Pallares, ``{Short term
  forecasting of solar radiation},'' in \emph{Industrial Electronics, 2008.
  ISIE 2008. IEEE International Symposium on}, 2008, pp. 1537--1541.

\bibitem{Holttinen:2012:00}
H.~Holttinen, M.~Milligan, E.~Ela, N.~Menemenlis, J.~Dobschinski, B.~Rawn,
  R.~J. Bessa, D.~Flynn, E.~Gomez-Lazaro, and N.~K. Detlefsen, ``Methodologies
  to determine operating reserves due to increased wind power,'' \emph{IEEE
  Trans. Sustain. Energy}, vol.~3, no.~4, pp. 713--723, Oct 2012.

\bibitem{Ela:2011:00}
E.~Ela, M.~Milligan, and B.~Kirby, ``Operating reserves and variable
  generation,'' \emph{Contract}, vol. 303, no. August, pp. 275--3000, 2011.

\bibitem{Rebours:2007:00}
Y.~G. Rebours, D.~S. Kirschen, M.~Trotignon, and S.~Rossignol, ``A survey of
  frequency and voltage control ancillary services---part i: Technical
  features,'' \emph{IEEE Transactions on power systems}, vol.~22, no.~1, pp.
  350--357, 2007.

\bibitem{CIGRE:2010:00}
CIGRE, ``Ancillary services: An overview of international practices technical
  brochure 435,'' CIGRE Working Group C5.06, Tech. Rep., October 2010.

\bibitem{Ela:2010:00}
E.~Ela, B.~Kirby, E.~Lannoye, M.~Milligan, D.~Flynn, B.~Zavadil, and
  M.~O'Malley, ``Evolution of operating reserve determination in wind power
  integration studies,'' in \emph{Power and Energy Society General Meeting,
  2010 IEEE}.\hskip 1em plus 0.5em minus 0.4em\relax IEEE, 2010, pp. 1--8.

\bibitem{ISO-NE:2017:04}
{ISO New England}, ``Iso new england reserve requirements,'' ISO New England
  Inc, Tech. Rep., 2017.

\bibitem{Frank:2012:00}
S.~Frank and S.~Rebennack, ``{A Primer on Optimal Power Flow: Theory,
  Formulation, and Practical Examples},'' \emph{{Colorado School of Mines,
  Tech. Rep}}, no. October, pp. 1--42, 2012.

\bibitem{Carpentier:1962:00}
J.~Carpentier, ``{Contribution to the economic dispatch problem},'' \emph{Bull.
  Soc. Franc. Electr.}, vol.~3, no.~8, pp. 431--447, 1962.

\bibitem{Stott:2009:00}
B.~Stott, J.~Jardim, and O.~Alsa\c{c}, ``{DC Power Flow Revisited},''
  \emph{IEEE TRANSACTIONS ON POWER SYSTEMS}, vol.~24, no.~3, pp. 1290--1300,
  2009.

\bibitem{ISO-New-England:2016:00}
\BIBentryALTinterwordspacing
{ISO New England}, ``{CELT} report: 2016--2025 forecast report of capacity,
  energy, loads, and transmission,'' ISO New England, Tech. Rep., 2016.
  [Online]. Available:
  \url{https://www.iso-ne.com/static-assets/documents/2016/05/2016_celt_report.xls}
\BIBentrySTDinterwordspacing

\bibitem{ISO-New-England:2016:01}
------, ``Renewable portfolio standards spreadsheet,'' ISO New England, Tech.
  Rep., 2016.

\bibitem{:2017:00}
\BIBentryALTinterwordspacing
(2017) Interconnection request queue. [Online]. Available:
  \url{https://www.iso-ne.com/system-planning/transmission-planning/interconnection-request-queue}
\BIBentrySTDinterwordspacing

\bibitem{iso-ne:2018:00}
\BIBentryALTinterwordspacing
{ISO NEWSWIRE}. (2018, {March}) {A regional first: New Englanders used less
  grid electricity midday than while they were sleeping on April 21}. [Online].
  Available:
  \url{http://isonewswire.com/updates/2018/5/3/a-regional-first-new-englanders-used-less-grid-electricity-m.html}
\BIBentrySTDinterwordspacing

\end{thebibliography}
\end{document}